\documentclass[%
reprint,
amsmath,amssymb,
aps,
prd,onecolumn,nofootinbib
]{revtex4-2}

\usepackage{lineno}
\usepackage[utf8]{inputenc}
\usepackage[top=3.cm,bottom=3.cm,left=3.cm,right=3.cm]{geometry}

\usepackage{titlesec}
\titlespacing*{\section}
{0pt}{2ex}{2ex}
\titlespacing*{\subsection}
{0pt}{1.5ex}{1.5ex}
\titlespacing*{\subsubsection}
{0pt}{1.5ex}{1.5ex}
\usepackage{titlesec}
\titlespacing*{\paragraph}{0pt}{10pt plus 2pt minus 1pt}{1em}

\usepackage[T1]{fontenc}
\usepackage[english]{babel}
\usepackage{graphicx}
\usepackage[font=scriptsize]{subcaption}
\usepackage[font=small]{caption}
\usepackage{pgf}
\usepackage{tikz}
\usepackage{physics}
\usetikzlibrary{arrows.meta}
\usepackage{mathtools}
\usepackage{sidecap}
\usepackage{xfrac}
\usepackage{tikz-feynman}
\usepackage{slashed}
\usepackage{hyperref}
\allowdisplaybreaks

\tikzset{
    ->-/.style={decoration={
  markings,
  mark=at position .6 with {\arrow{>}}},postaction={decorate}},
    -<-/.style={decoration={
  markings,
  mark=at position .5 with {\arrow{<}}},postaction={decorate}},
}

\definecolor{electron-purple}{RGB}{142,10,161}

\tikzset{%
	lineA/.style={%
		black,line width=0.5mm,
	},
    lineEl/.style={%
		electron-purple,line width=0.5mm,
	},
	lineB/.style={
	black!70!white, dotted, line width=0.4mm
},
	lineC/.style={
	draw= white, line width=2mm
},
        lineD/.style={
	black, line width=0.5mm, ->-
},
cutA/.style={
draw= red, line width=0.4mm
},
cutAB/.style={
	draw= red!50!black, line width=0.4mm
},
cutAC/.style={
	draw= red!75!black, line width=0.4mm
},
cutB/.style={
	draw=blue , line width=0.4mm
},
cutBB/.style={
	draw=blue!60!black , line width=0.4mm
},
cutBC/.style={
	draw=blue!80!black , line width=0.4mm
}
}

\usetikzlibrary{trees}
\usetikzlibrary{decorations.pathmorphing}
\usetikzlibrary{intersections,calc,shapes.geometric}
\usetikzlibrary{decorations.markings}
\usetikzlibrary{external}
\usetikzlibrary{intersections}
\usetikzlibrary{shapes,arrows}
\usetikzlibrary{arrows.meta}
\usetikzlibrary{calc}
\usetikzlibrary{shapes.misc}
\usetikzlibrary{decorations.text}
\usetikzlibrary{backgrounds}
\usetikzlibrary{intersections,calc}
\usetikzlibrary{decorations.pathreplacing,decorations.markings}
\usetikzlibrary{positioning}
\usetikzlibrary{shapes.misc}

\tikzset{
  main node/.style={
    shape=ellipse,                 
    fill=blue!20,
    draw,
    pattern=crosshatch dots,
    pattern color=black!20!white,
    minimum width=2cm,             
    minimum height=1.2cm,          
    inner sep=0pt,
    line width=0.4mm
  },
}

\newcommand{\pent}{\scalebox{0.5}{\begin{tikzpicture}[scale=0.01]
		\node[regular polygon,
		draw,
		regular polygon sides = 5,line width=0.03cm] (p) at (0,0) {};
\end{tikzpicture}}}
\renewcommand{\imath}{\mathrm{i}}

\newcommand{\q}{\ensuremath{\mathfrak{q}}}
\newcommand{\dPi}{\mathrm{d}\overline{\Pi}}

\newcommand{\emb}{(G,\wind)}
\newcommand{\embs}{\Gamma}
\newcommand{\embsI}{\Gamma}
\newcommand{\embsF}{\bar{\Gamma}}
\newcommand{\edges}{\mathcal{E}}
\newcommand{\vertices}{\mathcal{V}}

\newcommand{\allcycles}{\mathcal{C}_u}
\newcommand{\alldircycles}{\mathcal{C}}
\newcommand{\ZZ}{\mathbb{Z}}
\newcommand{\QQ}{\mathbb{Q}}
\newcommand{\RR}{\mathbb{R}}

\newcommand{\NN}{\mathbb{N}}
\newcommand{\transpose}[1]{#1^{\intercal}}
\newcommand{\partitle}[1]{\paragraph{\hspace{-2mm}#1}}
\newcommand{\undG}{G_u}
\newcommand{\dirG}{G}
\newcommand{\embdis}{\mathrm{Emb}_\mathrm{dis}}
\newcommand{\wind}{\mathbf{w}}
\newcommand{\cycle}{c}
\newcommand{\basis}{\mathcal{B}}

\newcommand{\basecycle}{b}
\newcommand{\orientedbasecycle}{d}

\DeclarePairedDelimiterX\Set[1]{\{}{\}}{#1}

\newcommand{\ObigLambda}{\mathcal{O}(\Lambda^2)}
\newcommand{\ObigPsq}{\mathcal{O}((p^2)^0)}

\usepackage{dsfont}

\usepackage{pdfpages}

\makeatletter
\AtBeginDocument{\let\LS@rot\@undefined}
\makeatother

\begin{document}

\title{Multi-partonic interactions, iterated discontinuities and the virtuality expansion in deep inelastic scattering}

\author{Zeno Capatti}
\email{zeno.capatti@unibe.ch}
\author{Lucien Huber}
\email{mail@lucien.ch}
\affiliation{Institute for Theoretical Physics, \\
 University of Bern, \\
Bern, Switzerland}

\author{Michael Ruf}
\email{mruf@slac.stanford.edu}
\affiliation{Stanford Linear Accelerator Center \\
Menlo Park, United States}

\date{\today}

\begin{abstract}
We introduce a perturbative model that accounts for the contribution of multi-partonic interactions to collider observables.
A key feature of this multi-parton model is that cross sections are organised in terms of building blocks that are separately infrared-finite in virtue of the KLN theorem.
We find compact expressions for these building blocks in terms of double discontinuities of Feynman-type integrals 
obtained by endowing a vacuum graph with additional topological information.
The framework is applied to the computation of next-to-leading order structure functions in deep inelastic scattering.
We comment on the scale dependence of these structure functions and discuss the possibility of relating them by a scheme change to those computed in the parton model.
Finally we lay down next steps for analogous computations
for deep inelastic scattering at NNLO and Drell-Yan at NLO.
\end{abstract}

\maketitle

\pagebreak

\tableofcontents

\newpage

\section{Introduction}

The parton model is a cornerstone of the theoretical framework used to predict the outcome of scattering events involving protons as initial states, a core element of the success of quantum chromodynamics as a theory of strong interactions. It finds its historical roots in the proposal of the \emph{naive} parton model~\cite{osti_4153889,Gribov:1968ia} and Bjorken's work on the scaling properties of deep inelastic scattering structure functions~\cite{Bjorken:1968dy}, and an attempt at rigorous incorporation in the QCD framework through the operator-product expansion arguments proposed in the Nobel-winning papers by Gross, Wilczek and Politzer~\cite{PhysRevD.8.3633,DAVIDPOLITZER1974129}. Their work gave strong foundations to the intuitive idea that scattering of hadrons can be understood in terms of their constituents by demonstrating that leading-twist Wilson coefficients of the operator product expansion of the deep inelastic cross-section are independent of the way the proton is modeled, as long as reasonable analyticity constraints are satisfied, and can be computed perturbatively. The non-perturbative information is encoded in the expectation value of operators in the proton state. Concurrent foundational developments include the parametrisation of scaling-violating effects in terms of renormalisation group equations~\cite{Parisi:1973nx,Altarelli:1977zs,Dokshitzer:1977sg,Gribov:1972ri} (see~\cite{Altarelli:1981ax,Forte:2025twm,Parisi:2025nob} for detailed historical accounts of the birth of the parton model, as well as~\cite{Sterman:1993hfp,Collins:2011zzd,Ellis:1996mzs} for classic textbook presentations of the subject).

These developments, and the interest in the application of the parton model to more general processes, including proton-proton scattering, eventually gave rise to the ``improved'' parton model that is today a staple of perturbative predictions for colliders such as the LHC. The scattering between two hadrons or one hadron and an electron is described in terms of the scattering of one constituent parton from each proton, weighted by a distribution, the parton distribution functions, expressing the probability of extracting a parton from the proton with a certain fraction of the proton's momentum. For leading-twist inclusive deep inelastic scattering, this picture is equivalent to the OPE construction, assuming a specific choice of state in the computation of Wilson coefficients. Thus, the multiplicity of initial-state massless particles is fixed to be the same as the number of protons involved in the collision. Conversely, one allows a varying multiplicity of final-state massless particles, which are clustered into jets.

This asymmetry in the treatment of initial and final states gives rise to a somewhat dissonant situation. The sum over final-state multiplicities allows one to leverage the KLN theorem~\cite{Kinoshita:1962ur,PhysRev.133.B1549}, which then guarantees that all final-state singularities are canceled (so that, for example, QCD corrections to leptonic collisions are infrared finite). However, for the same reason, fixing the number of initial-state massless particles breaks the cancellation of initial-state singularities that would otherwise be present if one accounted for all degenerate (collinear, soft) initial-states (see ref.~\cite{Agarwal:2021ais} for an account of this asymmetry and its historical/scientific roots). 

The initial-state singularities, once dimensionally-regulated, factorise, together with the logarithms that they produce~\cite{Ellis:1978ty,Altarelli:1979ub,Altarelli:1981ax,Collins:1983ju,Collins:1985ue,Bodwin:1981fv,Bodwin:1984hc,Sterman:1995fz}. The singularities are re-absorbed in a redefinition of the parton distribution functions and, by the same process, the logarithms are resummed. The parton distribution functions, which then parametrise the non-perturbative dynamics of the proton at leading twist, are fit to experiment. Thanks to the universality of the infrared poles and logarithms that they absorb, they are process independent, which ensures predictivity. There is no community-wide consensus that the ``improved'' parton model follows directly from first-principles QCD (in an analogous way as OPE proofs did for inclusive deep inelastic scattering), although much work has been done on providing a first-principles basis to the improved parton model. Regardless, the structure that underlies the factorisation of infrared singularities and infrared logarithms as well as its physical implications are seen by most as a strong-enough motivation for the correctness of the procedure.

For this precise reason factorisation of infrared singularities has been a core subject of research in the decades after the formulation of QCD. A decisive contribution in this direction was provided by the diagrammatic proofs of factorisation of Collins, Soper and Sterman~\cite{Collins:1981ta,Collins:1983ju,Collins:1983pk,Bodwin:1984hc,Sterman:1995fz}, cementing future studies. While the factorisation of collinear initial-state poles is, by-and-large, accepted and understood, the parton model also crucially relies on the disappearance of soft logarithms that describe the long-range interactions between initial-state protons as well as protons and final-state jets~\cite{Landshoff:1982ep,Bodwin:1981fv,Collins:1981ta,Rothstein:2016bsq,Aybat:2008ct,Diehl:2015bca}. It is in order to precisely understand these long-range interactions that one studies factorisation-breaking effects, associated with the non-cancellation of soft logarithms, especially beyond next-to-leading order~\cite{Catani:2011st,Rothstein:2016bsq,Caola:2020xup,Zeng:2015iba,Liu:2019tuy,Forshaw:2012bi,Bauer_2002,becher2024factorizationrestorationglaubergluons}. The detailed mechanics of factorisation have been and still are today the topic of wide-ranging research, as well as the motivation behind the development of systematic frameworks such as soft-collinear effective field theory~\cite{Bauer_2002,Beneke_2002,Becher_2016,Rothstein:2016bsq}. 

 Extensions of the parton model include the definition of transverse-momentum-dependent parton distribution functions~\cite{Collins:1981uk,Collins:1984kg,Ji:2004wu}, which also account for the transverse dynamics of the parton that is singled out of the proton. Other generalisations include the parametrisation of multi-partonic interactions~\cite{Radyushkin_1997,Ji_1997,Diehl_2003,Diehl_2011,Blok_2012,Diehl_2012,Blok_2014,Andersen:2023hzm}: instead of assuming that only one parton from each hadron enters the hard scattering, one considers the contribution from more than one constituent partons entering the scattering, all exactly collinear to the originating hadron. These effects are mostly argued to be power-suppressed, and only relevant in particular regions of phase-space. Roughly speaking, their contribution to the cross-section is written as an $n$-convolution integral in each of the momentum-fractions of the partons. These contributions are computed separately from the single parton case.

\begin{center}
*
\end{center}

In this paper, we take a different approach to the perturbative modelling of hadrons and multi-partonic interactions, following a remark made in a seminal paper by Sterman and Weinberg~\cite{Sterman:1977wj}. The underlying inspiration comes from the Kinoshita-Lee-Nauenberg theorem which, as previously mentioned, states that a cross-section is finite once all multiplicities of massless particles in the initial and final state are summed over. The KLN theorem naturally provides a framework to treat single and multi-parton contributions on the same footing, constraining heavily the allowed dynamics of the constituent partons in the process. The kinematic distribution of the multi-partonic initial-state is described by an initial-state density. In order for the cross-section to be infrared finite, the density must be infrared-safe. When one combines all diagrams following the KLN theorem and imposes the initial-state density to be such that cancellations occur, one is left with an infrared-finite quantity.

The infrared-safety constraint is not enough to completely define the kinematic properties of the constituent partons. However, as we describe in sect.~\ref{sec:MPI_overview}, the need to specify this additional information disappears once one defines a suggestive virtuality expansion within the model. Under the assumption that the invariant mass of the cluster of scattering partons is small with respect to the hard scale of the process, infrared-safety imposes any density of initial states to collapse to a function that only depends on the kinematics of the proton. What one obtains is a clustering choice that only imposes two constraints: (a) that the momenta of the initial state partons sum to the momentum of the cluster, given as an input, and (b) that the invariant mass of the cluster is lower than a certain scale $\Lambda^2$, subject to expansion. We refer to this clustering choice as \emph{the simplest clustering}. This choice varies from other ones made in analogous attempts in the literature by the fact that the phase-space of the scattering partons is extremely simple. Previous work such as~\cite{Axelrod:1985yi}, that details the computation of $F_2$ restricting to DIS diagrams that fall into class one and three embeddings in the classification of sect.~\ref{sec:four_types_of_embeddings}, uses a more traditional jet definition involving a soft and collinear scale. On the other hand, ref.~\cite{Hannesdottir:2022xki} details a calculation of Drell-Yan diagrams using a hemisphere-jet definition. Contrasted with the type of loop integrals obtained in these two works, ours never feature theta functions other than those entering the definition of plus/minus Dirac delta distributions, which makes them especially easy to perform. In other words, the objects at the core of our computation are ``simple'', manifestly-covariant, cut Feynman integrals.

Having defined a way to cluster multi-partonic initial-states into a hadron, one proceeds to list all the diagrams contributing to the cross-section. It should be noted that current approaches to the calculation of fixed-order collider cross-sections only include a subset of the diagrams that in principle participate in the S-matrix. Only connected amplitudes, with no spectators, are argued to contribute to the observable. Once one includes sums over initial-state multiplicities and relaxes the connectedness constraint (in part to realise the KLN theorem, which requires the inclusion of spectators, in part out of sheer curiosity), one finds what could be called a ``swampland'' of diagrams, whose exact treatment is dubious. This problem is known and has been investigated to some extent in the past~\cite{Lavelle_2006,Blazek:2021zoj,Frye:2018xjj,khalil2017completediagrammaticimplementationkinoshitaleenauenberg,Capatti:2023omc}. In sect.~\ref{sec:graph_embeddings}, we provide a strong criterion for understanding these diagrams: we will classify them in terms of their IR-cancellation property, naturally formalised in terms of vacuum graph embeddings. This in turn provides a top-down approach to generating the relevant graphs and identifying \emph{gauge-invariant and infrared finite} subclasses of them. 

Turning to the computation of these diagrams, we show in sect.~\ref{sec:MPI_and_discs} that, provided one describes initial states by the simplest clustering criterion, the interference diagrams identified by the same embedding can be entirely written in terms of iterated discontinuities of a Feynman-diagram-like object associated with the embedding itself. We call this result a ``doubled'' optical theorem. We believe this to be one of the key insights of our work, as the appropriate generalisation of the simplest clustering criterion to more complicated cases (such as hadron collisions) would ensure the scalability of the method that we propose here. Indeed a crucial aspect of the framework we propose is that cross-section calculations within it appear to be approachable by techniques available today, including higher orders. Even more, it seems possible that the calculation may be done either a) purely analytically or b) purely numerically in the Local Unitarity formalism~\cite{Soper:1998ye,Soper:1999xk,Capatti:2020xjc,Capatti:2021bsm,Capatti:2022tit}. 

The possibility of performing all integrals analytically is a byproduct, again, of the choice of simplest clustering. It enables one to write cross-sections in terms of simple cut integral, objects to which traditional methods (such as integration-by-parts identities and differential equations) can be applied. In particular, the type of integrals that we encounter can be written as cuts of Feynman diagram, with no additional distribution involved in their definition. This makes it so that they can be computed straight-forwardly within a Reverse Unitarity approach~\cite{Anastasiou:2002qz,Anastasiou:2002yz,Anastasiou:2003yy,Anastasiou:2013mca,Anastasiou:2013srw}. 

Conversely, the possibility of instead computing the cross-section by direct numerical integration is guaranteed by the infrared-finiteness of the cross-section. Indeed, one of the biggest conceptual and computational obstacles to the automation of the calculation of collider cross-sections is the treatment of their infrared singularities. Loop integrals are computed in dimensional regularisation, regulating both UV and IR poles, while phase-space integrals are computed numerically by introducing infrared counter-terms, since the presence of observables makes analytic integration unfeasible. The bare cross-section is anyway expected to have infrared poles, and becomes finite only after parton distribution functions are renormalised. Instead, the MP cross-section is infrared-finite from the start. Infrared-finite classes of interference diagrams can be combined together at the local level, giving an integrand without infrared singularities. This is the objective of the Local Unitarity program. 

\begin{center}
*
\end{center}
\newpage
A natural byproduct of a community-wide focus shift to so-called ``precision physics'' is a request for tight theoretical control of uncertainties associated with predictions~\cite{Bagnaschi:2014wea,Bonvini:2020xeo,Duhr:2021mfd,Tackmann:2024kci,Lim:2024nsk}. Theoretical sources of uncertainty include the error associated with the truncation of the perturbative series and the inability to systematically account for non-perturbative effects, as well as the myriad of assumptions that underlie the framework we use to make predictions. Current methods for estimating such effects, e.g. scale variation, have been proven to be inadequate in a host of situations. It seems unlikely that this problem can be addressed directly and completely in the near future, due to our lack of control over the assumptions. However, there exists a rather strong, indirect way to probe the effect of these assumptions, namely to construct a model for hadronic scattering that satisfies the same broad physical properties as the current one while being substantially different in its formulation. We think this is one of the strongest motivations of our work.

The results shown in sect.~\ref{sec:results} may be summarised as follows: at next-to-leading order, the model that we propose, for inclusive deep inelastic scattering, is equivalent to the parton model. The scale evolution of parton distribution functions is the same and the finite part that differs between the models may be re-absorbed in a renormalisation scheme re-definition. Ultimately, however, equivalence between the models should be established for a variety of processes. This would truly constitute an operational proof of the equivalence: the same scheme re-definition that translates our computation of deep inelastic scattering to the parton model's should be utilised in every other process. For this reason, in sect.~\ref{sec:beyond_NLO_DIS}, we develop the first steps toward generalising our method to higher orders as well as proton collisions.

\section{Overview: multi-partonic initial states in deep inelastic scattering}
\label{sec:MPI_overview}

\subsection{A scalar example}

We start by giving a concise example of the procedure carried out in this paper. We consider a massless scalar theory with interaction $-\lambda \phi^3/3!$. We wish to compute the inclusive process $\phi^\star \phi\rightarrow X$, where $X$ is any collection of massless scalars. $\phi^\star$ is an off-shell space-like scalar that has the same role as the photon in deep inelastic scattering. The analogue of the parton model (PM) partonic tensor reads
\begin{equation}
W_{\text{PM}}(p,q)=\sum_{n=1}^\infty \int \mathrm{d}\Pi_n |\bra{\phi(p_1')...\phi(p_n')} S \ket{\phi^\star(q)\phi(p)}|^2,
\end{equation}
where
\begin{equation}
\label{eq:n_particle_phase_space}
\mathrm{d}\Pi_n =\prod_{i=1}^n \frac{\mathrm{d}^d p_i'}{(2\pi)^{d}}\tilde{\delta}^+(p_i'), \quad \tilde{\delta}^\pm (q)=(2\pi \imath)\delta(q^2)\Theta(\pm q^0).
\end{equation}
In this paper we will consider the following extension:
\begin{equation}
\label{eq:scalar_W}
W(p,q)=\sum_{n,m=1}^\infty \int \mathrm{d}\Pi_n \mathrm{d}\Pi_m(2\pi)^4\delta\left(p-p_{1m}\right) |\bra{\phi(p_1')...\phi(p_n')} S \ket{\phi^\star(q)\phi(p_1)...\phi(p_m)}|^2,
\end{equation}
with $p_{1m}=p_1+...+p_m$. We cluster the initial-state massless particles $\phi(p_1),...,\phi(p_m)$ into a compound object of momentum $p$ by momentum-conservation alone (this is what we will soon introduce as the \emph{simplest clustering criterion}). The virtuality-integrated partonic tensor is obtained from $W(p,q)$ by first setting 
\begin{equation}
p=\xi(x)\hat{n}_++\frac{p^2}{4\xi(x)}\hat{n}_-, \quad \xi(x)=\frac{-q^2}{2x(n_+\cdot q)}\,.
\end{equation}
where $x$ is the Bjorken variable, $\hat{n}_{\pm}+$ are the light-cone vectors satisfying $\hat{n}_+\cdot \hat{n}_-=2$, and then integrating over $p^2$ up to the maximum value of $\Lambda^2$:
\begin{equation}
W(x,\Lambda^2,q^2)=\frac{1}{4\pi}\int_0^{\Lambda^2} \frac{\mathrm{d}p^2}{2\pi \imath} W(p,q),
\end{equation}
where $4\pi$ is the flux factor and the normalisation of $2\pi \imath$ for the integration in $p^2$ makes sure that integrating $W_{\text{PM}}(p,q)$ against the same measure gives the parton model's partonic tensor. 

Finally, we perform what we call the \emph{virtuality expansion} of this object, namely the expansion in $\Lambda^2$, and define the leading-virtuality contribution by the truncation $W(x,\Lambda^2,q^2)=W^{(0)}(x,\Lambda^2,q^2)+\ObigLambda$. Observe that, since $p$ is a sum of light-like vectors $p=p_1+...+p_m$ and since $p^2$, the virtuality of the cluster, is bounded from above by $\Lambda^2$, the limit $\Lambda^2\rightarrow 0$ imposes all momenta $p_1,...,p_m$ to be \emph{collinear} to each other and to the input direction $p$, and $p$ itself to be $p=\xi(x)\hat{n}_+$.

The squared amplitude with $m>1$ initial states and $n$ final states gives rise to many interference diagrams, some of which appear rather exotic. In this paper, we will provide a full treatment for them, but for now let us focus on a class of diagrams that arises from different cuts of the triangle diagram that is also well-known from the parton model:
\begin{align}
W_{\triangle}(p,q)=&\raisebox{-1cm}{\begin{tikzpicture}

    \node[inner sep=0pt] (U1) {};
    \node[inner sep=0pt] (U2) [above right = 1cm and 1cm of U1] {};
    \node[inner sep=0pt] (U3) [right = 2cm of U2] {};
    \node[inner sep=0pt] (U4) [below right = 1cm and 1cm of U3] {};

    \node[inner sep=0pt] (U12) [above right = 0.45cm and 0.45cm of U1] {};
    \node[inner sep=0pt] (U23) [right = 0.75cm of U2] {};

    \node[inner sep=0pt] (D2) [above = 0.5cm of U2]{};
    \node[inner sep=0pt] (D3) [above = 0.5cm of U3] {};

    \node[inner sep=0pt] (UP1) [left = 1cm of D2]{};
    \node[inner sep=0pt] (UP2) [right = 1cm of D3] {};

    \node[inner sep=0pt] (C1) [above right = 0.75cm and 1cm of U2]{};
    \node[inner sep=0pt] (C2) [below right = 1.25cm and 1cm of U2] {};

        \begin{feynman}
        \draw[-, thick, black, line width=0.5mm, momentum=\(p\)] (U1) to (U12);
        \draw[-, thick, black!70!white, dotted, line width=0.4mm, momentum=\(q\)] (UP1) to (U2);
    	
    	\end{feynman}
    	
        \draw[-, thick, black, line width=0.5mm] (U12) to (U2);
     \draw[-, thick, black, line width=0.5mm] (U2) to (U3);
     \draw[-, thick, black, line width=0.5mm] (U3) to (U4);
     
     \draw[-, thick, black, line width=0.5mm, out=-30, in=-90] (U12) to (U23);

     \draw[-, thick, black!70!white, dotted, line width=0.4mm] (UP2) to (U3);

     \draw[-, thick, blue, line width=0.4mm] (C1) to (C2);

	\path[draw=black, fill=black] (U2) circle[radius=0.05];
 \path[draw=black, fill=black] (U12) circle[radius=0.05];
 \path[draw=black, fill=black] (U23) circle[radius=0.05];
 \path[draw=black, fill=black] (U3) circle[radius=0.05];

\end{tikzpicture}}+\raisebox{-1cm}{\begin{tikzpicture}

    \node[inner sep=0pt] (U1) {};
    \node[inner sep=0pt] (U2) [above right = 1cm and 1cm of U1] {};
    \node[inner sep=0pt] (U3) [right = 2cm of U2] {};
    \node[inner sep=0pt] (U4) [below right = 1cm and 1cm of U3] {};

    \node[inner sep=0pt] (U12) [above right = 0.45cm and 0.45cm of U1] {};
    \node[inner sep=0pt] (U23) [right = 0.75cm of U2] {};

    \node[inner sep=0pt] (D2) [above = 0.5cm of U2]{};
    \node[inner sep=0pt] (D3) [above = 0.5cm of U3] {};

    \node[inner sep=0pt] (UP1) [left = 1cm of D2]{};
    \node[inner sep=0pt] (UP2) [right = 1cm of D3] {};

    \node[inner sep=0pt] (C1) [above right = 0.75cm and 0.4cm of U2]{};
    \node[inner sep=0pt] (C2) [below right = 1.25cm and 0.4cm of U2] {};

        \begin{feynman}
        \draw[-, thick, black, line width=0.5mm, momentum=\(p\)] (U1) to (U12);
        \draw[-, thick, black!70!white, dotted, line width=0.4mm, momentum=\(q\)] (UP1) to (U2);
    	
    	\end{feynman}
    	\draw[-, thick, black, line width=0.5mm] (U12) to (U2);
     \draw[-, thick, black, line width=0.5mm] (U2) to (U3);
     \draw[-, thick, black, line width=0.5mm] (U3) to (U4);
     
     \draw[-, thick, black, line width=0.5mm, out=-30, in=-90] (U12) to (U23);

     \draw[-, thick, black!70!white, dotted, line width=0.4mm] (UP2) to (U3);

     \draw[-, thick, blue, line width=0.4mm] (C1) to (C2);

	\path[draw=black, fill=black] (U2) circle[radius=0.05];
 \path[draw=black, fill=black] (U12) circle[radius=0.05];
 \path[draw=black, fill=black] (U23) circle[radius=0.05];
 \path[draw=black, fill=black] (U3) circle[radius=0.05];

\end{tikzpicture}} 
\nonumber\\
+&\raisebox{-1cm}{\begin{tikzpicture}

    \node[inner sep=0pt] (U1) {};
    \node[inner sep=0pt] (U2) [above right = 1cm and 1cm of U1] {};
    \node[inner sep=0pt] (U3) [right = 2cm of U2] {};
    \node[inner sep=0pt] (U4) [below right = 1cm and 1cm of U3] {};

    \node[inner sep=0pt] (U12) [above right = 0.45cm and 0.45cm of U1] {};
    \node[inner sep=0pt] (U23) [right = 0.75cm of U2] {};

    \node[inner sep=0pt] (D2) [above = 0.5cm of U2]{};
    \node[inner sep=0pt] (D3) [above = 0.5cm of U3] {};

    \node[inner sep=0pt] (UP1) [left = 1cm of D2]{};
    \node[inner sep=0pt] (UP2) [right = 1cm of D3] {};

    \node[inner sep=0pt] (C1) [above right = 0.75cm and 1.5cm of U2]{};
    \node[inner sep=0pt] (C2) [below right = 1.25cm and 1.5cm of U2] {};

    \node[inner sep=0pt] (UM1) [above right = 0.5cm and 0.5cm of U1] {};
    \node[inner sep=0pt] (UM2) [above left = 0.5cm and 0.5cm of U4] {};
    \node[inner sep=0pt] (E1) [above=0.3cm of U1] {};
    \node[inner sep=0pt] (E2) [above=0.3cm of U4] {};

        \begin{feynman}
        \draw[-, thick, black, line width=0.5mm, momentum=\(p_1\)] (E1) to (U2);
        \draw[-, thick, black, line width=0.5mm, reversed momentum=\(p-p_1\)] (U23) to (U1);
        \draw[-, thick, black!70!white, dotted, line width=0.4mm, momentum=\(q\)] (UP1) to (U2);
    	
    	\end{feynman}
    	
     \draw[-, thick, black, line width=0.5mm] (U2) to (U3);
     \draw[-, thick, black, line width=0.5mm] (U3) to (UM2);
     
     \draw[-, thick, black, line width=0.5mm] (U1) to (U23);

     \draw[-, thick, black, line width=0.5mm] (UM2) to (E2);
     \draw[-, thick, black, line width=0.5mm] (UM2) to (U4);

     \draw[-, thick, black!70!white, dotted, line width=0.4mm] (UP2) to (U3);

     \draw[-, thick, blue, line width=0.4mm] (C1) to (C2);

	\path[draw=black, fill=black] (U2) circle[radius=0.05];
 \path[draw=black, fill=black] (UM2) circle[radius=0.05];
 \path[draw=black, fill=black] (U23) circle[radius=0.05];
 \path[draw=black, fill=black] (U3) circle[radius=0.05];

\end{tikzpicture}} +\raisebox{-1cm}{\begin{tikzpicture}

    \node[inner sep=0pt] (U1) {};
    \node[inner sep=0pt] (U2) [above right = 1cm and 1cm of U1] {};
    \node[inner sep=0pt] (U3) [right = 2cm of U2] {};
    \node[inner sep=0pt] (U4) [below right = 1cm and 1cm of U3] {};

    \node[inner sep=0pt] (U12) [above right = 0.45cm and 0.45cm of U1] {};
    \node[inner sep=0pt] (U23) [right = 0.75cm of U2] {};

    \node[inner sep=0pt] (D2) [above = 0.5cm of U2]{};
    \node[inner sep=0pt] (D3) [above = 0.5cm of U3] {};

    \node[inner sep=0pt] (UP1) [left = 1cm of D2]{};
    \node[inner sep=0pt] (UP2) [right = 1cm of D3] {};

    \node[inner sep=0pt] (C1) [above right = 0.75cm and 0.4cm of U2]{};
    \node[inner sep=0pt] (C2) [below right = 1.25cm and -0.1cm of U2] {};

    \node[inner sep=0pt] (UM1) [above right = 0.5cm and 0.5cm of U1] {};
    \node[inner sep=0pt] (UM2) [above left = 0.5cm and 0.5cm of U4] {};
    \node[inner sep=0pt] (E1) [above=0.3cm of U1] {};
    \node[inner sep=0pt] (E2) [above=0.3cm of U4] {};

        \begin{feynman}
        \draw[-, thick, black, line width=0.5mm, momentum=\(p_1\)] (E1) to (U2);
        \draw[-, thick, black, line width=0.5mm, reversed momentum=\(p-p_1\)] (U23) to (U1);
        \draw[-, thick, black!70!white, dotted, line width=0.4mm, momentum=\(q\)] (UP1) to (U2);
    	
    	\end{feynman}
     \draw[-, thick, black, line width=0.5mm] (U2) to (U3);
     \draw[-, thick, black, line width=0.5mm] (U3) to (UM2);
     

     \draw[-, thick, black, line width=0.5mm] (UM2) to (E2);
     \draw[-, thick, black, line width=0.5mm] (UM2) to (U4);

     \draw[-, thick, black!70!white, dotted, line width=0.4mm] (UP2) to (U3);

     \draw[-, thick, blue, line width=0.4mm] (C1) to (C2);

	\path[draw=black, fill=black] (U2) circle[radius=0.05];
 \path[draw=black, fill=black] (UM2) circle[radius=0.05];
 \path[draw=black, fill=black] (U23) circle[radius=0.05];
 \path[draw=black, fill=black] (U3) circle[radius=0.05];

\end{tikzpicture}}\label{eq:triangle_diag_2nd_line}
\end{align}

The dashed line, in the above drawings, corresponds to the off-shell scalar $\phi^\star(q)$.
The momentum-conservation delta in eq.~\eqref{eq:scalar_W} forces the initial-state massless particles to have momenta that sum up to $p$, and the routing showed in the diagrams indeed satisfies this constraint. When only one initial-state parton is present, such as in the first two diagrams, its momentum is set to $p$, $p_1=p$. When two initial-state partons are present, their sum is set to be $p$, $p_1+p_2=p$, which leaves one independent momentum, say $p_1$, and constrains the other to be $p_2=p-p_1$. In the above diagrammatic representation, the independent variable is assumed to be integrated, cut particles are on-shell (including all initial-state solid lines) and their contribution to the integrand follows usual cutting rules. In particular, the integral corresponding to the second diagram of eq.~\eqref{eq:triangle_diag_2nd_line} is
\begin{equation}
\raisebox{-1cm}{\begin{tikzpicture}

    \node[inner sep=0pt] (U1) {};
    \node[inner sep=0pt] (U2) [above right = 1cm and 1cm of U1] {};
    \node[inner sep=0pt] (U3) [right = 2cm of U2] {};
    \node[inner sep=0pt] (U4) [below right = 1cm and 1cm of U3] {};

    \node[inner sep=0pt] (U12) [above right = 0.45cm and 0.45cm of U1] {};
    \node[inner sep=0pt] (U23) [right = 0.75cm of U2] {};

    \node[inner sep=0pt] (D2) [above = 0.5cm of U2]{};
    \node[inner sep=0pt] (D3) [above = 0.5cm of U3] {};

    \node[inner sep=0pt] (UP1) [left = 1cm of D2]{};
    \node[inner sep=0pt] (UP2) [right = 1cm of D3] {};

    \node[inner sep=0pt] (C1) [above right = 0.75cm and 0.4cm of U2]{};
    \node[inner sep=0pt] (C2) [below right = 1.25cm and 0.4cm of U2] {};

        \begin{feynman}
        \draw[-, thick, black, line width=0.5mm, momentum=\(p\)] (U1) to (U12);
        \draw[-, thick, black!70!white, dotted, line width=0.4mm, momentum=\(q\)] (UP1) to (U2);
    	
    	\end{feynman}
    	\draw[-, thick, black, line width=0.5mm] (U12) to (U2);
     \draw[-, thick, black, line width=0.5mm] (U2) to (U3);
     \draw[-, thick, black, line width=0.5mm] (U3) to (U4);
     
     \draw[-, thick, black, line width=0.5mm, out=-30, in=-90] (U12) to (U23);

     \draw[-, thick, black!70!white, dotted, line width=0.4mm] (UP2) to (U3);

     \draw[-, thick, blue, line width=0.4mm] (C1) to (C2);

	\path[draw=black, fill=black] (U2) circle[radius=0.05];
 \path[draw=black, fill=black] (U12) circle[radius=0.05];
 \path[draw=black, fill=black] (U23) circle[radius=0.05];
 \path[draw=black, fill=black] (U3) circle[radius=0.05];

\end{tikzpicture}}=\frac{\tilde{\delta}(p^2)}{(p+q)^2}\int \frac{\mathrm{d}^d k}{(2\pi)^d}\frac{\tilde{\delta}^+(k+q)\tilde{\delta}^+(p-k)}{k^2},
\end{equation}
where again $\tilde{\delta}(p^2)=2\pi \imath\delta(p^2)$. The last diagram of eq.~\eqref{eq:triangle_diag_2nd_line} reads
\begin{equation}
\raisebox{-1cm}{}=\frac{1}{p^2(p+q)^2}\int \frac{\mathrm{d}^d k}{(2\pi)^d}\tilde{\delta}^+(k+q)\tilde{\delta}^+(p-k)\tilde{\delta}^+(k),
\end{equation}
having renamed $p_1=k$ to align with the notation of the previous diagram. Although the four diagrams in eq.~\eqref{eq:triangle_diag_2nd_line} look different, they are related by the KLN theorem: their sum is infrared finite. This classification of diagrammatic contributions will be formalised in sect.~\ref{sec:graph_embeddings} in terms of vacuum graph embeddings. 

Writing down all four contributions using Feynman rules one finds:
\begin{align}
W_{\triangle}(p,q)
&=\tilde{\delta}(p^2)\tilde{\delta}((p+q)^2)T(p,q,-p-q)+\tilde{\delta}(p^2)\frac{\text{disc}_{(p+q)^2}T(p,q,-p-q)}{(p+q)^2} \nonumber \\
&+\tilde{\delta}((p+q)^2)\frac{\text{disc}_{p^2}T(p,q,-p-q)}{p^2}+\frac{\text{disc}_{p^2,(p+q)^2}T(p,q,-p-q)}{p^2(p+q)^2},
\end{align}
where we wrote the cut integrals in terms of discontinuities of the triangle diagram:
\begin{equation}
    T(p_1,p_2,p_3) = \int \frac{\mathrm{d}^dk}{(2 \pi) ^d} \frac{1}{k^2(k-p_1)^2(k+p_2)^2}\,.
\end{equation}
Observe that all four contributions contain infrared poles when taken individually. We may now evaluate these quantities explicitly in terms of a dimensional regulator, $\epsilon=(4-d)/2$. Integrating over $p^2$, expanding in $\Lambda^2$ and taking the $\epsilon\rightarrow 0$ limit gives
\begin{align}
W_{\triangle}(x,\Lambda^2,q^2)={}&-\frac{\alpha^2}{6(q^2)^2}\Bigg[\pi^2\delta(1-x)-6\frac{1+x^2}{1-x}\log(x)\nonumber \\
&-3\frac{1+x^2}{(1-x)_+}\log\left(\frac{\Lambda^2}{-q^2}\right)\Bigg]+\ObigLambda\label{eq:expanded_triangle}.
\end{align}
with $\alpha=\lambda^2/(4\pi)$. $W_{\triangle}(x,\Lambda^2,q^2)$ is now an infrared finite quantity that contains collinear logarithms in the ratio $\Lambda^2/(-q^2)$. It can be convolved with a parton distribution function to obtain the analogue of the hadronic tensor. The extension of this simple example is the topic at the heart of this paper.

\subsection{The multi-partonic (MP) cross-section and the ``simplest clustering'' criterion}
\label{sec:setup}

\partitle{The multi-partonic (MP) cross-section} We model the proton via the following density matrix:
\begin{align}
\label{eq:density_matrix}
\hat{\rho}_{\q}(p)=\frac{(2\pi)^4}{\text{av}(\q)}&\sum_{n\ge 1}\sum_{\substack{X\in \mathcal{X}_\q ^n 
}}\int \mathrm{d}\Pi_X \delta(p_{1n}-p)\ket{X_1(p_1)\cdots X_n(p_n)}\bra{X_1(p_1)\cdots X_n(p_n)},
\end{align}
where $p_{1n}=p_1+\dots+p_n$ and $\mathrm{d}\Pi_X$ is the $n$-particle phase-space
\begin{equation}
\mathrm{d}\Pi_X =\prod_{i=1}^{|X|} \frac{\mathrm{d}^d p_i}{(2\pi)^{d}}\tilde{\delta}^+(p_i), \quad \tilde{\delta}^\pm (q)=(2\pi \imath)\delta(q^2)\Theta(\pm q^0).
\end{equation}
while $\mathcal{X}_\q^n\subset \{q,\bar{q},g\}^n$ contains all combinations of quarks, anti-quarks and gluons such that the overall quark content is $\q$. To compute the quark content of a state $\ket{X_1\cdots X_n}$, we assign to $X_i$ the quark content 
\begin{equation}
\q_i=\begin{cases}
1 \quad &\text{if } X_i=q \\
-1\quad &\text{if } X_i=\bar{q} \\
0\quad &\text{if } X_i=g
\end{cases}.
\end{equation}
The quark content of the full-state is then $\q=\q_1+\dots+\q_n$. $\text{av}(\q)^{-1}$ denoted the combined spin and colour average: $\text{av}(\q=\pm 1)=2N_c$ for a quark or antiquark, $\text{av}(\q=0)=(N_c^2-1)(d-2)$ for any number of gluons and $\text{av}(\q=n)=(2N_c)^n$ for an $n$-uplet of quarks or anti-quarks accompanied by any number of gluons. 

$\hat{\rho}_{\q}$ describes the simplest model for the clustering of partons in an object of momentum $p$. The initial-state of deep inelastic scattering is thus:
\begin{equation}
\hat{\rho}_{\q\text{DIS}}=\hat{\rho}_{\q}(p)\otimes \ket{e^-(p_e)}\bra{e^-(p_e)}.
\end{equation}
The measurement is defined through a projector that expresses inclusivity with respect to all partonic final state radiation. We can write it as
\begin{equation}
\hat{P}_{\text{incl}}=\hat{\mathds{1}}_{\text{QCD}}\otimes \ket{e^-(p_e')}\bra{e^-(p_e')},
\end{equation}
with $\hat{\mathds{1}}_{\text{QCD}}$ being the completeness relation in terms of free states, i.e.
\begin{equation}
\hat{\mathds{1}}_{\text{QCD}}=\sum_{n\ge 1}\sum_{\substack{Y\in \{q,\bar{q},g\}^n 
}}\int \mathrm{d}\Pi_Y \ket{Y_1(p_1)\dots Y_n(p_n)}\bra{Y_1(p_1)\cdots Y_n(p_n)}\,.
\end{equation}
All spins and physical polarisations are implicitely assumed to be summed with same weight. Using the density matrix and projector, we can now construct the cross-section in the laboratory frame 
\begin{align}
\label{eq:MPI_cross-section}
\frac{\mathrm{d}\sigma_{\q}}{\mathrm{d}x\mathrm{d}p_e'}(\Lambda^2)&=\frac{1}{4\pi}\int \dPi \text{Tr}[\hat{\rho}_{\q\mathrm{DIS}}\hat{S}\hat{P}_{\mathrm{incl}}\hat{S}^\dagger]\,, \\
\dPi &= \frac{\mathrm{d}^dp}{2\pi \imath x}\delta\left(x^{-1}-\frac{(p+q)^2-q^2}{-q^2}\right)\delta^{d-2}(p_\perp)\Theta(\Lambda^2-p^2)\,.
\label{eq:MPmeasure}
\end{align}
where $q=p_e-p_e'$, $p_e=(E,0,0,\sqrt{E^2-m_e^2})$, $x$ is the Bjorken variable. Note that eq.~\eqref{eq:MPI_cross-section} refers to the cross-section prior to convolution with parton distribution function. $\Lambda^2$ is a scale that bounds the virtuality $p^2$. Since $p^2=p_{1n}^2\ge 0$ due to $p$ being the sum of light-like momenta all lying in the same forward cone, we have $0\le p^2\le \Lambda^2$. Any $d$-dimensional factors in the definition of $\dPi$ will turn out to not matter, since $\mathrm{d}\overline{\Pi}$ will be integrated against a finite distribution. The factor of $2\pi \imath$ makes sure that the diagrams with one initial-state parton have the same normalisation as in the parton model. We call eq.~\eqref{eq:MPI_cross-section} the \emph{multi-partonic (MP) cross-section }. 

\partitle{Modelling of scattering states in terms of the ``simplest clustering'' criterion} The idea that initial-state protons could be modelled in terms of jets goes back to the foundational paper by Sterman and Weinberg~\cite{Sterman:1977wj} in which it is originally put forward, and to the best of our knowledge finds a first concrete computational attempt in refs.~\cite{PhysRevD.25.2222} and~\cite{Axelrod:1985yi}, where the authors select a subset of all multi-partonic diagrams contributing to the deep inelastic scattering cross-sections and cluster the initial-state partons using two parameters, a soft and a collinear one, bounding the on-shell energy of gluons and the relative angle between partons and hadron, respectively. Another calculation in this direction, using a mass and hemisphere jet definition, was conducted for a Drell-Yan-like process in ref.~\cite{Frye:2018xjj}. The model that we propose here, while in line with the broad idea of clustering initial state partons into jets, is even simpler: it does not impose any constraint on the momenta of constituents aside from those that directly follow from constraints on the overall momentum of the hadron, $p$. As we will see, this choice uniquely translates in a natural power-counting in virtuality and in a remarkable identification of cross-sections with iterated discontinuities of Feynman diagrams, which itself allows for a simple computation of the result. We will also see how to extend it to the Drell-Yan process.

The density matrix $\hat{\rho}_\q$ of eq.~\eqref{eq:density_matrix} indeed represents the simplest way a cluster of particles can be identified with a ``macroscopic'' object of momentum $p$. The only physical principle that is used is momentum conservation, and any other information concerning the kinematic space of the constituents is assumed to be unavailable - a uniform distribution is taken. The choice of $\hat{\rho}_\q$ can be justified on theoretical ground following these principles: first, the density matrix $\hat{\rho}_\q$ should be diagonal in the basis of free states: this agrees with the usual generalisation of Fermi golden's rule to degenerate states, and is compeletely analogous to what is done with final states. This choice also enables the identification of squared amplitudes with cuts of forward-scattering diagrams or vacuum diagrams, and is a cornerstone of the KLN theorem in its diagrammatic form. Note that the use of any diagonal density matrix implies that we are not dealing with a pure state, and thus an amount of information on the exact quantum state is assumed to be lost, which might as well be the consequence of the adiabatic switch-off of the interaction. The second constraint is that the momenta of the constituent particles should sum up to an input momentum $p$.

Provided these motivations are strong enough for the reader, the most general density matrix satisfying this property reads:
\begin{align}
\label{eq:full_density_matrix}
\hat{\rho}(p)=&\sum_{n\ge 1}\sum_{\substack{X\in \{q,\bar{q},g\}^n}}\int \mathrm{d}\Pi_X (2\pi)^4 \delta(p_{1n}-p)f(p_1, \dots, p_n )\ket{X}\bra{X}\,,
\end{align}
where we introduced the short-hand $\ket{X}=\ket{X_1(p_1)\cdots X_n(p_n)}$. Additional constraints on the precise superposition of states, such as requiring certain transformation properties of the density matrix, is assumed to be part of the choice of the distribution $f$. We are interested in the $p^2\rightarrow 0$ limit of $\hat{\rho}(p)$; as we will see later, this limit is fundamental in defining a notion of virtuality expansion. When $p^2\rightarrow 0$, $p$ becomes light-like. The fundamental constituents are also taken to be on-shell massless, which implies, since $p_{1n}=p_1+\dots +p_n=p$, that in the limit $p^2\rightarrow 0$ \emph{all fundamental constituents become collinear to each other and to the input momentum $p$}:
\begin{equation}
p_i=\xi_i p+\mathcal{O}(p^2), \quad \xi_i\in(0,1), \ \sum_{i=1}^n \xi_i=1\,.
\end{equation}
If we impose $f$ to be an infrared safe density, as required by infrared finiteness, then, by the usual degeneracy arguments, we immediately find that $f$ depends only on $p$:
\begin{align}
\hat{\rho}(p)=&f(p)\sum_{n\ge 1}\sum_{\substack{X\in \{q,\bar{q},g\}^n}}\int \mathrm{d}\Pi_X (2\pi)^4 \delta(p_{1n}-p)\ket{X}\bra{X}+\mathcal{O}(p^2),
\end{align}
which matches our original definition of density matrix, up to an overall normalisation. $f(p)$ does not depend on $p^2$ but it can depend on the momentum fraction, making it a natural candidate for a parton distribution function. In summary, the request of infrared finiteness and momentum conservation, together with an expansion in virtuality, allow us to conclude that the internal dynamics of the constituents should be treated uniformly. In practical cases, the expansion in virtuality has to be approached carefully as it involves the expansion of distributions and non-analytic functions, and we will make sure to do so in the following.

\partitle{Partonic and leptonic tensors} QCD corrections to the cross-section can be written in terms of a partonic and a leptonic tensor:
\begin{equation}
\text{Tr}[\hat{\rho}_{\q\mathrm{DIS}}\hat{S}\hat{P}_{\mathrm{incl}}\hat{S}^\dagger]|_{\text{QCD}}=\frac{(-\imath)^2}{(q^2)^2}L_{\mu\nu}(p_e,p_e')W_\q^{\mu\nu}(p,q),
\end{equation}
where the leptonic tensor, if we are interested in QCD corrections, only has a tree level contribution, 
\begin{equation}
L_{\mu\nu}(p_e,p_e')=(-\imath e)^2\text{Tr}[(\slashed{p}_e+m_e)\gamma^\mu(\slashed{p}_e'+m_e)\gamma^\nu],
\end{equation} 
while the partonic tensor reads
\begin{align}
W^{\mu\nu}_\q(p,q)=\sum_{n,m\ge 1}\sum_{\substack{X\in \mathcal{X}_\q^n  \\ Y\in\{q,\bar{q},g\}^m}}\frac{(2\pi)^4}{\text{av}(\q)}\int \mathrm{d}\Pi_X \mathrm{d}\Pi_Y \delta(p_{1n}-p) |\bra{X\gamma^\star(q)} S \ket{Y}|^2,\label{eq:amplitudes}
\end{align}
where, again, $\ket{X\gamma^\star(q)}=\ket{X_1(p_1)\cdots X_n(p_n)\gamma^\star(q)}$, $\ket{Y}=\ket{Y_1(p_1')\cdots Y_m(p_m')}$ and $\mathrm{d}\Pi_X$, $\mathrm{d}\Pi_Y$ are the Lorentz-invariant phase-space measures for the integrations in $p_1,\dots,p_n$ and $p_1',\dots,p_m'$ respectively. It is useful to divide the partonic tensor into the parton model (PM) partonic tensor and contributions from higher multiplicity (HM) partonic initial states:
\begin{align}
W^{\mu\nu}_\q(p,q)=\underbrace{\sum_{m\ge 1}\sum_{\substack{ Y\in\{q,\bar{q},g\}^m}}\frac{ \tilde{\delta}(p^2)}{\text{av}(\q)}\int \mathrm{d}\Pi_Y  |\bra{\q(p)\gamma^\star(q)} S \ket{Y}|^2}_{=W_{\q\text{PM}}^{\mu\nu}} + W^{\mu\nu}_{\q\text{HM}}(p,q).
\end{align}
This expression implicitly defines the value of $W^{\mu\nu}_{\q\text{HM}}(p,q)$ to include all the terms of the sum in $n$ with $n\ge 2$. $W_{\q\text{PM}}^{\mu\nu}$ is instead the ordinary parton model partonic tensor. The full partonic cross-section, in terms of leptonic and partonic tensors, reads
\begin{align}
\label{eq:x_sec_decomp}
\frac{\mathrm{d}\sigma_{\q}}{\mathrm{d}x\mathrm{d}p_e'}(\Lambda^2)&=\frac{(-\imath)^2}{(q^2)^2}L_{\mu\nu}(p_e,p_e') W^{\mu\nu}_\q (x,\Lambda^2,q^2), \\
W_\q^{\mu\nu}(x,\Lambda^2,q^2)&=\frac{1}{4\pi}\int \dPi W_\q^{\mu\nu}(p,q),
\end{align}
where we defined the \emph{virtuality-integrated partonic tensor} $W^{\mu\nu}_\q (x,\Lambda^2,q^2)$ by a slight abuse of notation. In the following, we will also simply call $W^{\mu\nu}_\q (x,\Lambda^2,q^2)$ the \emph{integrated partonic tensor}. Since the partonic tensor satisfies the Ward identity $q_\mu W_\q^{\mu\nu}=0$ and excluding weak bosons, $W_\q^{\mu\nu}$ admits the Lorentz decomposition
\begin{equation}
W_\q^{\mu\nu}(p,q)=W_{1\q}(p^2,q^2,(p+q)^2)\left(g^{\mu\nu}-\frac{q^\mu q^\nu}{q^2}\right)+W_{2\q }(p^2,q^2,(p+q)^2)P^\mu P^\nu,
\end{equation}
where
$P^\mu=p^\mu-\frac{(p\cdot q)}{q^2}q^\mu$. $W_{1\q }$ and $W_{2\q}$ are obtained by contraction of the partonic tensor with appropriately constructed projectors:
\begin{align}
W_{1\q}(p^2,q^2,(p+q)^2)&=\frac{1}{2-d}\left(g_{\mu\nu}-\frac{P_\mu P_\nu}{P^2}\right)W_\q^{\mu\nu}=\mathds{P}^1_{\mu\nu}(p,q)W_\q^{\mu\nu}, \\
W_{2\q}(p^2,q^2,(p+q)^2)&=\frac{1}{P^2}\left[\frac{1}{2-d}g_{\mu\nu}+\frac{1-d}{2-d}\frac{P_\mu P_\nu}{P^2}\right]W_\q^{\mu\nu}=\mathds{P}^2_{\mu\nu}(p,q)W_\q^{\mu\nu}.
\end{align}
\newpage
Using these definitions we define the structure functions:
\begin{align}
F_{1\q }(x,\Lambda^2,q^2)&=\frac{1}{4\pi}\int \dPi W_{1\q}\left(p,q\right)\label{eq:F1},\\
F_{2\q}(x,\Lambda^2,q^2)&=-\frac{q^2}{8\pi x}\int \dPi W_{2\q}\left(p,q\right)\label{eq:F2},\\
F_{L\q}(x,\Lambda^2,q^2)&=F_{1\q}(x,\Lambda^2,q^2)-2x F_{2\q}(x,\Lambda^2,q^2) \label{eq:FL}.
\end{align}
These will be the objects of our computation in the following. Finally, we note that the full hadronic tensor, namely the partonic tensor convolved with parton distribution functions, also derives its structure by close analogy with the parton model:
\begin{equation}
W_H^{\mu\nu}(x,q^2)=\sum_{\q=-\infty}^{\infty}\int_x^1 \frac{\mathrm{d}\xi}{\xi} f_\q(\xi,\Lambda^2)W_\q^{\mu\nu}(x/\xi,\Lambda^2,q^2),
\end{equation}
where the $f_\q$ functions are the analogues of the parton distribution functions. However, in general, in the MP model the hadronic tensor receives contributions from any quark content, and not only $\q=-1,0,1$.

\subsection{The virtuality expansion}

We are interested in the leading term of the MP cross-section in an expansion in the highest allowed virtuality $\Lambda^2$ for the cluster of initial state partons:
\begin{equation}
\frac{\mathrm{d}\sigma_{\q}}{\mathrm{d}x\mathrm{d}p_e'}(\Lambda^2)=\frac{\mathrm{d}\sigma_{\q}^{(0)}}{\mathrm{d}x\mathrm{d}p_e'}(\Lambda^2)+\ObigLambda.
\end{equation}
All other scales are of order $Q^2=-q^2=\mathcal{O}(\Lambda^0)$\footnote{Our use of the big O notation, like often in the literature, includes logarithmic terms, so that $\mathcal{O}(\delta^m)=\mathcal{O}(\delta^m\log^n(\delta))$ for any $n$.}. 

Because of the decomposition of eq.~\eqref{eq:x_sec_decomp}, studying the virtuality expansion of the cross-section is equivalent to studying the expansion of the integrated partonic tensor:
\begin{equation}
W_\q^{\mu\nu}(x,\Lambda^2,q^2)=\left(W_\q^{(0)}(x,\Lambda^2,q^2)\right)^{\mu\nu}+\ObigLambda.
\end{equation}
This expansion can be equivalently carried out before integration in $\dPi$, provided that we are careful about the expansion of the non-analytic terms involved in the integrand. In particular, let us consider the usual light-cone decomposition
\begin{equation}
p=\underbrace{\xi \hat{n}_+}_{=p_+} + p_\perp + \frac{p^2-p_\perp^2}{4 \xi }\hat{n}_-,
\end{equation}
with $\hat{n}_\pm=\eta^0\pm \eta^3$, where $\eta^\mu$ defines the lab frame, $p_\perp\cdot \hat{n}_\pm=0$, and $\hat{n}_+\cdot \hat{n}_-=2$. We have
\begin{equation}
\mathrm{d}^d p=\frac{\mathrm{d}p^2\mathrm{d}\xi  \mathrm{d}^{d-2}p_\perp }{\xi}.
\end{equation}

In these coordinates we can define the leading-virtuality expansion of the higher-multiplicity part of the partonic tensor by the following truncation
\begin{align}
W^{\mu\nu}_{\q\text{HM}}(p,q)=\left(W^{(0)}_{\q\text{HM}}(p,q)\right)^{\mu\nu}+\ObigPsq.
\end{align}
 As well-known from expansion by regions arguments, $W^{\mu\nu}_{\q\text{HM}}(p,q)$ is not an analytic function of $p^2$, hence $W^{(0)}_{\q\text{HM}}(p,q)$ will still depend on $p^2$. 
Aside from being careful when expanding the partonic tensor, the rest follows rather smoothly. At leading-virtuality, the measure (eq.~\eqref{eq:MPmeasure}) reads
\begin{align}
\dPi&=\dPi^{(0)}+\mathcal{O}(p^2), \\ \dPi^{(0)}&=\frac{\mathrm{d}p^2\mathrm{d}\xi  \mathrm{d}^{d-2}p_\perp }{2 \pi \imath \xi x} \delta^{d-2}(p_\perp) \delta\left(x^{-1}-\xi \frac{2n_+\cdot q}{q^2}\right)\Theta(\Lambda^2-p^2).
\end{align}
In particular, taking into account all these facts, we re-write the leading-virtuality integrated partonic tensor as
\begin{align}
\label{eq:integrated_W}
\left(W_\q^{(0)}(x,\Lambda^2,q^2)\right)^{\mu\nu}&=\frac{1}{4\pi}\int_0^{\Lambda^2} \frac{\mathrm{d}p^2}{2\pi \imath}\left(W_{\q\text{PM}}(p,q)+W^{(0)}_{\q\text{HM}}(p,q)\right)^{\mu\nu},
\end{align}
with
\begin{equation}
p=\xi(x,q)\hat{n}_++\frac{p^2}{4\xi(x,q^2)}\hat{n}_-
\end{equation}
and $\xi(x,q^2)=-q^2/(2 x n_+\cdot q)$. Note that $W^{\mu\nu}_{\q\text{PM}}$ only depends on $p^2$ through the $\delta(p^2)$ distribution it absorbs in its definition, meaning the the virtuality is localised at its vanishing value, hence why it is not subject to the virtuality expansion.

It may be interesting to study the relation between the virtuality expansion presented here with the twist expansion in the sense of the operator-product expansion, since the inclusion of multi-partonic initial-states in the density matrix allows to uniquely probe higher-twist operators in the OPE. We further note that expansions in virtuality are also frequent in soft-collinear effective field theory, and sometimes called ``twist expansions'' in the corresponding literature, by a slight abuse. Finally, power-corrections in $\Lambda^2$ may indeed also be computed within the model, although in principle they would require the knowledge of the $p^2$ derivative of the initial-state density in eq.~\eqref{eq:full_density_matrix}.

\subsection{The KLN theorem and overview of NLO results}

\partitle{Infrared finiteness} The sum over degenerate initial and final-state multiplicites makes it so that the full partonic tensor defined above is infrared finite - its Laurent expansion in the dimensional regulator $\epsilon=\frac{4-d}{2}$ starts with a finite term $\epsilon^0$. In particular, if we think of the partonic tensor as a distribution, we have that
\begin{equation}
\label{eq:KLNp}
W^{\mu\nu}(p,q)=\sum_{n=0}^\infty a_n^{\mu\nu}(p,q) \epsilon^n, 
\end{equation}
where, again, the coefficients $a_n^{\mu\nu}(p,q)$ should be thought of as distributions to be integrated against a test function. Of course, eq.~\eqref{eq:KLNp} only holds after UV renormalisation has been performed by standard methods. 

\partitle{Collinear factorisation} After integration in $p^2$, the leading-virtuality structure functions $F_{\q 1}$ and $F_{\q 2}$, while finite, still have logarithmic corrections in ratios of $\Lambda^2/Q^2$, a reflection of the non-analiticity of the integrand in $p^2$. At next-to-leading order, we will find in sect.~\ref{sec:results} that the evolution in $\Lambda^2/Q^2$ factorises and is driven by the usual Altarelli-Parisi splitting kernels:
\begin{align}
F_{2q}(x,\Lambda^2,q^2)&=\frac{\alpha_s}{2\pi}P_{qq}(x) \log\left(\frac{Q^2}{\Lambda^2}\right)+C_{2q}(x)+\ObigLambda\,, \\
F_{2g}(x,\Lambda^2,q^2)&=\frac{\alpha_s}{2\pi}P_{qg}(x) \log\left(\frac{Q^2}{\Lambda^2}\right)+C_{2g}(x)+\ObigLambda\,, \\
F_{Lq}(x,\Lambda^2,q^2)&=C_{Lq}(x)+\ObigLambda \,,\\
F_{Lg}(x,\Lambda^2,q^2)&=C_{Lg}(x)+\ObigLambda\,.
\end{align}
This result is somewhat expected: if we assume that the parton model cross-section factorises, and if KLN ensures that the overall partonic tensor (parton model + higher multiplicity contributions) is finite and free of collinear logarithms in the dimensional regularisation scale $\mu$, \emph{then} also the overall partonic tensor must factorise. In other words, the MP cross-section has factorisation properties that are \emph{at least as good as} those of the parton model cross-section.

\partitle{Matching to parton model} One of the most important questions one may ask related to factorisation is whether the MP cross-section can be related to the parton model cross-section by a scheme change. It is easy to see that a necessary and sufficient condition for it to happen is 
\begin{equation}
\label{eq:matching_PM}
F_{L\q}=F_{L\q \text{PM}},
\end{equation}
i.e., if the longitudinal structure functions are the same in the MP model and in the parton model. Eq.~\eqref{eq:matching_PM} is precisely what we find in our computation of the longitudinal structure function.

\section{Classification of diagrammatic contributions by graph embeddings}
\label{sec:graph_embeddings}

Given equation~\eqref{eq:amplitudes}, one generates diagrams contributing to the amplitudes with varying multiplicity, and then glues them appropriately to form ``doubly-cut'' interference diagrams. However, the KLN cancellation mechanism imposes a natural classification of these diagrammatic contributions: in particular, the set of all interference diagrams can be subdivided in non-overlapping classes of interference diagrams such that the sum of all diagrams in a class is separately infrared finite. Instead of generating all interference diagrams and then grouping them into the relevant classes, it is easier, both for conceptual and computational reasons, to directly generate the classes themselves. These classes can be uniquely characterised in terms of embeddings of vacuum diagrams. In this section, we will study how to generate vacuum diagram embeddings.

The observation that interference diagrams can be obtained from double cuts of vacuum diagrams is far from being new, and in fact goes back to the work of Kinoshita~\cite{Kinoshita:1962ur}, and is restated in classic texbooks such as Sterman's~\cite{Sterman:1993hfp}; some more work on this identification has been done in the context of Local Unitarity~\cite{Capatti:2020xjc,Capatti:2023omc}, followed by later work~\cite{Ramirez-Uribe:2024rjg,LTD:2024yrb}, but is also strongly implied by classic work on Reverse Unitarity~\cite{Anastasiou:2002qz,Anastasiou:2002yz,Anastasiou:2003yy,Anastasiou:2013mca,Anastasiou:2013srw}. The intuitive idea that the embedding of vacuum diagrams on ``cylinders'' relates to IR-finiteness as realised within the KLN theorem also has a history; recently, it has been mentioned in~\cite{Frye:2018xjj} and expanded upon in \cite{Blazek:2021zoj,Capatti:2023omc}. More broadly, understanding the diagrammatics of the KLN theorem is a notoriously subtle task and has been the subject of a few more papers~\cite{PhysRevD.25.2222,Akhoury1997,khalil2017completediagrammaticimplementationkinoshitaleenauenberg}, with a particular focus on the appearance and treatment of spectator particles. We will provide here for the first time a systematic approach to generating vacuum graph embeddings for a realistic process, as well as a recipe for recognising when they are isomorphic. This constitutes an important step in turning the idea of vacuum graph embeddings from intuition to concrete computational tool.

\subsection{Motivation: embeddings and the KLN theorem}

A feature of any diagrammatic implementation of the KLN theorem is that it requires the inclusion of disconnected diagrams, on top of a sum over initial and final-state multiplicities. In order to broadly characterise the diagrammatic contributions, let us first observe with Kinoshita~\cite{Kinoshita:1962ur}, that contributions to a cross-section can be identified with interference diagrams, which themselves can be seen as double cuts on a vacuum graph. Consider for example, the following interference diagram contributing to the parton model
\begin{equation}
\label{eq:PM_triangle1}
\begin{tikzpicture}

    \node[inner sep=0pt] (U1) {};
    \node[inner sep=0pt] (U2) [above right = 1cm and 1cm of U1] {};
    \node[inner sep=0pt] (U3) [right = 2cm of U2] {};
    \node[inner sep=0pt] (U4) [below right = 1cm and 1cm of U3] {};

    \node[inner sep=0pt] (U12) [above right = 0.45cm and 0.45cm of U1] {};
    \node[inner sep=0pt] (U23) [right = 0.75cm of U2] {};

    \node[inner sep=0pt] (D2) [above = 0.5cm of U2]{};
    \node[inner sep=0pt] (D3) [above = 0.5cm of U3] {};

    \node[inner sep=0pt] (UP1) [left = 1cm of D2]{};
    \node[inner sep=0pt] (UP2) [right = 1cm of D3] {};

    \node[inner sep=0pt] (C1) [above right = 0.75cm and 1cm of U2]{};
    \node[inner sep=0pt] (C2) [below right = 1.25cm and 1cm of U2] {};

        \begin{feynman}
    	
    	\end{feynman}
    	\draw[-, thick, black, line width=0.5mm] (U1) to (U2);
     \draw[-, thick, black, line width=0.5mm] (U2) to (U3);
     \draw[-, thick, black, line width=0.5mm] (U3) to (U4);
     \draw[-, thick, electron-purple, line width=0.5mm] (UP1) to (UP2);
     
     \draw[-, thick, black, line width=0.5mm, out=-30, in=-90] (U12) to (U23);

     \draw[-, thick, black!70!white, dotted, line width=0.4mm] (U2) to (D2);
     \draw[-, thick, black!70!white, dotted, line width=0.4mm] (U3) to (D3);

     \draw[-, thick, blue, line width=0.4mm] (C1) to (C2);

	\path[draw=black, fill=black] (U2) circle[radius=0.05];
 \path[draw=black, fill=black] (U12) circle[radius=0.05];
 \path[draw=black, fill=black] (U23) circle[radius=0.05];
 \path[draw=black, fill=black] (U3) circle[radius=0.05];

\end{tikzpicture}
\,.
\end{equation}
Due to the cyclic identification of the lines an equivalent representation is the  doubly-cut vacuum diagram
\begin{equation}
\vcenter{\hbox{\scalebox{0.5}{\includegraphics[]{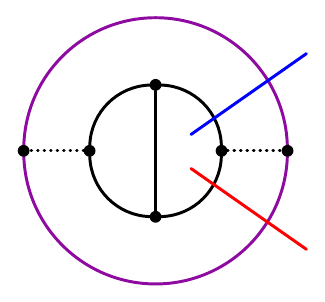}}}}
\end{equation}

where the purple line, in deep inelastic scattering, is identified with the electron, the dotted line with the photon. Here and in the following, we choose to work with the scalar skeleton of the diagrams so as not to overcrowd the visualizations. 
The solid black lines therefore represent either gluons or quarks.

Another observation of Kinoshita is that whichever diagrams we plan to add to that of eq.~\eqref{eq:PM_triangle1} in order to make it finite, should also be seen as another double-cut of the same vacuum diagram. In other words, interference diagrams that cancel each other about some infrared singularity can be obtained one from the other by ``moving'' the Cutkosky cuts that identify initial and final states. In the following, we will argue that not only that must be true, but both doubly-cut vacuum diagrams must also be consistent with the same graph ``embedding''. 

To give a simplified example contextual to deep inelastic scattering, let us consider the following non-planar one-loop diagram appearing in the parton model
\begin{equation}
\label{eq:PM_x_box}
\vcenter{\hbox{\begin{tikzpicture}

    \node[inner sep=0pt] (U1) {};
    \node[inner sep=0pt] (U2) [above right = 1cm and 1cm of U1] {};
    \node[inner sep=0pt] (U3) [right = 2cm of U2] {};
    \node[inner sep=0pt] (U4) [below right = 1cm and 1cm of U3] {};

    \node[inner sep=0pt] (U12) [above right = 0.45cm and 0.45cm of U1] {};
    \node[inner sep=0pt] (U4M) [above left = 0.45cm and 0.45cm of U4] {};
    \node[inner sep=0pt] (U23) [right = 0.75cm of U2] {};

    \node[inner sep=0pt] (D2) [above = 0.5cm of U2]{};
    \node[inner sep=0pt] (D3) [above = 0.5cm of U3] {};

    \node[inner sep=0pt] (UP1) [left = 1cm of D2]{};
    \node[inner sep=0pt] (UP2) [right = 1cm of D3] {};

    \node[inner sep=0pt] (C1) [above right = 0.75cm and 0.65cm of U2]{};
    \node[inner sep=0pt] (C2) [below right = 1.25cm and 0.65cm of U2] {};

    \node[inner sep=0pt] (CC1) [above left = 0.75cm and 0.9cm of U2]{};
    \node[inner sep=0pt] (CC2) [below left = 1.25cm and 0.9cm of U2] {};

        \begin{feynman}
        \draw[-, thick, black, line width=0.5mm, momentum=\(p\)] (U1) to (U12);
        \draw[-, thick, black, line width=0.5mm, momentum=\(k\)] (U12) to (U2);
     \draw[-, thick, black, line width=0.5mm] (U2) to (U4M);
     \draw[-, thick, black, line width=0.5mm, momentum=\(k-p\)] (U3) to (U12);
     \draw[-, thick, black, line width=0.5mm] (U3) to (U4);
     \draw[-, thick, electron-purple, line width=0.5mm] (UP1) to (UP2);
    	
    	\end{feynman}


     \draw[-, thick, black!70!white, dotted, line width=0.4mm] (U2) to (D2);
     \draw[-, thick, black!70!white, dotted, line width=0.4mm] (U3) to (D3);

     \draw[-, thick, blue, line width=0.4mm] (C1) to (C2);
     \draw[-, thick, red, line width=0.4mm] (CC1) to (CC2);

	\path[draw=black, fill=black] (U2) circle[radius=0.05];
 \path[draw=black, fill=black] (U12) circle[radius=0.05];
 \path[draw=black, fill=black] (U4M) circle[radius=0.05];
 \path[draw=black, fill=black] (U3) circle[radius=0.05];

\end{tikzpicture}}}\,.
\end{equation}
It has a $t$-channel collinear singularity when $k=xp$, for $x\in(0,1)$. The singularity is removed if we add to it the diagram obtained by ``moving'' the initial-state cut across the collinear singularity:
\begin{equation}
\begin{tikzpicture}

    \node[inner sep=0pt] (U1) {};
    \node[inner sep=0pt] (U2) [above right = 1cm and 1cm of U1] {};
    \node[inner sep=0pt] (U3) [right = 2cm of U2] {};
    \node[inner sep=0pt] (U4) [below right = 1cm and 1cm of U3] {};

    \node[inner sep=0pt] (U12) [above right = 0.45cm and 0.45cm of U1] {};
    \node[inner sep=0pt] (U4M) [above left = 0.45cm and 0.45cm of U4] {};
    \node[inner sep=0pt] (U23) [right = 0.75cm of U2] {};

    \node[inner sep=0pt] (D2) [above = 0.5cm of U2]{};
    \node[inner sep=0pt] (D3) [above = 0.5cm of U3] {};

    \node[inner sep=0pt] (UP1) [left = 1cm of D2]{};
    \node[inner sep=0pt] (UP2) [right = 1cm of D3] {};

    \node[inner sep=0pt] (C1) [above right = 0.75cm and 0.65cm of U2]{};
    \node[inner sep=0pt] (C2) [below right = 1.25cm and 0.65cm of U2] {};
    \node[inner sep=0pt] (C2P) [below right = 1.25cm and 0.55cm of U2] {};
    \node[inner sep=0pt] (CC) [below  = 0.15cm of U2] {};

    \node[inner sep=0pt] (CC1) [above left = 0.75cm and 0.9cm of U2]{};
    \node[inner sep=0pt] (CC2) [below left = 1.25cm and 0.9cm of U2] {};

        \begin{feynman}
        \draw[-, thick, black, line width=0.5mm, momentum=\(p\)] (U1) to (U12);
        \draw[-, thick, black, line width=0.5mm, momentum=\(k\)] (U12) to (U2);
     \draw[-, thick, black, line width=0.5mm] (U2) to (U4M);
     \draw[-, thick, black, line width=0.5mm, momentum=\(k-p\)] (U3) to (U12);
     \draw[-, thick, black, line width=0.5mm] (U3) to (U4);
     \draw[-, thick, electron-purple, line width=0.5mm] (UP1) to (UP2);

     \draw[-, dashed, black!25!white, line width=0.5mm, out=-135,in=-45, looseness=0.3] (U1) to (U4);

     \draw[-, dashed, electron-purple!25!white, line width=0.5mm, out=155,in=25, looseness=0.3] (UP1) to (UP2);
    	
    	\end{feynman}


     \draw[-, thick, black!70!white, dotted, line width=0.4mm] (U2) to (D2);
     \draw[-, thick, black!70!white, dotted, line width=0.4mm] (U3) to (D3);

     \draw[-, thick, red, line width=0.4mm, out=-90, in=180] (CC1) to (CC);
     \draw[-, thick, red, line width=0.4mm, out=90, in=0] (C2P) to (CC);

     \draw[-, thick, blue, line width=0.4mm] (C1) to (C2);

	\path[draw=black, fill=black] (U2) circle[radius=0.05];
 \path[draw=black, fill=black] (U12) circle[radius=0.05];
 \path[draw=black, fill=black] (U4M) circle[radius=0.05];
 \path[draw=black, fill=black] (U3) circle[radius=0.05];
 \path[draw=red, fill=red] (CC) circle[radius=0.019];

\end{tikzpicture}
\end{equation}
where the parton on the left with momentum $p$ is now intended to be ``glued'' with the corresponding partner on the right, and not on-shell: the dashed grey lines help with this identification. A traditional visualisation of this diagram would be
\begin{equation}
\label{eq:box_cancelling}
\begin{tikzpicture}

    \node[inner sep=0pt] (U1) {};
    \node[inner sep=0pt] (U11) [above=0.25cm of U1] {};
    \node[inner sep=0pt] (U2) [above right = 1cm and 1cm of U1] {};
    \node[inner sep=0pt] (U3) [right = 2cm of U2] {};
    \node[inner sep=0pt] (U4) [below right = 1cm and 1cm of U3] {};

    \node[inner sep=0pt] (U12) [above right = 0.45cm and 0.45cm of U1] {};
    \node[inner sep=0pt] (U4M) [above left = 0.45cm and 0.45cm of U4] {};
    \node[inner sep=0pt] (U44M) [right = 0.45cm of U4M] {};

    \node[inner sep=0pt] (F1) [right = 4.5cm of U1] {};
    \node[inner sep=0pt] (F2) [right = 4.5cm of U11] {};

    \node[inner sep=0pt] (U23) [right = 0.75cm of U2] {};

    \node[inner sep=0pt] (D2) [above = 0.5cm of U2]{};
    \node[inner sep=0pt] (D3) [above = 0.5cm of U3] {};

    \node[inner sep=0pt] (UP1) [left = 1cm of D2]{};
    \node[inner sep=0pt] (UP2) [right = 1.5cm of D3] {};

    \node[inner sep=0pt] (C1) [above right = 0.75cm and 0.65cm of U2]{};
    \node[inner sep=0pt] (C2) [below right = 1.25cm and 0.65cm of U2] {};
    \node[inner sep=0pt] (C2P) [below right = 1.25cm and 0.55cm of U2] {};
    \node[inner sep=0pt] (CC) [below  = 0.15cm of U2] {};

    \node[inner sep=0pt] (CC1) [above left = 0.75cm and 0.9cm of U2]{};
    \node[inner sep=0pt] (CC2) [below left = 1.25cm and 0.9cm of U2] {};

        \begin{feynman}
        \draw[-, thick, black, line width=0.5mm] (U11) to (U2);
     \draw[-, thick, black, line width=0.5mm] (U2) to (U4M);
     \draw[-, thick, black, line width=0.5mm] (U3) to (U1);
     \draw[-, thick, black, line width=0.5mm] (U3) to (U4M);
     \draw[-, thick, black, line width=0.5mm] (U44M) to (U4M);
     \draw[-, thick, black, line width=0.5mm] (U44M) to (F1);
     \draw[-, thick, black, line width=0.5mm] (U44M) to (F2);
     \draw[-, thick, electron-purple, line width=0.5mm] (UP1) to (UP2);
    	
    	\end{feynman}


     \draw[-, thick, black!70!white, dotted, line width=0.4mm] (U2) to (D2);
     \draw[-, thick, black!70!white, dotted, line width=0.4mm] (U3) to (D3);

     \draw[-, thick, red, line width=0.4mm] (CC1) to (CC2);

     \draw[-, thick, blue, line width=0.4mm] (C1) to (C2);

	\path[draw=black, fill=black] (U2) circle[radius=0.05];
 \path[draw=black, fill=black] (U44M) circle[radius=0.05];
 \path[draw=black, fill=black] (U4M) circle[radius=0.05];
 \path[draw=black, fill=black] (U3) circle[radius=0.05];

\end{tikzpicture}\,.
\end{equation}
At the left of the blue cut, the diagram is written as a connected component plus a spectator particle, travelling from the initial to the final state without interacting with anything. The inclusion of spectator particles is crucial to the implementation of the KLN theorem. 

Another useful way to represent interference diagrams that has been suggested in refs.~\cite{Frye:2018xjj,Blazek:2021zoj} is to draw them on a cylinder. In this paper, we prefer to embed vacuum diagrams in $\mathbb{R}^2\setminus\{0\}$, the punctured plane, for ease of visualisation. In this picture, cuts are ``rays'' that start at the puncture and unwind the vacuum diagram. For example, the triangle diagram of eq.~\eqref{eq:PM_triangle1} is represented as follows:
\begin{equation}
    \vcenter{\hbox{\scalebox{0.7}{\includegraphics[]{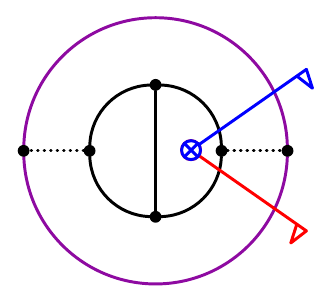}}}}=
    \vcenter{\hbox{\scalebox{0.7}{\includegraphics[]{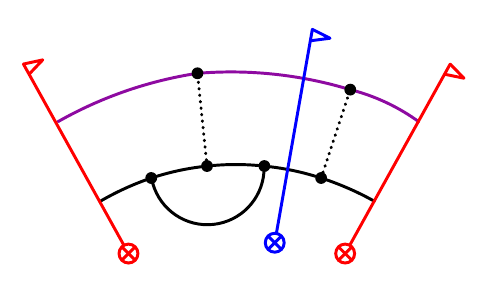}}}}.
\end{equation}
The initial and final-state cuts of the triangle are rays that start at the same puncture, denoted by the symbol $\otimes$. On the right-hand side of the equation, we see how the cuts allow to open-up the vacuum diagram into an interference diagram. The flag on top of the cut specifies the ``time-direction'', which is in principle ambiguous in the first representation if not specified. The flag can then be removed in the interference diagram, as it is assumed that time flows from left to right, giving exactly the diagram of eq.~\eqref{eq:PM_triangle1}.

In much the same way, we can draw the diagrams of eq.~\eqref{eq:box_cancelling} and eq.~\eqref{eq:PM_x_box}.
Both of them belong to the same ``embedding'' of the underlying vacuum diagram, namely to the same positioning of the puncture:
\begin{equation}
\vcenter{\hbox{\scalebox{0.6}{\begin{tikzpicture}

    \node[inner sep=0pt] (U1) {};
    \node[inner sep=0pt] (U2) [above right = 1cm and 1cm of U1] {};
    \node[inner sep=0pt] (U3) [right = 2cm of U2] {};
    \node[inner sep=0pt] (U4) [below right = 1cm and 1cm of U3] {};

    \node[inner sep=0pt] (U12) [above right = 0.45cm and 0.45cm of U1] {};
    \node[inner sep=0pt] (U4M) [above left = 0.45cm and 0.45cm of U4] {};
    \node[inner sep=0pt] (U23) [right = 0.75cm of U2] {};

    \node[inner sep=0pt] (D2) [above = 0.5cm of U2]{};
    \node[inner sep=0pt] (D3) [above = 0.5cm of U3] {};

    \node[inner sep=0pt] (UP1) [left = 1cm of D2]{};
    \node[inner sep=0pt] (UP2) [right = 1cm of D3] {};

    \node[inner sep=0pt] (C1) [above right = 0.75cm and 0.65cm of U2]{};
    \node[inner sep=0pt] (C2) [below right = 1.25cm and 0.65cm of U2] {};

    \node[inner sep=0pt] (CC1) [above left = 0.75cm and 0.9cm of U2]{};
    \node[inner sep=0pt] (CC2) [below left = 1.25cm and 0.9cm of U2] {};

        \begin{feynman}
        \draw[-, thick, black, line width=0.5mm] (U1) to (U12);
        \draw[-, thick, black, line width=0.5mm] (U12) to (U2);
     \draw[-, thick, black, line width=0.5mm] (U2) to (U4M);
     \draw[-, thick, black, line width=0.5mm] (U3) to (U12);
     \draw[-, thick, black, line width=0.5mm] (U3) to (U4);
     \draw[-, thick, electron-purple, line width=0.5mm] (UP1) to (UP2);
    	
    	\end{feynman}


     \draw[-, thick, black!70!white, dotted, line width=0.4mm] (U2) to (D2);
     \draw[-, thick, black!70!white, dotted, line width=0.4mm] (U3) to (D3);

     \draw[-, thick, blue, line width=0.4mm] (C1) to (C2);
     \draw[-, thick, red, line width=0.4mm] (CC1) to (CC2);

	\path[draw=black, fill=black] (U2) circle[radius=0.05];
 \path[draw=black, fill=black] (U12) circle[radius=0.05];
 \path[draw=black, fill=black] (U4M) circle[radius=0.05];
 \path[draw=black, fill=black] (U3) circle[radius=0.05];

\end{tikzpicture}}}}=\vcenter{\hbox{\scalebox{0.6}{\includegraphics[]{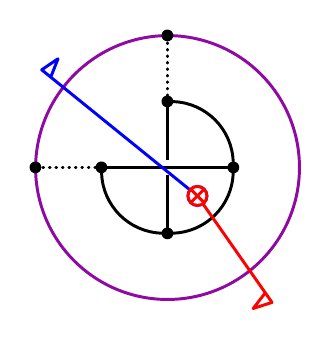}}}}, \quad \vcenter{\hbox{\scalebox{0.6}{}}}=\vcenter{\hbox{\scalebox{0.6}{\includegraphics[]{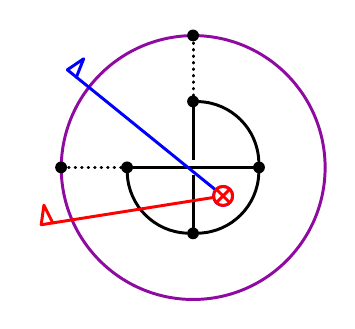}}}}\nonumber.
\end{equation}
In particular, initial and final-state cuts are ``anchored'' to the same point, and one choice of cuts is obtained from the other by rotating the cuts about the anchor. \emph{In other words, we find that diagrams that cancel each other's singularities through the KLN theorem are identified by the same placement of the puncture}. We will formalise the idea of ``anchored'' cuts by the notion of graph embedding. 

\subsection{Embeddings and cuts}

\label{sec:embeddings}

In the context of topological graph theory a graph embedding usually refers to representations of graphs on 2D-surfaces such that no edge of the graph intersects any other edge. A natural question is what are all of the topologically inequivalent surfaces that can embed a specific graph? Clearly, there always exists at least one surface (for finite graphs), since any graph can be drawn without crossings in Euclidean $\RR^3$ and can thus be embedded in the surface consisting of the ``fattened up'' graph, where edges are blown up into tubes.
In the following, we will ask a different question. 
Within the list of all topologically inequivalent surfaces that embed a graph, where can one hole/puncture be placed?

In fact, we will try to enumerate the possibilities of a single puncture placement with respect to the graph itself. Since trees (undirected graphs with no cycles) can be embedded in the sphere (the plane), all puncture placements are equivalent. For graphs with cycles, each cycle can host a puncture, and in fact each cycle can wind around that puncture multiple times. Since, for our purposes we do not really care what surface the puncture is a part of, we will focus on the properties of the graph in the presence of a single puncture, allowing the rest of the graph to embed in whatever surface can host it. In fact, to characterize the puncture placement, it is sufficient to look at all the winding numbers of the cycles. 
\subsubsection{Definition of an embedding}
In order to be more precise we need to introduce a modicum of graph theory terminology. 
More information can be obtained in any graph theory textbook. 
A standard reference is \cite{bondy_graph_2008}. The topic of directed cycles and bases is reviewed in \cite{kavitha_cycle_2009}.

Let $\undG=(\edges_u,\vertices)$ be an undirected graph where $\vertices$ is a ground set of vertices and $\edges_u$ is a collection of unordered pairs of vertices.
Any undirected graph can be endowed with an orientation, a choice per edge of a tail and a head vertex, turning it into a directed graph or digraph $\dirG=(\edges,\vertices)$, where pairs of edges are now ordered. This orientation will later be identified with the momentum flow in the edges, and can be fixed arbitrarily as long as it is kept consistent throughout the procedure. 

\partitle{Undirected cycles} An undirected cycle in $G_u$ is a subset of edges for which all vertices in the subgraph have degree 2, i.e. the edges form a non-self-intersecting closed path. We may identify the cycle by a characteristic vector $\cycle \in \ZZ_2^\abs{\edges}$, where $\cycle_e=1$ if the edge $e$ belongs to the cycle, and $c_e=0$ otherwise. Let $\allcycles$ denote the set of all such cycles. Cycles can be concatenated by adding their respective characteristic vectors in $\ZZ_2$.

There exists a subset of cycles 
\begin{equation}
\basis=\{\basecycle_1,\dots,\basecycle_L\} \subset \allcycles,
\end{equation}
where $L$ is the number of loops of the graph, such that any other cycle can be written as a unique linear combination of the elements of $\basis$, with coefficients in $\ZZ_2$. $\basis$ is an \emph{undirected cycle basis}. In particular, defining the $\abs{\edges}\times L$ matrix $B$ with components $B_{ec}=(b_c)_e$ that collects as columns all the cycle basis elements, each cycle may be written
\begin{equation}
\cycle = B \cycle_\basis,
\end{equation}
where $\cycle_\basis \in \ZZ_2^L$ are the coefficients of the decomposition of the cycle in the basis. Here and in the following, we will be interested in fundamental bases only, namely those that are constructed from a spanning tree and such that each cycle in them is identified by an edge not in the tree.   

\partitle{Directed Cycles} An analogous story holds for directed graphs. However, for our purposes, we need to additionally endow the cycles with a consistent orientation, so that they can be identified with directed closed paths that can be walked through in a consistent direction. In order to encode this information, we now identify cycles of the directed graph by a characteristic vector $\cycle \in \{-1,0,1\}^{|\edges|}$, where $\cycle_e=1$ implies that the edge $e$ is selected with its original orientation in $G$, $\cycle_e=0$ that the edge is not part of the cycle and $\cycle_e=-1$ that the edge $e$ should be thought to be part of the cycle but with reversed orientation. For each cycle of the undirected graph, there are two possible ways to lift it to an oriented cycle in the directed graph. In particular, for a given oriented cycle $\cycle$, $-\cycle$ is also an oriented cycle, with opposite orientation. We collect all the oriented cycles in a set $\alldircycles$ with double the size with respect to the set $\allcycles$. 

In the same way as before, we can find a basis of oriented cycles such that any other oriented cycle is written as a linear combination of them, with coefficients, now, in $\{-1,0,1\}$. Moreover, such a basis can be found by taking the undirected counter-part $\basis$ and lifting all the oriented cycles in it by choosing an arbitrary orientation for each of them. For example, if the cycle basis derives from a spanning tree, then one may simply require the orientation of a cycle to agree with that of the edge not contained in the tree identifying it. If $\mathcal{D}=\{\orientedbasecycle_1,\dots,\orientedbasecycle_L\}$ is such a lifted cycle basis, then again is a basis for all cycles $d \in \alldircycles$, meaning that $d$ can be written as a linear combination of the elements of $\mathcal{D}$ with coefficients in $\{-1,0,1\}$.

We can straightforwardly lift all the notation and definitions from undirected cycles to directed ones. We will write $D \in \ZZ^\abs{\edges} \times \ZZ^L$ for the directed analog of the cycle matrix $B$, meaning that $D_{ec}=(d_c)_e$ and write $c_\mathcal{D} \in \ZZ^L$ for the components of $c \in \alldircycles$ in the basis $\mathcal{D}$. Finally, for future convenience, we also define the support of a vector $v$ as $\sup(v)=\{e\in\edges \, | \, v_e\neq 0\}$.

\partitle{Embeddings and winding numbers}
Now given such a directed cycle basis $\mathcal{D}=\{\orientedbasecycle_1,\dots,\orientedbasecycle_\ell\}$ we can assign  a winding number $\wind_i \in \ZZ$ to each basis cycle $\orientedbasecycle_i$.
This yields a vector $\wind \in \ZZ^L$. 
The winding number $\wind_i$ corresponds to the number of times $\orientedbasecycle_i$, winds around the hole. 
For any other cycle $\cycle \in \alldircycles$, we can straightforwardly obtain the induced winding number

\begin{equation}
\omega_\cycle= \cycle_\mathcal{D} \cdot \wind.
\label{eq:indwind}
\end{equation}
In much the same way that the fundamental group of a manifold identifies the genus of a surfaces without the necessity of an ambient space, associating a winding number to each basis cycle (and thus by extension all cycles) identifies the ``location'' of the hole without needing to embed it in any surface. 
The thruple $(\dirG,\mathcal{D},\wind)$ is then what we call an \textit{embedding}. 
We will omit the dependence on $\mathcal{D}$ as it is implied by the definition of $\wind$ and therefore write $\emb$. Note that we are only interested in winding signatures $\wind$ up to an overall sign, i.e. $\wind \equiv -\wind$.

 \subsubsection{Embeddings as equivalence classes of rooted cuts}
 \label{sec:emb_cuts}

An embedding can be equivalently characterised by the collections of edges that need to be deleted from it in order to unwind it from the hole. We call these collections rooted cuts. Of all rooted cuts, a specific subset is of interest for us, namely simple rooted cuts: these are the collection of edges that specifically open up the embedding into a connected forward-scattering diagram.

\partitle{Rooted cuts} We now introduce the notion of anchored or rooted cut. Given an embedding $\emb$ and the associated cycle matrix $D$, a rooted cut is a vector $r \in \ZZ^\abs{\edges}$ that solves the following linear system:
\begin{equation}
 \transpose{D}r = \wind.
\label{eq:rooted_cut}
\end{equation}

This is an affine constraint on a vector space and therefore the set of all such cuts $R_{\emb}$
forms an affine space. 
The associated translation vector space is given by the null space of $\transpose{D}$, and can be though of as the anchored cuts anchored to the puncture outside of the graph (such that $\wind = 0$). 
These are true cuts in the graph theory sense (they disconnect the graph, bi-partitioning the vertices) and in fact this space is usually called the bond-space. Let $R_{\emb}$ be the collection of all rooted cuts. 
Any representative rooted cut $r \in R_{\emb}$ can be mapped to any other rooted cut by translating by a full cut $t \in R_{(G,0)}$. 

\partitle{Simple rooted cuts} We now define \textit{simple rooted cuts}: a rooted cut $r$ is simple if $G \setminus \sup(r)$, the graph obtained by removing the edges such that $r_e \neq 0$, is still connected. Intuitively, a simple rooted cut identifies a ray starting at the hole and ending at infinity, crossing a number of edges of the graph in the process. It opens up the vacuum graph into a forward scattering graph: the cut edges are doubled into external edges carrying the same momentum, and their energy flow is determined by the specific sign of $r_e$.

For any non-trivial embedding $\wind \neq 0$, one can always find simple rooted cuts that satisfy \eqref{eq:rooted_cut}. We collect all such vectors in a set, an equivalence class of cuts $[r]_{\emb} \subset R_{\emb}$. Note that it is not an affine space anymore, but a set of equivalent cuts, that can be related to each other by a translating cut $t \in R_{(G,0)}$. 
It is this class of cuts that define the set of forward scattering diagrams that, when summed together, are infrared finite by the KLN theorem. 

\partitle{Quark number} Assume now that the edges of the graph have an additional label establishing whether they are a fermion, an anti-fermion or a boson. Typically, this characterisation, which is usually understood in terms of fermion flow, is encoded in a sign $\q(e)\in\{-1,0,1\}$. If $\q(e)=0$, then the edge corresponds to the propagation of a boson, if $\q(e)=1$ it corresponds to the propagation of a fermion with flow aligned to the underlying orientation of the digraph, if $\q(e)=-1$ it corresponds to the propagation of a fermion with flow opposite to the underlying orientation of the digraph (recall that the underlying orientation is understood in terms of momentum flow). Then, to each cut we can assign the quark number 
\begin{equation}
\q(r)=\sum_{e\in\edges} r_e \q(e).
\end{equation}
Since fermions, in QCD vacuum diagrams, always appear in closed loops, we have that $\q(t)=0$ for any translation cut $t\in R_{(G,0)}$. Hence, $\q(r)=\q(r')$ for all $r,r'\in R_{\emb}$. In other words, the quark number is a conserved quantity across all cuts in the same embedding, and we denote it by $\q_{\emb}$.



\subsubsection{Generation pipeline} 
\label{sec:emb_generation_pip}

Now that we have defined an embedding and explored its relationship with equivalence classes of rooted cuts, we are able to outline a procedure to generate them.  Since $\ZZ$ is not finite, the number of embeddings will never be. In particular, we can increase indefinitely the number of windings for each basis cycle. Thus, we need to restrict and order our enumeration. Additionally, we want to impose physically-sensitive restrictions on the allowed embeddings.

\partitle{Generating embeddings} In the previous section, we identified an embedding with an equivalence class of simple rooted cuts. One can be translated into the other by eq.~\eqref{eq:rooted_cut}. We thus generate all equivalence classes of simple rooted cuts, which in turn gives us all possible embeddings. In order to do so, we enumerate all non-disconnecting edge sets, namely all subsets $\mathcal{R} \subset \mathcal{E}$ such that, deleting from the graph the edges in $\mathcal{R}$, leaves one connected component. From each subset $\mathcal{R}$ we construct all vectors $r\in \ZZ_n^{|\edges|}$ such that $r_e\neq 0$ for all $e\in\mathcal{R}$ and $r_e=0$ otherwise, i.e. all vectors $r$ such that $\sup(r)=\mathcal{R}$. $r$ is then a rooted cut satisfying eq.~\eqref{eq:rooted_cut}, for some choice of winding numbers. By using $\ZZ_n$ we bound the possible winding numbers to be less than $n$. We collect all of these rooted cuts in $R_{\mathrm{tot}}$.

Once the set $R_{\mathrm{tot}}$ is determined, we can use eq.\eqref{eq:rooted_cut} to classify the elements $R_{\mathrm{tot}}$ into different equivalence classes, each associated to a winding signature. We essentially quotient by the kernel of $\transpose{D}$:
\begin{equation}
   \mathrm{Emb}= R_{\mathrm{tot}} \mathbin{/} \ker{\transpose{D}} = \{[r_1]_{(G,\wind_1)},[r_2]_{(G,\wind_2)},\dots\},
    \label{eq:classificationquotient}
\end{equation}
obtaining all embeddings corresponding to those cuts.

\partitle{Filtering embeddings} Generating from anchored cuts is nice for multiple reasons. For one, we can open up the graph along the anchored cut, obtaining a forward scattering graph. This allows us to validate the generation of cuts by comparing to a diagram generator.
Second, it enables physically-motivated filters on the types of embeddings we want. Since we are interested in QCD corrections to the DIS process, the vacuum digraphs $G$ under consideration will always have two electron edges, $a_1,a_2$, and two photon edges, $a_3,a_4$. In our case we only consider a subset $R_\text{DIS} \subset R_\text{tot}$, where we only allow cuts $r\in R_\text{DIS}$ that pass the following criterion:
\begin{itemize}
    \item Each cut $r$ must not have support on either photon edge, $a_3,a_4\notin \sup(r)$, i.e. the cut is excluded if it crosses a photon edge. In other words, we are not interested in initial or final-state cuts that include a photon.
    \item The cut $r$ must cross one electron once. In other words either $r_{a_1}=1$ and $r_{a_2}=0$, or $r_{a_1}=0$ and $r_{a_2}=1$. Initial and final-state cuts must contain one electron each.
    \item The cut $r$ does not leave a disconnected partonic tensor i.e. $G \setminus (\sup(r)\cup\{a_1,a_2\}) $ is connected. This constraint can be rephrased by simply saying that the cut leaves one connected component upon deletion of its edges plus all electron edges.
\end{itemize}

Once $R_\text{DIS}$ has been determined in this way, we may construct all equivalence classes of them, and thus enumerate all embeddings, $R_{\text{DIS}}\mathbin{/} \ker{\transpose{D}}$. We then impose further filters on the embeddings contained in $R_{\text{DIS}}\mathbin{/} \ker{\transpose{D}}$. Given an equivalence class $[r]_{\emb}\in R_{\text{DIS}}\mathbin{/} \ker{\transpose{D}}$ we only allow it if:
\begin{itemize}
    \item There exists two cuts $r_1,r_2\in[r]_{\emb}$ that cut different electrons, namely $(r_1)_{a_1}=(r_2)_{a_2}=1$ and $(r_1)_{a_2}=(r_2)_{a_1}=0$. This makes sure that each embedding has both an initial and a final-state cut.
    \item There exists a cut in $[r]_{\emb}$ such that $\abs{r_e} \leq 1$ for all edges $e \in \edges$.
\end{itemize}

Let us call the subset of embeddings thus obtained $\embdis$. The set of embeddings in $\embdis$ forms a gauge-invariant subset. Any embedding that does not satisfy the last property is dubbed a ``higher-winding embedding''. This last property is the least physically motivated constraint. We briefly discuss their contribution concomitantly with the results in sect.~\ref{sec:results}. 




\subsubsection{Automorphism invariant embeddings}

An unfortunate fact about the above definition for an embedding is that it depends on the choice of fundamental cycle basis. This in turn means that it is also not invariant under graph automorphisms. 

\partitle{Graph Automorphisms} An isomorphism of a graph $G=(\edges,\vertices)$ is a tuple of bijections $\phi:\edges\leftrightarrow\edges'$, $\theta: \vertices\leftrightarrow\vertices'$, such that the adjacency is preserved. In other words, for an edge $e=(v_1,v_2)$, then $\phi(e) = (\theta(v_1),\theta(v_2))$. If $\edges'=\edges$ and $\vertices=\vertices'$, then we talk about automorphism of a graph and the bijections are just permutations of the edge and vertex labels. Applying an automorphism to an embedding will inevitably change the cycle subgraphs and thus the winding signature.

\partitle{Canonical embeddings} Thus we need an automorphism invariant characterisation of the embedding $\emb$. Instead of focusing on the winding numbers of the cycle basis, we consider the winding numbers $\omega_c$ for all cycles $c \in \alldircycles$. This set is clearly label independent, and basis independent. 
Given an embedding signature $\wind$ we define the following vector 
\begin{equation}
\omega =\frac{1}{2} \sum_{c \in \alldircycles} \omega_c c   \in \ZZ^\abs{\edges},
\label{eq:emb_circulation}
\end{equation}
which we call the circulation of the embedding\footnote{In particular there exists an orientation reassignment of $G$ such that $\omega$ satisfies ``momentum conservation'' at each vertex. Vectors that satisfy such constraints are called circulations in the technical literature.}. Eq.~\eqref{eq:emb_circulation} sums over pairs of cycles $c,-c$ with corresponding winding numbers $\omega_c=-\omega_{-c}$, which explains the factor of $\frac{1}{2}$ that undoes such double counting. 
Any choice of cycle basis and winding signature $\wind$ that identify the same embedding will be mapped on to the same circulation. 
However we are still sensitive to the chosen orientation $G$ of the underlying graph $\undG$.
Since the orientation choice is arbitrary, we can reorient the graph in such a way that $\omega_e \geq 0$. 
The embedding $\emb$ is equivalently represented by the undirected graph $G_\omega$, obtained by adding the weight  $\abs{\omega_e}$ to each edge $e$ in $\undG$.
At this point we can can create a canonical representative of the class of isomorphic embeddings, by choosing a canonical form of $G_\omega$ (using nauty for example \cite{mckay_practical_2014}). The set of embeddings of eq.~\eqref{eq:classificationquotient} can be further reduced by only considering those emdeddings that have different canonical forms, and choosing a representative for each set of isomorphic embeddings. We call this set, $\mathrm{Emb}\mathbin{/}\mathrm{Aut}$, and thus the subset relevant for our DIS computation is
\begin{equation}
\label{eq:embdis_quot}
   \overline{\embdis}= \embdis \mathbin{/}\mathrm{Aut},
\end{equation}
as long as adequate symmetry factors are correctly taken into account. In particular we assign to the embedding with canonical form $G_\omega$ the symmetry factor $1/\mathrm{sym}_{G_\omega}=\abs{\mathrm{Aut}(G_\omega)}/\abs{\mathrm{Aut}(\undG)} \leq 1$. 

\partitle{NLO DIS generation} For DIS at NLO, there are only two vacuum diagrams that contribute after performing color algebra, the ``self-energy'', and the ``double-triangle'' diagrams:

\begin{equation}
G_\text{DT}=\raisebox{-1.2cm}{\scalebox{0.5}{\includegraphics[]{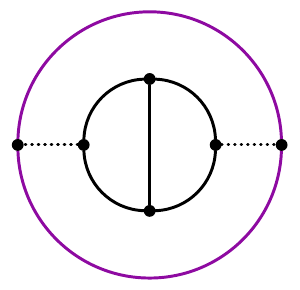}}}, \quad
G_\text{ST}=\raisebox{-1.2cm}{\scalebox{0.5}{\includegraphics[]{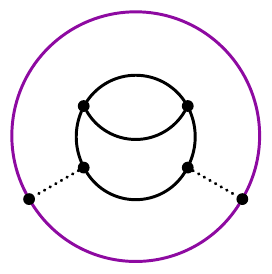}}}.
\end{equation}
For $G_\text{SE}$ we classified 124 rooted cuts into 34 embeddings and for $G_\text{DT}$ we classified 70 cuts into 20 embeddings. In total, we find 54 non-isomorphic embeddings. The symmetry factors for all embeddings evaluate to one. The winding numbers for all these embeddings can be found in appendix~\ref{sec:emb_tables}.

\partitle{Code}
The code for the generation pipeline that we developed is hosted on github at this \href{https://github.com/lcnbr/dis}{link}. The canonicalisation was done using \href{https://symbolica.io}{Symbolica}'s function \href{https://docs.rs/symbolica/latest/symbolica/graph/struct.Graph.html#method.canonize}{\texttt{canonise()}}, that implements the nauty algorithm in Rust. We validated the embedding generation at NLO by comparing the forward-scattering diagrams given by our cut generation procedure with an independent forward-scattering \href{https://github.com/alphal00p/gammaloop/releases/tag/v0.3.3}{diagram generator}.

\subsection{Examples}
\label{sec:embedding_examples_1}

We will now go through an example generation of an embedding from a cut.
We start from the following vacuum graph, for which we have chosen an edge orientation, corresponding to a choice of momentum flow:
\begin{equation*}
    G=\raisebox{-23mm}{\includegraphics[width=4cm]{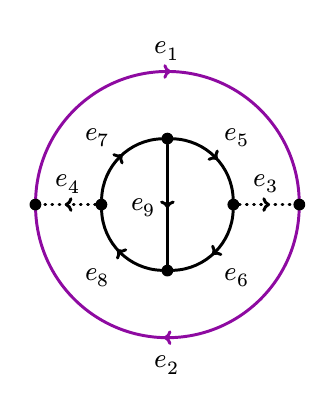}}.
\end{equation*}
The purple line, in deep inelastic scattering, is identified with the electron, while the dotted line is identified with the photon. In the following we will denote by $e_i \in \ZZ_2^\abs{\edges}$ the single edge subgraph containing only the edge $e_i$:
\begin{equation}
e_i=(\underbrace{0,\ldots,0}_{\substack{i-1 \\ \text{times}}},1,\underbrace{0,\ldots,0}_{\substack{|\edges|-i\\ \text{times}}})\,.
\end{equation}
The first thing is to choose a cycle basis. To obtain one, we simply choose a rooted spanning tree: the cycles in the basis will be in one-to-one relation with the edges not contained in it. One possibility is the following spanning tree 
\begin{align}
T&=-e_3-e_5+e_9+e_8+e_4 = \transpose{\begin{pmatrix*}[r]
0,0,-1,1,-1,0,0,1,1
\end{pmatrix*}} 
=\raisebox{-9mm}{\includegraphics[]{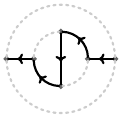}
}.    
\end{align}
Notice that those edges coming with minus signs have a reversed orientation with respect to the original graph.
The edges not in the tree are $e_1,e_2,e_6,e_7$, and the cycles in the induces cycle basis, with orientation aligned with their corresponding edge, is: 
\begin{align*}
\mathcal{D}&=\{
        e_1+T, e_2-T, e_6-e_9+e_5,e_7+e_9+e_8
    \}=\{d_1,d_2,d_3,d_4\}\\&=\left\{\raisebox{-9mm}{\includegraphics[]{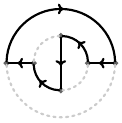}},\raisebox{-9mm}{\includegraphics[]{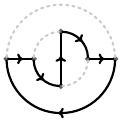}},\raisebox{-9mm}{\includegraphics[]{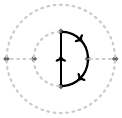}},\raisebox{-9mm}{\includegraphics[]{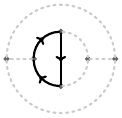}}\right\}.
\end{align*}
All other cycles can be expressed as linear combinations of these 4 cycles. For example $e_1 + e_2= d_1 + d_2$ and $e_5+e_6+e_7+e_8=d_3+d_4$. The cycle matrix is then:
\begin{equation}
\transpose{D} = 
\begin{pmatrix*}[r]
\:1 & \phantom{-}0 & -1 &  \phantom{-}1 & -1 & \phantom{-}0 & \phantom{-}0 &  1 &  1 \\
0 & 1 &  1 & -1 &  1 & 0 & 0 & -1 & -1 \\
0 & 0 &  0 &  0 &  1 & 1 & 0 &  0 & -1 \\
0 & 0 &  0 &  0 &  0 & 0 & 1 &  1 &  1 \\
\end{pmatrix*}=\begin{pmatrix*}[r]
\transpose{d_1}\\
\transpose{d_2}\\
\transpose{d_3}\\
\transpose{d_4}
\end{pmatrix*}.
\end{equation}
\begin{figure}
    \begin{minipage}{\textwidth}
        \centering
        \begin{minipage}{0.55\textwidth}
            \includegraphics[width=0.8\linewidth]{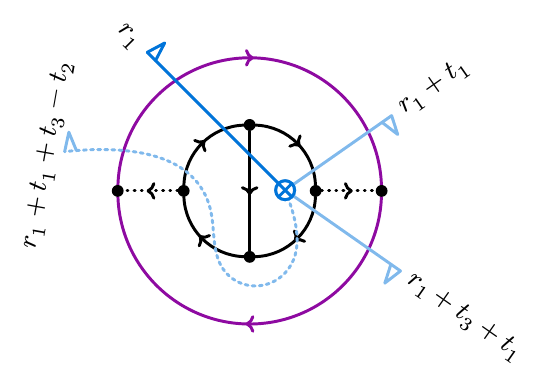}
        \end{minipage}%
        \hfill
        \begin{minipage}{0.45\textwidth}
            \subcaption{The cut $r_1=e_1+e_7-e_9$, along with 3 other cuts obtained by translation. The cut that is dashed is an example of a cut that would not be in $R_\text{DIS}$ because it would create a disconnected partonic tensor. }
        \end{minipage}
    \end{minipage}
    \vspace{1em}

    \begin{minipage}{\textwidth}
        \centering
        \begin{minipage}{0.55\textwidth}
            \includegraphics[width=0.8\linewidth]{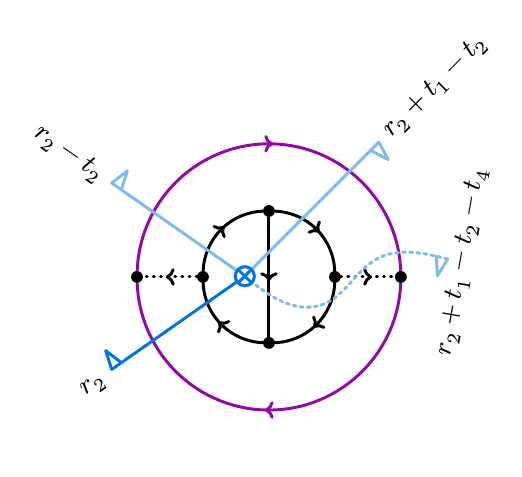}
        \end{minipage}%
        \hfill
        \begin{minipage}{0.45\textwidth}
            \subcaption{The cut $r_2=e_2+e_8$, along with 3 other cuts obtained by translation. The cut that is dashed is an example of a cut that would not be in $R_\text{DIS}$ because it cuts a photon edge (dotted in the picture).}
        \end{minipage}
    \end{minipage}
    \vspace{1em}

    \begin{minipage}{\textwidth}
        \centering
        \begin{minipage}{0.55\textwidth}
            \includegraphics[width=0.8\linewidth]{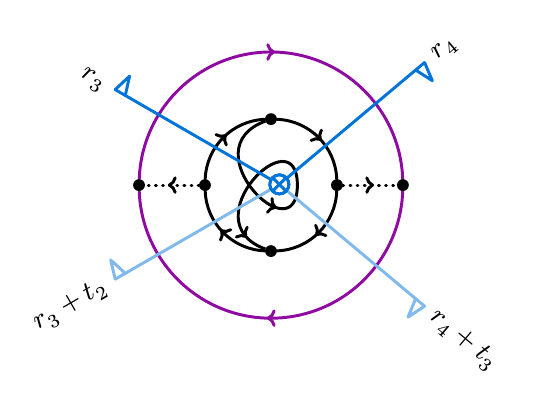} 
        \end{minipage}%
        \hfill
        \begin{minipage}{0.45\textwidth}
            \subcaption{The cuts $r_3=e_1+e_7-2e_9$ and $r_4=e_1+e_5-e_9$, belonging to the same embedding. $r_3$ and $r_4$ share the same root.}
        \end{minipage}
    \end{minipage}

    \caption{
    In the above graph, the purple lines denote the deep inelastic scattering electrons, while the dotted lines the photons. The cuts $R_\text{gen}$ are visualised in dark blue, along with a selection of other cuts (in light blue) in the same equivalence classes, obtained from $r_1,r_2,r_3$ and $r_4$ by translation with elements of the basis of $\ker(\transpose{D})$. The translation cuts $t_i$, $i=1,...,5$ are shown in fig.~\ref{fig:example_cut_translations}. All cuts are visualised as lines cutting through edges with a root, denoted by the crossed circle and a flag at the other end. This is the puncture or hole characterising the embedding. The flag on each cut represents the orientation of the cut. If the flow of a cut edge $e$ is aligned with this flag, then $r_e>0$.}
\label{fig:example_cuts}
\end{figure}

\begin{figure}
\centering
\includegraphics[height=6cm]{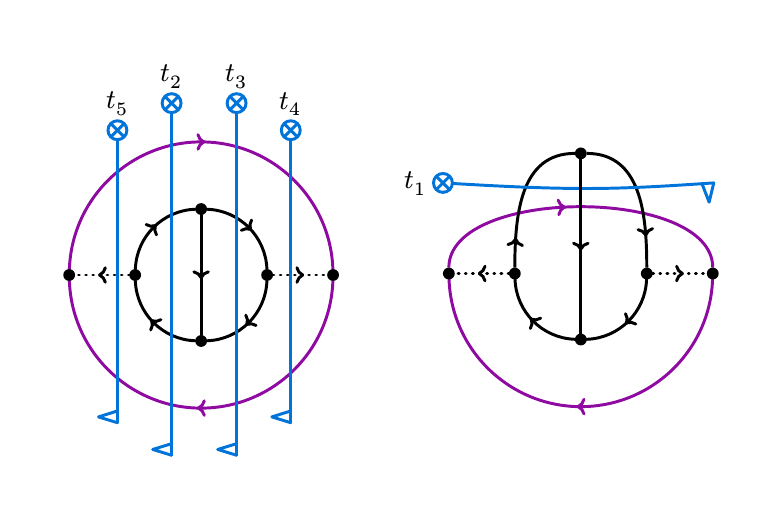}
\caption{The 5 base cut translations. Notice that all of them disconnect the graph, and bipartition the vertices.}
\label{fig:example_cut_translations}
\end{figure}

Now, let us imagine that we have a set of rooted cuts $R_\text{gen}$, given by our generation pipeline. As described in \ref{sec:emb_generation_pip}, these cuts $r\in R_\text{gen}$ are generated by first taking all subsets of the edges $\mathcal{R}\subset\mathcal{E}$, such that the graph obtained by deleting these edges, $G \setminus \mathcal{R}$, is still connected. We make sure that each of these sets contains only one of the two electron lines ($e_1$ and $e_2$), and no photon line ($e_4$ and $e_3$).
Some valid subsets are, for example, $\mathcal{R}_1=\{e_1,e_7,e_9\}$ and $\mathcal{R}_2=\{e_2,e_8\}$. Finally, we can lift these subsets of edges to rooted cuts by constructing all vectors $r\in \ZZ_n$ such that $\sup(r)=\mathcal{R}$. Let us take the following set of cuts: 
\begin{equation}
    R_\text{gen}= \{e_1+e_7-e_9,e_2+e_8,e_1+e_7-2e_9,e_1+e_5-e_9,\dots\}=\{r_1,r_2,r_3,r_4,\dots\}. 
\end{equation}
Notice $(r_3)_9=-2$ tells us this edge is cut twice in the reverse orientation compared to the one in the underlying graph. We have, for example, $\sup(r_3)=\{e_1,e_7,e_9\}$, which indeed satisfies the properties stated above. The set $R_\text{gen}$ contains many cuts, but for the purposes of this example we focus on the first four, and carry the classification procedure in terms of embeddings for them only. Fig.~\ref{fig:example_cuts} contains drawings for these four cuts.
  By drawing we directly notice that $r_3$ and $r_4$ must be in fact part of the same equivalence class. This can be confirmed by computing the winding numbers.
Using the cycle matrix $D$ we can straightforwardly compute them:
\begin{align}
&\transpose{D} r_1 = \begin{pmatrix*}[r]
        0\\
        1\\
        1\\
        0
    \end{pmatrix*}=\wind^{(1)}, \quad &\transpose{D} r_2 = \begin{pmatrix*}[r]
        1\\
        0\\
        0\\
        1
    \end{pmatrix*}=\wind^{(2)},\\ \quad &\transpose{D} r_3 = \begin{pmatrix*}[r]
        -1\\
        2\\
        2\\
        -1
    \end{pmatrix*}=\wind^{(3)}, \quad &\transpose{D} r_4 = \begin{pmatrix*}[r]
        -1\\
        2\\
        2\\
        -1
    \end{pmatrix*}=\wind^{(3)}.
\end{align}
We can see that the 4 cuts are now classified into three winding classes: 
\begin{equation}
    r_1 \in [r_1]_{(G,\wind^{(1)})}, \quad r_2 \in [r_2]_{(G,\wind^{(2)})}, \quad r_3,r_4 \in [r_3]_{(G,\wind^{(3)})}.
\end{equation}
We can also explore the other cuts in the same embedding by translating them by vector contained in the null space of $\transpose{D}$. These cuts are also elements of $R_\text{gen}$, of course. The null space of $\transpose{D}$ is spanned by the following five vectors:
\begin{equation}
    \{\underbrace{e_5-e_7+e_9}_{t_1},\underbrace{-e_1+e_2-e_7+e_8}_{t_2},\underbrace{-e_1+e_2-e_5+e_6}_{t_3},\underbrace{-e_1+e_2+e_4}_{t_4},\underbrace{e_1-e_2+e_3}_{t_5}\},
\end{equation}
which are drawn in fig.~\ref{fig:example_cut_translations}. 
\begin{figure}
\centering
\includegraphics[height=2.5cm]{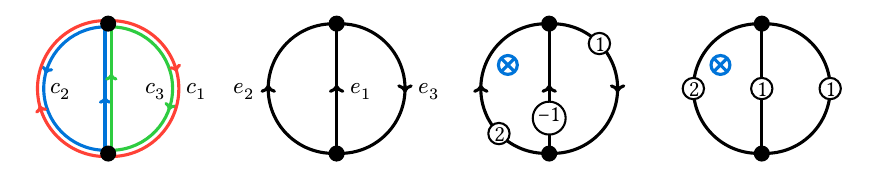}
\caption{
An isomorphism invariant characterisation of an embedding of the sunrise graph.}
\label{fig:sunrise}
\end{figure}

We classified these four cuts in terms of their respective embeddings, which can be in turn seen as an embedding-generation procedure: carrying out the same exact procedure for all cuts in $R_\text{gen}$ will provide all embeddings. 

However, two of the embeddings under consideration are isomorphic: $(G,\wind^{(1)}) \simeq (G,\wind^{(2)})$. To compute the isomorphism independent representation of the embedding we need to enumerate all the cycles of $G$. Since this is a four loop graph, there are quite a few (twelve) cycles, so let us go through the procedure for a simpler graph, the sunrise. The edge labels and orientations are shown in fig.~\ref{fig:sunrise}. We have six directed cycles,
\begin{equation}
    c_1 = e_2+e_3, \quad c_2 = e_1-e_2, \quad c_3= e_1+e_3,
\end{equation}
and their reversed orientations $-c_1,-c_2,-c_3$. We show $c_1,c_2,c_3$ in fig.~\ref{fig:sunrise}. Taking a cycle basis, e.g. $c_1,c_2$, with winding numbers $\wind_1,\wind_2$, all other winding numbers can be derived using eq.~\eqref{eq:indwind}: 
\begin{align}
    \omega_{c_1}=\wind_1=1=-\omega_{-c_1},\quad \omega_{c_2}=\wind_2=-1=-\omega_{-c_2},\nonumber \\
\omega_{c_3}=\omega_{c_1+c_2}=\omega_{c_1}+\omega_{c_2}=0=\omega_{-c_3}.
\end{align}
We form the vector that sums all cycles weighted by their winding number (see eq.~\eqref{eq:emb_circulation}):
\begin{equation}    
\omega=\frac{1}{2}\sum_{\sigma \in \{\pm 1\}} \sigma(\omega_{\sigma c_1} c_1+\omega_{ \sigma c_2} c_2+\omega_{\sigma c_3} c_3)=c_1-c_2= e_2+e_3-e_1+e_2=
\begin{pmatrix*}[r]
    -1\\
    2\\
    1\\
\end{pmatrix*}.
\end{equation}
The isomorphism invariant characterisation is now just the undirected graph with weights $\abs{\omega}$, and this is also shown in fig.~\ref{fig:sunrise}. We can now repeat this procedure for $(G,\wind^{(1)})$ and $(G,\wind^{(2)})$ and we obtain:
\begin{equation}
    G_{\omega^{(1)}}=\raisebox{-20mm}{\includegraphics[width=4cm]{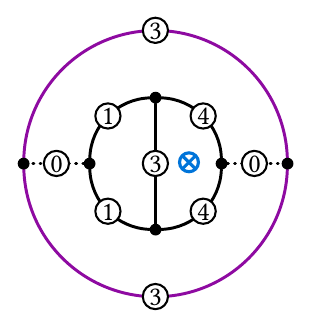}},\quad G_{\omega^{(2)}}=\raisebox{-20mm}{\includegraphics[width=4cm]{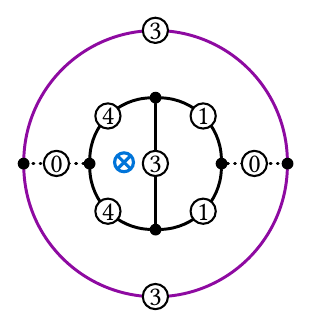}},
\end{equation}
notice that it is now clear that these two are infact the same embedding, since one can be rotated into the other. Thus there exists an automorphism of the graph that maps one to the other. Therefore where we started with 4 different cuts, we obtain just 2 equivalence classes:
\begin{equation}
    r_1,r_2 \in [r_1]_{(G,\wind^{(1)})}, \quad r_3,r_4 \in [r_3]_{(G,\wind^{(3)})},
\end{equation}
where we have chosen $(G,\wind^{(1)})$ as the canonical representative between $(G,\wind^{(1)})$ and $(G,\wind^{(2)})$.
Notice that the while the automorphism group  of $\undG$ has size $8$, that of $G_{\omega^{(1)}}$ and $(G,\wind^{(3)})$ both have size $4$. 
Thus we assign a symmetry factor of $\frac{1}{2}$ to $(G,\wind^{(1)})$ and $(G,\wind^{(3)})$.

\section{Integrands of embeddings}
\label{sec:integrands_of_embeddings}

In the previous section, we defined embeddings. In this one, we will show how to associate a loop integral to an embedding, obtained by applying Feynman rules. The only effect on the integrand of the additional topological structure introduced by the puncture is encoded in modified momentum conservation conditions consistent with the simplest clustering criterion. In other words, once an embedding is defined, it is then routed according to momentum conservation conditions that follow from its topological structure, and its corresponding loop integral is then determined by ordinary application of Feynman rules on the routed object.

\subsection{Simplest clustering and momentum conservation conditions}

We have seen that an embedding $\emb$ defines an equivalence class of cuts $[r]_{\emb}$ that unwind it from the puncture. In deep inelastic scattering, since initial and final states have the same definition, $e^-+X\rightarrow e^-+Y$, with $X,Y$ being collections of partons, there is an additional symmetry that forces us to partition the set of cuts $[r]_{\emb}$ into initial and final-state cuts. Of course, there is no reason to prefer one partition over the other, which implies that when computing the cross-section we will have to include the contribution of both partitions.

Let $a_1,a_2\in\edges$ are the edge labels for the two electrons in $G$. We can choose $a_1$ to be the initial-state electron. The embedding plus this choice is represented as a thruple $\embs=(G,\wind,a_1)$. In the future, for compactness, we will still refer to $\Gamma$ as an embedding, although it also assumes a choice of initial-state electron. The complementary choice is denoted as $\bar{\Gamma}=(G,\wind,a_2)$. The choice of initial-state electron induces a bipartition of the cuts. We let $[r]_{\embsI}\cup [r]_{\embsF}=[r]_{\emb}$ be the bipartition of the cuts such that the cuts in $[r]_{\embsI}$ cut the edge $a_1$ but not $a_2$ and the cuts in $[r]_{\embsF}$ cut the edge $a_2$ but $a_1$. 

Now focus on $\embsI$. We know that initial-state cuts will only belong to the set $[r]_{\embsI}$, and will cut through the edge $a_1$. Recall that the simplest clustering criterion forces the momenta of the partons in the initial-state cut of an interference diagram to sum up to $p$. Thus, implementing the simplest clustering criterion diagrammatically is equivalent to imposing that the embedding $\embsI$ is routed according to the following momentum conservation conditions:
\begin{align}
&\sum_{e\in \edges}r_eq_e=p+p_{a_1}, \quad \forall r'\in[r]_{\embsI}, \label{eq:mom_cons_c}\\
&\sum_{e\in\edges} a_{ve} q_e=0,\quad \forall v\in\mathcal{V}, \label{eq:mom_cons_v}
\end{align}
where $a_{ve}$ are the components of the adjacency matrix of the digraph $G$. In other words, on top of ordinary momentum-conservation conditions at each vertex (eq.~\eqref{eq:mom_cons_v}), in eq.~\eqref{eq:mom_cons_c} we impose the momentum flowing through each cut in $[r]_{\embsI}$ to be equal to $p+p_{a_1}$, $p_{a_1}$ being the momentum of the electron. These constraints are linearly dependent. It is easy to see that a linearly independent basis of constraints is 
\begin{align}
&\sum_{e\in \edges }r_e q_e=p+p_{a_1}, \label{eq:mom_cons_c2}\\
&\sum_{e\in\edges} a_{ve} q_e=0,\quad \forall v\in\mathcal{V}\setminus\{v^\circ\}, \label{eq:mom_cons_v2}
\end{align}
where $v^\circ$ is any arbitrarily chosen vertex and the additional momentum-conservation condition is now imposed at one cut only (which we chose to coincide with the representative of $[r]_{\embsI}$). Since simplest clustering only amounts to a momentum conservation constraints, once the graph has been labelled according to the solution of eq.~\eqref{eq:mom_cons_c2} and eq.~\eqref{eq:mom_cons_v2} we are essentially finished and we are ready to build the integrand.

\subsection{Embeddings as Feynman diagrams}

Given an embedding $\embsI$ and its associated equivalence class of cuts $[c]_{\embsI}$ with representative $c$, we define its loop integral to be
\begin{align}
(2\pi)^{dL}\tilde{I}_{\embsI}(p,p_{a_1},p_{a_2})=&\int \left[\prod_{e\in\edges}\frac{\mathrm{d}^d q_e}{(q_e^2-m_e^2+i\varepsilon)}\right]\delta\left(p+p_{a_1}-\sum_{e\in \edges }r_e q_e\right)\times \nonumber \\
\times&\prod_{v\in \vertices\setminus\{v^\circ\}}\delta\left(\sum_{e\in\edges}a_{ve}q_e\right)\tilde{\mathcal{N}}(\{q_e\}_{e\in\edges};p,q).
\end{align}
In the above, $p_{a_1}$ and $p_{a_2}$ are the momenta of the two electron edges of the vacuum diagram. The same momenta appear on the right-hand side but under the generic notation of $q_e$ (more specifically, $q_{a_1}=p_{a_1}$ and $q_{a_2}=p_{a_2}$). $\mathcal{\tilde{N}}$ is a polynomial numerator obtained by Feynman rules. 

The first Dirac delta function imposes that the parton momenta in the cut $r$ sum up to $p$, i.e. the simplest clustering criterion. The product of Dirac delta functions instead imposes momentum conservation at each vertex $v$ aside from a root $v^\circ$ which can be chosen arbitrarily. These Dirac delta functions simply impose the constraints of eq.~\eqref{eq:mom_cons_c} and eq.~\eqref{eq:mom_cons_v}. 

In the following, we will denote by $I_{\embsI}(p,q)$ the truncated version of the integrand that arises by extracting the leptonic tensor $L^{\mu\nu}$ and the photon propagators. In other words, the set $\mathcal{E}$ will be associated with partonic edges only, and the numerator $\mathcal{N}$ will similarly only be constructed from the partonic part of the embedding:
\begin{equation}
\tilde{I}_{\embsI}(p,p_{a_1},p_{a_2})=\frac{(-\imath)^2}{(q^2)^2}L_{\mu\nu}(p_{a_1},p_{a_2})I_{\embsI}^{\mu\nu}(p,q=p_{a_1}-p_{a_2}).
\end{equation}
We will also suppress the Lorentz indices of $I_{\embsI}^{\mu\nu}$ to keep a lighter notation. After solving momentum conservation condition, the integral will reduce in dimensionality and take the form
\begin{equation}
\label{eq:emb_routed}
I_{\embsI}(p,q)=\int \left[\prod_{i=1}^L\frac{\mathrm{d}^d k_i}{(2\pi)^d }\right]\frac{\mathcal{N}(\{q_e(\{k_j\}_{j=1}^L;p,q)\}_{e\in\edges};p,q)}{\prod_{e\in \edges}q_e(\{k_j\}_{j=1}^L;p,q)^2}.
\end{equation}
where $q_e(\{k_j\}_{j=1}^L;p,q)$ is a linear combination of the input vectors \emph{with coefficients that need not be in }$\{\pm 1,0\}$. In particular the coefficients will be, in general, integers. Eq.~\eqref{eq:emb_routed} gives \emph{the integral obtained by straight-forwardly applying Feynman rules on the routed embedding}.

Because of the particularity of the momentum conservation condition imposed on the cut $r$, the identification of the integral $I_{\embsI}$ with a Feynman integral $V_{\embsI}$ for a given diagram $G$ is not immediate. At next-to-leading order, we find that all embedding integrands correspond to one loop Feynman graphs with at most five loop propagators:
\begin{align}
I_{\embsI}(p,q)&=y\frac{V_{\embsI}(p,q)}{[p^2]^n [(p+q)^2]^m} \label{eq:embedding_feynman}\\
V_{\embsI}(p,q)&=\int \frac{\mathrm{d}^dk}{(2\pi)^d}\frac{\mathcal{N}}{\prod_{i=1}^5 [(k+p_i)^2]^{\nu_i}},
\end{align}
where $p_i$ is a linear combination of $p$ and $q$, and $y\in\QQ$ is a rational constant. $n,m\in\NN_+ $ are positive or vanishing integers. 

One convenient aspect of defining the embedding integrand in this way is that now the integrand for any interference diagram can be straightforwardly obtained from it by simply changing propagators into on-shell Dirac delta functions according to standard cutting rules. In other words, consider an embedding $\embsI$ and two cuts $r_1\in[r]_{\embsI}$ and $r_2\in[r]_{\embsF}$ that define an interference diagram such that $r_1$ is the initial-state cut, while $r_2$ is the final-state cut. Let $r_{12}=r_1+r_2$ be the sum of the two rooted cuts. $\sup(r_{12})$ gives the set of all cut edges. We then define the cut signs to be
\begin{equation}
\sigma(r)_e=\begin{cases}
1 \quad &\text{if } r_e>0 \\
-1 \quad &\text{if } r_e<0 \\
0 \quad &\text{if } r_e=0
\end{cases}.
\end{equation}
Of course, we have $\sup(\sigma(r))=\sup(r)$. In terms of these vector of signs, the integrand for the interference diagram will be
\begin{equation}
I_{\embsI,(r_1,r_2)}(p,q)=\int \left[\prod_{i=1}^L\frac{\mathrm{d}^d k_i}{(2\pi)^d }\right]\frac{\mathcal{N}}{\prod_{e\in \edges\setminus\sup(r_{12})}q_e^2}\prod_{e\in \sup(r_{12})} \tilde{\delta}^{\sigma(r_{12})_e}(q_e).
\end{equation}
where $q_e$ should again be thought of as a linear function of the loop momenta $k_j$ and the input momenta $p$ and $q$ as in eq.~\eqref{eq:emb_routed}. This implies that all interference diagram that belong to the same embedding equivalence class inherit their momentum routing from the embedding itself. 

We now define the contribution of the embedding $\Gamma$ to the partonic tensor as the sum of the interference diagrams arising from the embedding:
\begin{equation}
\label{eq:w_interference}
    W_\Gamma(p,q) = \frac{1}{\text{av}(\q)} \sum_{(r_1,r_2) \in R_\Gamma} I_{\embsI,(r_1,r_2)}(p,q),
\end{equation} 
where the interefence diagrams are identified with the couplets of cuts in the set: 
\begin{equation}
    R_\Gamma = \left\{(r_1,r_2) \in [r]_{\embsI} \times[r]_{\embsF} \,\big{|}\,  G \setminus \sup(r_{12}) \text{ has two connected components}\right\}.
\end{equation}
$R_\Gamma$ selects only the initial and final state cuts that leave two connected components upon deletion up to spectators.
Of course, we have that the partonic tensor we will compute for a specified quark content $\q$ expresses as the sum of $W_\Gamma(p,q)$ for all embeddings $\Gamma$ that have the quark content $\mathfrak{q}$: 
\begin{equation}
W_\mathfrak{q}(p,q)=\sum_{\substack{\Gamma=(G,\mathbf{w},a) \\ \mathfrak{q}_\Gamma=\mathfrak{q}}}\frac{W_{\Gamma}(p,q)}{\mathrm{sym}_{(G,\mathbf{w})}} ,
\end{equation}
where the sum over $\Gamma$ includes all $(G,\mathbf{w})$ in $\overline{\embdis}$ (as defined in eq.~\eqref{eq:embdis_quot}) and all choices of initial-state electrons $a$ such that there exists $r'\in[r]_\Gamma$ with $r'\in \{-1,0,1\}^{|\edges|}$. $\q_\Gamma$ is the quark content associated with the embedding (see the discussion of sect.~\ref{sec:emb_cuts}). This set of embeddings is, of course, infrared-finite, and it is additionally gauge invariant. While this is a good choice of ``base'' embeddings, one should still explain why all higher-winding embeddings are excluded. In sect.~\ref{sec:results}, we discuss this aspect and the need for further physical constraints to define the exact set of embeddings contributing to the partonic tensor. The embeddings we computed, together with their corresponding integrand, are tabulated in sect.~\ref{sec:emb_tables}.

\subsection{Four types of integrands}
\label{sec:four_types_of_embeddings}
\begin{figure}
\centering
\begin{subfigure}[t]{0.4\textwidth}
\centering
\input{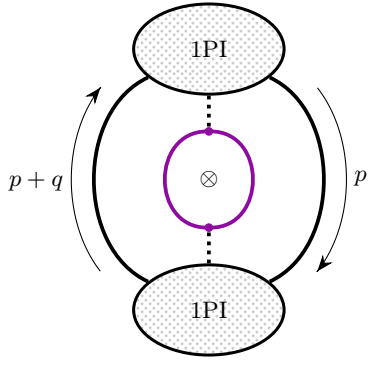}
\caption{$n,m\ge 0$ in eq.~\eqref{eq:embedding_feynman} .
Class one includes all embeddings that, after routing, posess both a $p$ and a $p+q$ propagator. In order for this to happen, there must be two cuts cutting one parton only. Furthermore, these two cuts must go through different electron lines.}
\label{fig:emb_type1}
\end{subfigure}\hspace{0.3cm}
\begin{subfigure}[t]{0.4\textwidth}
\centering
\input{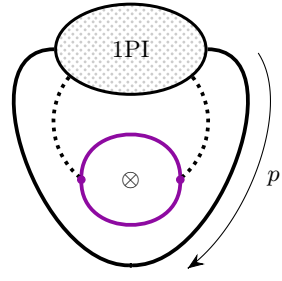}
\caption{$n\ge 0$, $m=0$ in eq.~\eqref{eq:embedding_feynman}.
Class two includes all embeddings that have a $p$ propagator after routing but no $p+q$ propagator. The embedding must allow for an initial-state cut crossing one parton only.}
\label{fig:emb_type2}
\end{subfigure}

\begin{subfigure}[t]{0.4\textwidth}
\centering
\input{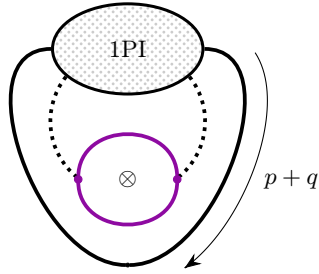}
\caption{$m\ge 0$, $n=0$ in eq.~\eqref{eq:embedding_feynman}. Class three includes all embeddings that have a $p+q$ propagator after routing but no $p$ propagator. The embedding must allow for a unique (possibly degenerate) final-state cut crossing one parton only.}
\label{fig:emb_type3}
\end{subfigure}\hspace{0.3cm}
\begin{subfigure}[t]{0.4\textwidth}
\centering
\input{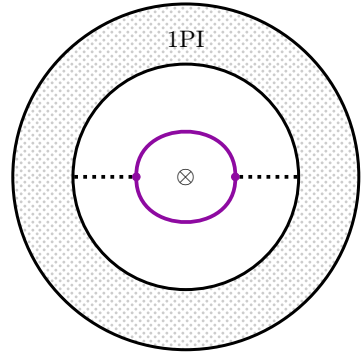}
\caption{$n=m=0$ in eq.~\eqref{eq:embedding_feynman}. Class four includes all embeddings that have no $p$ or $p+q$ propagator after routing. Every cut of these embeddings goes through more than one partonic line.}
\label{fig:emb_type3}
\end{subfigure}
\caption{Embeddings may be classified in four different types. If their integrand has both a $p$ and a $p+q$ propagator, then they fall in class one. If they have a $p$ propagator but no $p+q$ propagator, then they fall in class two. If they have a $p+q$ propagator but no $p$ propagator, then they fall in class three. Finally, if they have neither a $p$ nor a $p+q$ propagator, then they fall in class four. This request in turn forces a one-particle irreducible decomposition of the vacuum graph embedding, shown in the four diagrams above. Parton model interference diagrams all belong to embeddings in class one and two: they correspond to the choice of initial-state cut crossing the parton carrying momentum $p$. In deep inelastic scattering the purple line is identified with the electron, and the dotted line with the photon.}
\label{fig:four_types}
\end{figure}
We collected forward-scattering diagrams in equivalence classes based on a topological criterion. We now provide further classification: given an embedding together with a choice of an initial state electron, $\embsI$, we will say that
\begin{itemize}
\item It belongs to the first class, if the corresponding equivalence classes of cuts, 
$[r]_{\embsI}$ and $[r]_{\embsF}$, contain one cut each, $r_1\in [r]_{\embsI}$ and $r_2\in [r]_{\embsF}$, crossing one parton only. In other words, $\embsI$ belongs to the first class if its associated loop integral of eq.~\eqref{eq:embedding_feynman} has $n,m\ge 1$. 
\item It belongs to the second class, if it is not in the first class and if 
$[r]_{\embsI}$ contains one cut $r'$ crossing one parton only. Again, we can simply say that it will belong to class two if the loop integral of loop integral of eq.~\eqref{eq:embedding_feynman} has $m=0$ and $n\ge 1$.
\item It belongs to the third class, if it is not in the first class and if the corresponding equivalence class of cuts, 
$[r]_{\embsF}$, contains one cut $r'$ crossing one parton only, or equivalently, if the integral of eq.~\eqref{eq:embedding_feynman} has $n=0$ and $m\ge 1$.
\item It belongs to the fourth class if it does not belong to the other three classes. Its integrand, defined through eq.~\eqref{eq:embedding_feynman}, will have $n=m=0$.  
\end{itemize}
This classification is relevant for many reasons: one is that it makes it easier for us to draw the exact relationship with the parton model. Parton 
model diagrams, by definition, only have one initial-state parton. Hence, the corresponding interference diagrams will appear in embeddings
that belong to the first or second class. On the contrary, embeddings in the third and fourth class cannot contain interference diagrams 
that also appear in the parton model. Correspondingly, we denote the contributions from interference diagrams arising from an embedding of class $i$ as
\begin{equation}
W_{\q}(p,q)|_{\text{class-}1}, W_{\q}(p,q)|_{\text{class-}2}, W_{\q}(p,q)|_{\text{class-}3}, W_{\q}(p,q)|_{\text{class-}4}.
\end{equation}
According to the previous discussion, the partonic tensor for the parton model is contained in the sum of the partonic tensors for classes one and two; schematically,
\begin{equation}
W_{\q\text{PM}}(p,q)\subset W_{\q}(p,q)|_{\text{class-}1}+W_{\q}(p,q)|_{\text{class-}2}.
\end{equation}
The equation above is not an equality because, while parton model interference diagrams all belong to embeddings in class one or two, such embeddings also contain other interference diagrams that are not in the parton model (for example, all those needed to cancel the initial-state singularities of the parton model diagram themselves). Computing $W_{\q}(p,q)|_{\text{class-}1}$ and $W_{\q}(p,q)|_{\text{class-}2}$ amounts to selecting only the minimal set of interference diagrams needed to make the parton model finite. This is for example the choice performed in refs.~\cite{PhysRevD.25.2222,Axelrod:1985yi} for the quark contributions, $\mathfrak{q}=q$.

The second reason this classification is important is that the way the cross-section is written down, in each of these classes, is slightly different. For example, we will see that embeddings in class four have new branch-cuts that are absent in the other three classes. Similarly, the precise realisation of the KLN theorem within each of them will be different.

\subsection{Examples}
\label{sec:embedding_examples}
\begin{center}
\begin{figure}
\centering
\begin{subfigure}[t]{0.3\textwidth}
\centering
\scalebox{0.75}{\includegraphics[]{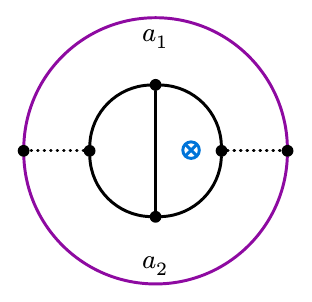}}
\caption{$(G,\mathbf{w}_1)$}
\label{fig:emb1_ex}
\end{subfigure}%
\begin{subfigure}[t]{0.3\textwidth}
\centering
\scalebox{0.75}{\includegraphics[]{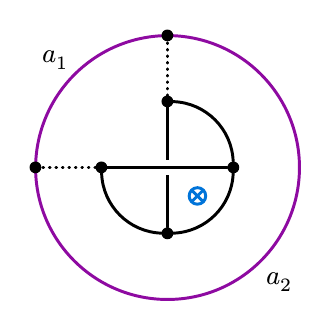}}
\caption{$(G,\mathbf{w}_2)$}
\label{fig:emb2_ex}
\end{subfigure}%
\begin{subfigure}[t]{0.3\textwidth}
\centering
\scalebox{0.75}{\includegraphics[]{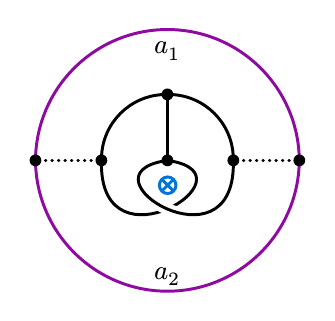}}
\caption{$(G,\mathbf{w}_3)$}
\label{fig:emb3_ex}
\end{subfigure}
\caption{Three example embeddings, with the purple line denoting the electron, and dotted line denoting the photon. Removing the puncture ($\otimes$), we find that they identify the same vacuum diagram $G$. However, once the puncture is placed, the additional topological information makes them inequivalent. This is expressed by the different winding numbers $\mathbf{w}_1,\mathbf{w}_2,\mathbf{w}_3$. }
\label{fig:example_embeddings}
\end{figure}
\end{center}

\begin{center}
\begin{figure}
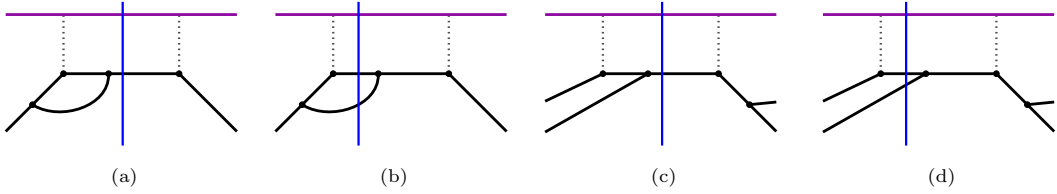

\centering
\begin{subfigure}[t]{0.23\textwidth}
\centering
\scalebox{0.75}{\input{Submission/diagrams/embeddings/examples/interferences_triangle/interference1}}
\caption{}
\label{fig:emb1_interference_1}
\end{subfigure}%
\begin{subfigure}[t]{0.23\textwidth}
\centering
\scalebox{0.75}{\input{Submission/diagrams/embeddings/examples/interferences_triangle/interference2}}
\caption{}
\label{fig:emb1_interference_2}
\end{subfigure}%
\begin{subfigure}[t]{0.23\textwidth}
\centering
\scalebox{0.75}{\input{Submission/diagrams/embeddings/examples/interferences_triangle/interference3}}
\caption{}
\label{fig:emb1_interference_3}
\end{subfigure}
\begin{subfigure}[t]{0.23\textwidth}
\centering
\scalebox{0.75}{\input{Submission/diagrams/embeddings/examples/interferences_triangle/interference4}}
\caption{}
\label{fig:emb1_interference_4}
\end{subfigure}
\caption{Interference diagrams belonging to the equivalence class identified by the embedding $(G,\mathbf{w}_1)$, together with the choice of an initial-state electron, $a_2$, which can be collected in the thruple $\embs_1=(G,\wind_1,a_2)$. The blue cut identifies the final states, and cuts through the electron labelled by $a_1$. Swapping the role of initial states and final states provides a class that we treat as distinct. The sum of the four interference diagrams in infrared-finite.}
\label{fig:emb1_interference}
\end{figure}
\end{center}

We will discuss three different embeddings of the same vacuum diagram; for each we will collect all the interference diagrams associated to that embedding and we will provide a routing for the embedding (which of course implies a routing for the individual interference diagrams). The three embeddings are given in fig.~\ref{fig:example_embeddings}. They correspond to different placement of the puncture for the same vacuum diagram:
\begin{equation}
\raisebox{-1cm}{\scalebox{0.5}{\includegraphics[]{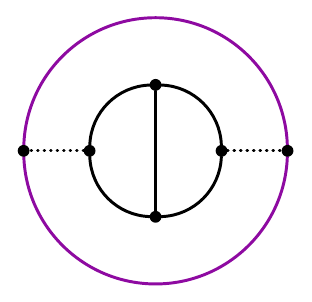}}}.
\end{equation}
By the identification explained in this section, each embedding, together with a choice of initial-state electron, identifies a set of interference diagrams, where the initial-state cut crosses the initial-state electron. $(G,\mathbf{w}_1,a_2)$ identifies the interference diagrams given in fig.~\ref{fig:emb1_interference}, $(G,\mathbf{w}_2,a_2)$ those given in fig.~\ref{fig:emb2_interference}, $(G,\mathbf{w}_3,a_1)$ those given in fig.~\ref{fig:emb3_interference}. 

\begin{center}
\begin{figure}
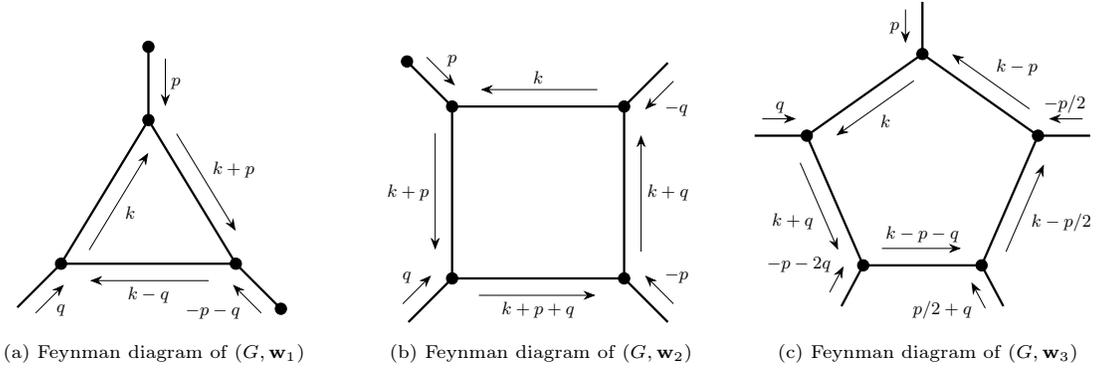

\centering
\begin{subfigure}[t]{0.33\textwidth}
\centering
\scalebox{0.75}{\input{Submission/diagrams/discs_and_cuts/triangle_ex/triangle_labelled}}
\caption{Feynman diagram of $(G,\mathbf{w}_1)$}
\label{fig:emb1_feyn}
\end{subfigure}%
\begin{subfigure}[t]{0.33\textwidth}
\centering
\scalebox{0.75}{\input{Submission/diagrams/discs_and_cuts/box_ex/box_labelled}}
\caption{Feynman diagram of $(G,\mathbf{w}_2)$}
\label{fig:emb2_feyn}
\end{subfigure}%
\begin{subfigure}[t]{0.33\textwidth}
\centering
\scalebox{0.75}{\input{Submission/diagrams/discs_and_cuts/pentagon_ex/pentagon_labelled}}
\caption{Feynman diagram of $(G,\mathbf{w}_3)$}
\label{fig:emb3_feyn}
\end{subfigure}
\caption{After routing, we can build the integrand for the three embeddings $\Gamma_i$, $i=1,2,3$, and map it to an ordinary one-loop Feynman diagram. Whenever an external edge is delimited by two dots, it is assumed that the corresponding propagator is included. If the external edge does not have a dotting identifying an external vertex, then it is truncated. Interference diagrams may be identified with cuts of these Feynman diagrams, and hence they inherit their routing.}
\label{fig:feynman_diags_ex}
\end{figure}
\end{center}

Finally, each embedding, and by association each interference diagram identified by the same embedding, can be given a routing consistent with the simplest clustering criterion. This allows to apply Feynman rules at the embedding level, and then mapping the corresponding integral to a Feynman diagram; for the three example embeddings of fig.~\ref{fig:example_embeddings}, these Feynman diagrams are given in fig.~\ref{fig:feynman_diags_ex}.  

\begin{center}
\begin{figure}
\centering
\begin{subfigure}[t]{0.23\textwidth}
\centering
\scalebox{0.75}{\input{Submission/diagrams/embeddings/examples/interferences_box/interference1}}
\caption{}
\label{fig:emb2_interference_1}
\end{subfigure}%
\begin{subfigure}[t]{0.23\textwidth}
\centering
\scalebox{0.75}{\input{Submission/diagrams/embeddings/examples/interferences_box/interference2}}
\caption{}
\label{fig:emb2_interference_2}
\end{subfigure}%
\begin{subfigure}[t]{0.23\textwidth}
\centering
\scalebox{0.75}{\reflectbox{\input{Submission/diagrams/embeddings/examples/interferences_box/interference2}}}
\caption{}
\label{fig:emb2_interference_3}
\end{subfigure}
\caption{Interference diagrams belonging to the equivalence class identified by the embedding thruple $\embs_2=(G,\mathbf{w}_2,a_2)$. The blue cut identifies the final states, and cuts through the electron labelled by $a_1$. Swapping the role of initial states and final states provides a class that we treat as distinct. The sum of the three interference diagrams in infrared-finite.}
\label{fig:emb2_interference}
\end{figure}
\end{center}
\begin{center}
\begin{figure}
\centering
\begin{subfigure}[t]{0.23\textwidth}
\centering
\scalebox{0.75}{\input{Submission/diagrams/embeddings/examples/interferences_pentagon/interference1}}
\caption{}
\label{fig:emb3_interference_1}
\end{subfigure}%
\begin{subfigure}[t]{0.23\textwidth}
\centering
\scalebox{0.75}{\input{Submission/diagrams/embeddings/examples/interferences_pentagon/interference2}}
\caption{}
\label{fig:emb3_interference_2}
\end{subfigure}%
\begin{subfigure}[t]{0.23\textwidth}
\centering
\scalebox{0.75}{\input{Submission/diagrams/embeddings/examples/interferences_pentagon/interference3}}
\caption{}
\label{fig:emb3_interference_3}
\end{subfigure}
\begin{subfigure}[t]{0.23\textwidth}
\centering
\scalebox{0.75}{\input{Submission/diagrams/embeddings/examples/interferences_pentagon/interference4}}
\caption{}
\label{fig:emb3_interference_4}
\end{subfigure}

\begin{subfigure}[t]{0.23\textwidth}
\centering
\scalebox{0.75}{\input{Submission/diagrams/embeddings/examples/interferences_pentagon/interference5}}
\caption{}
\label{fig:emb3_interference_5}
\end{subfigure}%
\begin{subfigure}[t]{0.23\textwidth}
\centering
\scalebox{0.75}{\input{Submission/diagrams/embeddings/examples/interferences_pentagon/interference6}}
\caption{}
\label{fig:emb3_interference_6}
\end{subfigure}
\caption{Interference diagrams belonging to the equivalence class identified by the embedding thruple $\embs_3=(G,\mathbf{w}_3,a_1)$. The blue cut identifies the final states, and cuts through the electron labelled by $a_2$. Swapping the role of initial states and final states provides a class that we treat as distinct. The sum of the six interference diagrams in infrared-finite.}
\label{fig:emb3_interference}
\end{figure}
\end{center}

Let us work out the routing assignment for the embedding of fig.~\ref{fig:emb1_ex} and fig.~\ref{fig:emb3_ex}. The four interference diagrams of fig.~\ref{fig:emb1_interference} corresponds to couplets of (red,blue) cuts chosen out of the following:
\begin{equation}
\label{eq:embedding1_with_cuts}
\vcenter{\hbox{{
\scalebox{0.75}{
\includegraphics[]{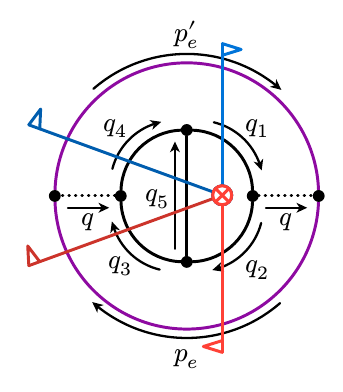}}}}}.
\end{equation}
The red cuts all cross the chosen initial-state electron. The simplest-clustering inspired momentum-conservation conditions impose that the sum of the momenta in each \emph{red} cut is $p+p_{a_2}$, where $p_{a_2}$ is the momentum of the initial-state electron according to the definition of the embedding. In other words, the sum of all parton's momenta in each red cut must sum up to $p$. 

\newpage
We can thus write down all such conservation conditions plus all the usual ones that are imposed at each vertex
\begin{align}
\text{cuts:    }&q_2=p \ (q_2+p_{a_2}=p+p_{a_2}), \ q_3+q_5=p \ (q_3+q_5+p_{a_2}=p+p_{a_2}), \\
\text{vertices:    }&q_3+q-q_4=0, \ q_4+q_5-q_1, \ q_1+q-q_2, \\ 
&q_2-q_3-q_5=0, \ p_{a_1}-p_{a_2}+q=0.
\end{align}
These constraints are linearly dependent. Regardless, they admit the solution: $q=p_{a_1}-p_{a_2}'$ (in the following, we will often keep the dependence of $q$ on the electron's momenta implicit), $q_2=p$, $q_1=p+q$, $q_3=-k$, $q_5=p+k$, $q_3=-k+q$, where we dubbed the independent loop momentum $q_3=-k$. The routing can be visualized on the embedding:
\begin{equation}
\label{eq:embedding1_with_cuts}
\raisebox{-1.5cm}{
\scalebox{0.75}{
\includegraphics[]{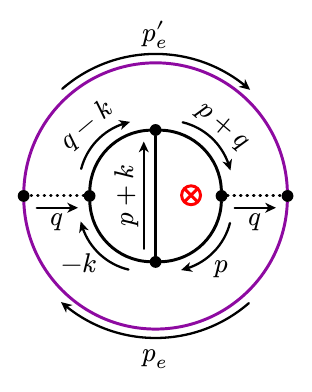}}}.
\end{equation}
In summary, forgetting the ``electron'' and ``photon'' propagators, i.e. focusing on the partonic part of the integrand, we can write down the loop integral
\begin{equation}
I_{\embsI_1}=\frac{1}{p^2(p+q)^2}\int \frac{\mathrm{d}^d k}{(2\pi)^d }\frac{1}{k^2 (k+p)^2 (k-q)^2}.
\end{equation}
It is a triangle diagram multiplied by the propagators $(p^2)^{-1}$ and $((p+q)^2)^{-1}$. The corresponding drawing is given in fig.~\ref{fig:emb1_feyn}. This routing is inherited by all the interference diagrams of fig.~\ref{fig:emb1_interference}. For example, the diagram of fig.~\ref{fig:emb1_interference_3} reads
\begin{equation}
\raisebox{-1cm}{\scalebox{0.8}{\begin{tikzpicture}

    \node[inner sep=0pt] (U1) {};
    \node[inner sep=0pt] (U1p) [above = 0.5cm of U1] {};
    \node[inner sep=0pt] (U2) [above right = 1cm and 1cm of U1] {};
    \node[inner sep=0pt] (U3) [right = 2cm of U2] {};
    \node[inner sep=0pt] (U4) [below right = 1cm and 1cm of U3] {};
    \node[inner sep=0pt] (U34) [below right = 0.5cm and 0.5cm of U3] {};
    \node[inner sep=0pt] (U4p) [above = 0.5cm of U4] {};

    \node[inner sep=0pt] (U12) [above right = 0.45cm and 0.45cm of U1] {};
    \node[inner sep=0pt] (U23) [right = 1cm of U2] {};

    \node[inner sep=0pt] (D2) [above = 1cm of U2]{};
    \node[inner sep=0pt] (D3) [above = 1cm of U3] {};

    \node[inner sep=0pt] (UP1) [left = 1cm of D2]{};
    \node[inner sep=0pt] (UP2) [right = 1cm of D3] {};

    \node[inner sep=0pt] (C1) [above right = 1.25cm and 1.7cm of U2]{};
    \node[inner sep=0pt] (C2) [below right = 1.25cm and 1.7cm of U2] {};

        \begin{feynman}

    	\draw[-, thick, black, line width=0.5mm, momentum=\(-k\)] (U1p) to (U2);
     \draw[-, thick, black, line width=0.5mm, momentum=\(q-k\)] (U2) to (U23);
     \draw[-, thick, black, line width=0.5mm, momentum=\(p+q\)] (U23) to (U3);
     \draw[-, thick, black, line width=0.5mm, momentum=\(p\)] (U3) to (U34);
     \draw[-, thick, black, line width=0.5mm] (U4) to (U34);
     \draw[-, thick, black, line width=0.5mm] (U4p) to (U34);
     \draw[-, thick, electron-purple, line width=0.5mm, momentum=\(p_e\)] (UP1) to (D2);
     \draw[-, thick, electron-purple, line width=0.5mm, momentum=\(p_{e}'\)] (D2) to (D3);
     \draw[-, thick, electron-purple, line width=0.5mm] (D3) to (UP2);
     
     \draw[-, thick, black, line width=0.5mm, reversed momentum=\(p+k\)] (U23) to (U1);

     \draw[-, thick, black!70!white, dotted, line width=0.4mm,reversed momentum=\(q\)] (U2) to (D2);
     \draw[-, thick, black!70!white, dotted, line width=0.4mm, reversed momentum=\(q\)] (D3) to (U3);

     \draw[-, thick, blue, line width=0.4mm] (C1) to (C2);
     \end{feynman}

	\path[draw=black, fill=black] (U2) circle[radius=0.05];
 \path[draw=black, fill=black] (U23) circle[radius=0.05];
 \path[draw=black, fill=black] (U34) circle[radius=0.05];
 \path[draw=black, fill=black] (U3) circle[radius=0.05];
 \path[draw=black, fill=black] (D2) circle[radius=0.05];
 \path[draw=black, fill=black] (D3) circle[radius=0.05];

\end{tikzpicture}}}=\frac{\tilde{\delta}((p+q)^2)}{p^2}\int \frac{\mathrm{d}^d k}{(2\pi)^d }\frac{\tilde{\delta}^-(k)\tilde{\delta}^+(k+p)}{ (k-q)^2}.
\end{equation}
where again we excluded the ``photon'' and ``electron'' propagators. We see that the initial-state partons have momenta that sum up to $p$. Finally, let us classify this embedding in line with sect.~\ref{sec:four_types_of_embeddings}. From eq.~\eqref{eq:embedding1_with_cuts}, we see that there are two cuts cutting a single parton. Furthermore, isolating these two cut parton momenta, we see that they are $p$ and $p+q$. In other words, the embedding $I_{\embsI_1}$ has a propagator $(p^2)^{-1}$ and a propagator $((p+q)^2)^{-1}$. This implies that it belongs to class one.

The case of the embedding of fig.~\ref{fig:emb2_ex} does not present additional difficulties for what concerns the routing. Again, we start by drawing all cuts:
\begin{equation}
\vcenter{\hbox{{
\scalebox{0.75}{
\includegraphics[]{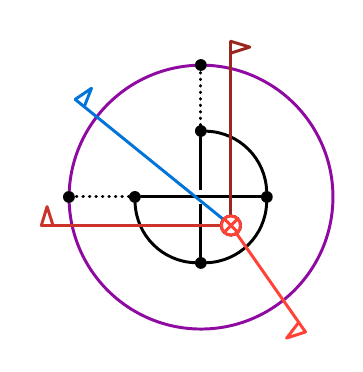}}}}}\,.
\end{equation}

The red and blue cuts are distinguished by the electron they cut: all ref cuts cut the same electron, the blue cut cuts the other one. This set of cuts identifies two subclasses of infrared-finite diagrams: if we choose the red cut to identify our choice of initial states, then we obtain the three interference diagrams of fig.~\ref{fig:emb2_interference}, each corresponding to a different initial-state cut.

In order to find the routing for this embedding, for pedagogical purposes, let us directly work at the level of the forward-scattering diagram (or interference diagram). We start from the interference diagram of fig.~\ref{fig:emb2_interference_2}:
\begin{center}
\scalebox{0.75}{
\begin{tikzpicture}

    \node[inner sep=0pt] (U1) {};
    \node[inner sep=0pt] (U11) [above=0.25cm of U1] {};
    \node[inner sep=0pt] (U2) [above right = 1cm and 1cm of U1] {};
    \node[inner sep=0pt] (U3) [right = 2cm of U2] {};
    \node[inner sep=0pt] (U33) [above right = 0.5cm and 1.5cm of U1] {};
    \node[inner sep=0pt] (U4) [below right = 1cm and 1cm of U3] {};

    \node[inner sep=0pt] (U12) [above right = 0.45cm and 0.45cm of U1] {};
    \node[inner sep=0pt] (U4M) [above left = 0.45cm and 0.8cm of U4] {};
    \node[inner sep=0pt] (U44M) [right = 0.45cm of U4M] {};

    \node[inner sep=0pt] (F1) [right = 4.5cm of U1] {};
    \node[inner sep=0pt] (F2) [right = 4.5cm of U11] {};

    \node[inner sep=0pt] (U23) [right = 0.75cm of U2] {};

    \node[inner sep=0pt] (D2) [above = 1cm of U2]{};
    \node[inner sep=0pt] (D3) [above = 1cm of U3] {};

    \node[inner sep=0pt] (UP1) [left = 1cm of D2]{};
    \node[inner sep=0pt] (UP2) [right = 1.5cm of D3] {};

    \node[inner sep=0pt] (C1) [above right = 1.25cm and 0.9cm of U2]{};
    \node[inner sep=0pt] (C2) [below right = 1.25cm and 0.9cm of U2] {};
    \node[inner sep=0pt] (C2P) [below right = 1.25cm and 0.55cm of U2] {};
    \node[inner sep=0pt] (CC) [below  = 0.15cm of U2] {};

    \node[inner sep=0pt] (CC1) [above left = 0.75cm and 0.9cm of U2]{};
    \node[inner sep=0pt] (CC2) [below left = 1.25cm and 0.9cm of U2] {};

        \begin{feynman}
        \draw[-, thick, black, line width=0.5mm, momentum=\(q_2\)] (U11) to (U2);
     \draw[-, thick, black, line width=0.5mm, reversed momentum=\(q_3\)] (U4M) to (U2);
     \draw[-, thick, black, line width=0.5mm, reversed momentum=\(q_1\)] (U33) to (U1);
     \draw[-, thick, black, line width=0.5mm] (U33) to (U3);
     \draw[-, thick, black, line width=0.5mm, momentum=\(q_5\)] (U3) to (U4M);
     \draw[-, thick, black, line width=0.5mm, reversed momentum=\(q_4\)] (U44M) to (U4M);
     \draw[-, thick, black, line width=0.5mm] (U44M) to (F1);
     \draw[-, thick, black, line width=0.5mm] (U44M) to (F2);
     \draw[-, thick, electron-purple, line width=0.5mm,momentum=\(p_e\)] (UP1) to (D2);
     \draw[-, thick, electron-purple, line width=0.5mm,momentum=\(p_e'\)] (D2) to (D3);
     \draw[-, thick, electron-purple, line width=0.5mm] (D3) to (UP2);
    \draw[-, thick, black!70!white, dotted, line width=0.4mm, reversed momentum=\(q\)] (U2) to (D2);
     \draw[-, thick, black!70!white, dotted, line width=0.4mm, momentum=\(q\)] (U3) to (D3);
    	\end{feynman}



     \draw[-, thick, blue, line width=0.4mm] (C1) to (C2);

	\path[draw=black, fill=black] (U2) circle[radius=0.05];
 \path[draw=black, fill=black] (U44M) circle[radius=0.05];
 \path[draw=black, fill=black] (U4M) circle[radius=0.05];
 \path[draw=black, fill=black] (U3) circle[radius=0.05];
 \path[draw=black, fill=black] (U33) circle[radius=0.025];
 \path[draw=black, fill=black] (D2) circle[radius=0.05];
 \path[draw=black, fill=black] (D3) circle[radius=0.05];

\end{tikzpicture}}\,.
\end{center}
Momentum conservation is applied by imposing that the sum of initial-state momenta is $p+p_e$, plus momentum conservation at all but one vertex (the constraint at this last vertex is linearly dependent). In other words, we have $q_1+q_2=p$ for the initial state constraint, plus $q_3=q_2+q$, $q_1=q+q_5$, $q_3+q_5=q_4$, for momentum-conservation at each vertex. These four constraints are linearly independent, and they are solved by $q_1=-k$, $q_2=k+p$, $q_3=k+p+q$, $q_5=k+q$, $q_4=p$. Using these facts, we may write the integrand for the interference diagram
\begin{equation}
\label{eq:non_planar}
\raisebox{-1.cm}{\scalebox{0.75}{
}}=\frac{1}{p^2}\int \frac{\mathrm{d}^dk}{(2\pi)^d} \frac{\tilde{\delta}^-(k)\tilde{\delta}^+(p+k)\tilde{\delta}^+(k+p+q)}{(k+q)^2}\,.
\end{equation}
We can now infer the integral for the full embedding and all other interference diagrams directly from this one. In particular, we have:
\begin{equation}
I_{\embsI_2}=\frac{1}{p^2}\int \frac{\mathrm{d}^dk}{(2\pi)^d} \frac{1}{k^2(k+p)^2(k+q)^2(k+p+q)^2},
\end{equation}
which is the box diagram of fig.~\ref{fig:emb2_feyn}. This embedding has one cut going through a single partonic line, which receives the momentum $p$. No propagator $((p+q)^2)^{-1}$ is present instead, which implies this embedding is in class two. From it we can also deduce the routing for the interference diagram of fig.~\ref{fig:emb2_interference_1}:
\begin{equation}
\raisebox{-1.cm}{\scalebox{0.75}{\begin{tikzpicture}

    \node[inner sep=0pt] (U1) {};
    \node[inner sep=0pt] (U2) [above right = 1cm and 1cm of U1] {};
    \node[inner sep=0pt] (U3) [right = 2cm of U2] {};
    \node[inner sep=0pt] (U33) [above right = 0.62cm and 1.225cm of U1] {};
    \node[inner sep=0pt] (U4) [below right = 1cm and 1cm of U3] {};

    \node[inner sep=0pt] (U44) [above left = 0.62cm and 1.225cm of U4] {};

    \node[inner sep=0pt] (U12) [above right = 0.45cm and 0.45cm of U1] {};
    \node[inner sep=0pt] (U4M) [above left = 0.45cm and 0.45cm of U4] {};
    \node[inner sep=0pt] (U23) [right = 0.75cm of U2] {};

    \node[inner sep=0pt] (D2) [above = 1cm of U2]{};
    \node[inner sep=0pt] (D3) [above = 1cm of U3] {};

    \node[inner sep=0pt] (UP1) [left = 1cm of D2]{};
    \node[inner sep=0pt] (UP2) [right = 1cm of D3] {};

    \node[inner sep=0pt] (C1) [above right = 1.25cm and 0.65cm of U2]{};
    \node[inner sep=0pt] (C2) [below right = 1.25cm and 0.65cm of U2] {};

        \begin{feynman}
        \draw[-, thick, black, line width=0.5mm, momentum=\(p\)] (U1) to (U12);
        \draw[-, thick, black, line width=0.5mm, momentum=\(k+p\)] (U12) to (U2);
     \draw[-, thick, black, line width=0.5mm, reversed momentum=\(k+p+q\)] (U4M) to (U44);
     \draw[-, thick, black, line width=0.5mm] (U44) to (U2);
     \draw[-, thick, black, line width=0.5mm] (U3) to (U33);
     \draw[-, thick, black, line width=0.5mm, momentum=\(k\)] (U33) to (U12);
     \draw[-, thick, black, line width=0.5mm, reversed momentum=\(k+q\)] (U3) to (U4M);
     \draw[-, thick, black, line width=0.5mm] (U4M) to (U4);
     \draw[-, thick, black, line width=0.5mm, momentum=\(p_e\)] (UP1) to (D2);
     \draw[-, thick, black, line width=0.5mm, momentum=\(p_e'\)] (D2) to (D3);
     \draw[-, thick, black, line width=0.5mm] (D3) to (UP2);
     \draw[-, thick, black!70!white, dotted, line width=0.4mm, momentum=\(q\)] (D2) to (U2);
     \draw[-, thick, black!70!white, dotted, line width=0.4mm, momentum=\(q\)] (U3) to (D3);
    	
    	\end{feynman}


     \draw[-, thick, blue, line width=0.4mm] (C1) to (C2);

	\path[draw=black, fill=black] (U2) circle[radius=0.05];
    \path[draw=black, fill=black] (U33) circle[radius=0.02];
    \path[draw=black, fill=black] (U44) circle[radius=0.02];
 \path[draw=black, fill=black] (U12) circle[radius=0.05];
 \path[draw=black, fill=black] (U4M) circle[radius=0.05];
 \path[draw=black, fill=black] (U3) circle[radius=0.05];
 \path[draw=black, fill=black] (D3) circle[radius=0.05];
\path[draw=black, fill=black] (D2) circle[radius=0.05];

\end{tikzpicture}}}=\tilde{\delta}(p^2)\int \frac{\mathrm{d}^dk}{(2\pi)^d} \frac{\tilde{\delta}^-(k)\tilde{\delta}^+(k+p+q)}{(k+p)^2(k+q)^2}.
\end{equation}
Finally, let us discuss the embedding of fig.~\ref{fig:emb3_ex}. We draw the initial and final state cuts separately for the sake of graphical clarity. We also draw initial-state cuts and final-state cuts in different shades of red and blue respectively:
\begin{align}
\scalebox{0.75}{
\includegraphics[]{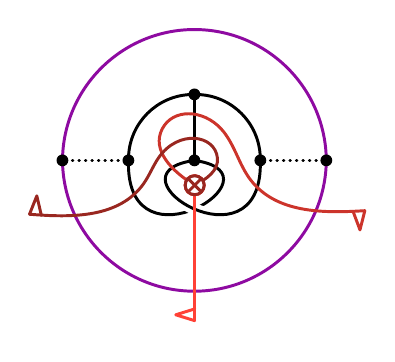}}\scalebox{0.75}{
\includegraphics[]{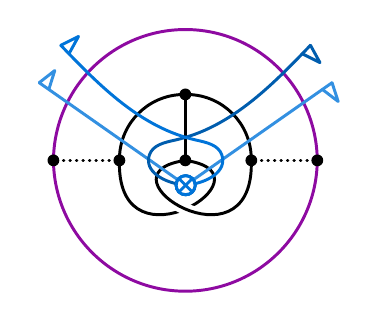}}\,.
\end{align}
All interference diagrams correspond to couplets of a red and blue cut from the pictures above. However, the reverse is not true: only couplets that divide the graph into two connected components should be considered. We construct the integrand by solving the momentum-conservation condition for an interference diagram, much like we did for the box. We start from the interference diagram of fig.~\ref{fig:emb3_interference_1}:
\begin{center}
\scalebox{0.75}{
\begin{tikzpicture}

    \node[inner sep=0pt] (U1) {};
    \node[inner sep=0pt] (U2) [above right = 1cm and 1cm of U1] {};
    \node[inner sep=0pt] (U3) [right = 2cm of U2] {};
    \node[inner sep=0pt] (U4) [below right = 1cm and 1cm of U3] {};

    \node[inner sep=0pt] (U11) [below = 0.5cm of U1] {};

    \node[inner sep=0pt] (U44) [right = 4cm of U11] {};

    \node[inner sep=0pt] (UM) [right = 1cm of U2] {};
    \node[inner sep=0pt] (UMM) [below = 1cm of UM] {};

    \node[inner sep=0pt] (U12) [above right = 0.45cm and 0.45cm of U1] {};
    \node[inner sep=0pt] (U4M) [above left = 0.45cm and 0.45cm of U4] {};
    \node[inner sep=0pt] (U23) [right = 0.75cm of U2] {};

    \node[inner sep=0pt] (D2) [above = 1cm of U2]{};
    \node[inner sep=0pt] (D3) [above = 1cm of U3] {};

    \node[inner sep=0pt] (UP1) [left = 1cm of D2]{};
    \node[inner sep=0pt] (UP2) [right = 1cm of D3] {};

    \node[inner sep=0pt] (C1) [below right = 0.2cm and 0.2cm of U11]{};
    \node[inner sep=0pt] (C2) [above = 3cm of C1] {};

    \node[inner sep=0pt] (CC1) [right = 1.5cm of C1]{};
    \node[inner sep=0pt] (CC2) [right = 1.5cm of C2] {};
   
        \begin{feynman}
        \draw[-, thick, black, line width=0.5mm, momentum=\(q_1\)] (U1) to (U2);
        
     \draw[-, thick, black, line width=0.5mm, reversed momentum=\(q_3\)] (UM) to (U2);
     \draw[-, thick, black, line width=0.5mm, momentum=\(q_4\)] (UM) to (U3);

     \draw[-, thick, black, line width=0.5mm] (U44) to (U3);
     
     \draw[-, thick, electron-purple, line width=0.5mm, momentum=\(p_e\)] (UP1) to (D2);
     \draw[-, thick, electron-purple, line width=0.5mm, momentum=\(p_e'\)] (D2) to (D3);
     \draw[-, thick, electron-purple, line width=0.5mm] (D3) to (UP2);

     \draw[-, thick, black, line width=0.5mm, momentum=\(q_5\)] (UM) to (UMM);

     \draw[-, thick, black, line width=0.5mm, momentum=\(q_2\)] (U11) to (UMM);
     \draw[-, thick, black, line width=0.5mm] (U4) to (UMM);
     \draw[-, thick, black!70!white, dotted, line width=0.4mm,momentum=\(q\)] (D2) to (U2);
     \draw[-, thick, black!70!white, dotted, line width=0.4mm,reversed momentum=\(q\)] (D3) to (U3);
    	
    	\end{feynman}

     \draw[-, thick, blue, line width=0.5mm] (CC1) to (CC2);

	\path[draw=black, fill=black] (U2) circle[radius=0.05];
 \path[draw=black, fill=black] (UM) circle[radius=0.05];
 \path[draw=black, fill=black] (UMM) circle[radius=0.05];
 \path[draw=black, fill=black] (U3) circle[radius=0.05];
 \path[draw=black, fill=black] (D2) circle[radius=0.05];
 \path[draw=black, fill=black] (D3) circle[radius=0.05];

\end{tikzpicture}}
\end{center}
Momentum conservation conditions for this interference diagram read: $q_1+q_2=p$ for the cut and $q_3=q_1+q$, $q_3=q_4+q_5$, $q_2+q_5-q_1=0$ for all vertices but one. Solving these momentum conservation conditions gives $q_1=k$, $q_2=p-k$, $q_5=2k-p$, $q_3=k+q$, $q_4=p+q-k$. The embedding integrand is
\begin{equation}
I_{\embsI_3}=\frac{1}{2^4}\int \frac{\mathrm{d}^dk}{(2\pi)^d} \frac{1}{k^2 (p-k)^2 (2k-p)^2 (k+q)^2 (p+q-k)^2}.
\end{equation}
We see the appearance of a propagator $((2k-p)^2)^{-1}$: the coefficient in front of $k$ is not in $\{-1,0,1\}$. The factor of $1/2^4$ arises from the jacobian associated with solving the momentum-conservation conditions: it is also a consequence of the appearance of general integer coefficients in front of $k$. In order to map this integral to a Feynman diagram, we extract the factor of $2$ in front of the loop momentum, by rewriting $(2k-p)^2=4(k-p/2)^2$. We obtain 
\begin{align}
I_{\embsI_3}&=\frac{1}{2^6}\int \frac{\mathrm{d}^d k}{(2\pi)^d} \frac{1}{k^2(p-k)^2(k+q)^2 (p/2-k)^2(k-p-q)^2} \nonumber\\
&=\frac{1}{2^6}P(p,q,-p-2q,q+p/2,-p/2).
\end{align}
The pentagon is shown in fig.~\ref{fig:emb3_feyn}. Again, we may now write the integrand for any other interference diagram. Let us give two significative examples. For the interference diagram of fig.~\ref{fig:emb3_interference_1}, we have
\begin{align}
\raisebox{-1.3cm}{\scalebox{0.75}{
\begin{tikzpicture}

    \node[inner sep=0pt] (U1) {};
    \node[inner sep=0pt] (U2) [above right = 1cm and 1cm of U1] {};
    \node[inner sep=0pt] (U3) [right = 2cm of U2] {};
    \node[inner sep=0pt] (U4) [below right = 1cm and 1cm of U3] {};

    \node[inner sep=0pt] (U11) [below = 0.5cm of U1] {};

    \node[inner sep=0pt] (U44) [right = 4cm of U11] {};

    \node[inner sep=0pt] (UM) [right = 1cm of U2] {};
    \node[inner sep=0pt] (UMM) [below = 1.cm of UM] {};

    \node[inner sep=0pt] (U12) [above right = 0.45cm and 0.45cm of U1] {};
    \node[inner sep=0pt] (U4M) [above left = 0.45cm and 0.45cm of U4] {};
    \node[inner sep=0pt] (U23) [right = 0.75cm of U2] {};

    \node[inner sep=0pt] (D2) [above = 1cm of U2]{};
    \node[inner sep=0pt] (D3) [above = 1cm of U3] {};

    \node[inner sep=0pt] (UP1) [left = 1cm of D2]{};
    \node[inner sep=0pt] (UP2) [right = 1cm of D3] {};

    \node[inner sep=0pt] (C1) [below right = 0.2cm and 0.2cm of U11]{};
    \node[inner sep=0pt] (C2) [above = 3cm of C1] {};

    \node[inner sep=0pt] (CC1) [right = 1.5cm of C1]{};
    \node[inner sep=0pt] (CC2) [right = 1.5cm of C2] {};
   
        \begin{feynman}
        \draw[-, thick, black, line width=0.5mm, momentum=\(k\)] (U1) to (U2);
        
     \draw[-, thick, black, line width=0.5mm, reversed momentum=\(k+q\)] (UM) to (U2);
     \draw[-, thick, black, line width=0.5mm, momentum=\(p+q-k \ \ \ \ \ \)] (UM) to (U3);

     \draw[-, thick, black, line width=0.5mm] (U44) to (U3);
     
     \draw[-, thick, electron-purple, line width=0.5mm, momentum=\(p_e\)] (UP1) to (D2);
     \draw[-, thick, electron-purple, line width=0.5mm, momentum=\(p_e'\)] (D2) to (D3);
     \draw[-, thick, electron-purple, line width=0.5mm] (D3) to (UP2);

     \draw[-, thick, black, line width=0.5mm, momentum=\(2k-p\)] (UM) to (UMM);

     \draw[-, thick, black, line width=0.5mm, reversed momentum=\(p-k\)] (UMM) to (U11);
     \draw[-, thick, black, line width=0.5mm] (U4) to (UMM);
     \draw[-, thick, black!70!white, dotted, line width=0.4mm, reversed momentum=\(q\)] (U2) to (D2);
     \draw[-, thick, black!70!white, dotted, line width=0.4mm,reversed momentum=\(q\)] (D3) to (U3);
    	
    	\end{feynman}

     \draw[-, thick, blue, line width=0.5mm] (CC1) to (CC2);

	\path[draw=black, fill=black] (U2) circle[radius=0.05];
 \path[draw=black, fill=black] (UM) circle[radius=0.05];
 \path[draw=black, fill=black] (UMM) circle[radius=0.05];
 \path[draw=black, fill=black] (U3) circle[radius=0.05];
 \path[draw=black, fill=black] (D2) circle[radius=0.05];
 \path[draw=black, fill=black] (D3) circle[radius=0.05];

\end{tikzpicture}}}=\frac{1}{2^6}\int \frac{\mathrm{d}^d k}{(2\pi)^d} \frac{\tilde{\delta}^+(k)\tilde{\delta}^+(p-k)\tilde{\delta}^+(k+q)}{(k-p/2)^2 (p+q-k)^2}.
\end{align}
An interesting case is given by the interference diagram of fig.~\ref{fig:emb3_interference_3}, because of the extra winding propagator. Although the pictorial representation includes a propagator being cut multiple times, the Feynman rule is the same as if it has been cut one time only. 
\begin{align}
\raisebox{-1.3cm}{\scalebox{0.75}{
\begin{tikzpicture}

    \node[inner sep=0pt] (U1) {};
    \node[inner sep=0pt] (U2) [above right = 1cm and 1cm of U1] {};
    \node[inner sep=0pt] (U3) [right = 1.5cm of U2] {};
    \node[inner sep=0pt] (U4) [
right = 4cm of U1] {};
    \node[inner sep=0pt] (U4M) [below right = 0.5cm and 0.5cm of U3] {};

    \node[inner sep=0pt] (U11) [below = 0.7cm of U1] {};
    \node[inner sep=0pt] (U111) [below = 0.25cm of U1] {};

    \node[inner sep=0pt] (U44) [right = 4cm of U11] {};

    \node[inner sep=0pt] (U444) [right = 4cm of U111] {};

    \node[inner sep=0pt] (UM) [right = 0.75cm of U2] {};
    \node[inner sep=0pt] (UMM) [below = 1cm of UM] {};

    \node[inner sep=0pt] (U12) [above right = 0.45cm and 0.45cm of U1] {};
    \node[inner sep=0pt] (U23) [right = 0.75cm of U2] {};

    \node[inner sep=0pt] (D2) [above = 1cm of U2]{};
    \node[inner sep=0pt] (D3) [above = 1cm of U3] {};

    \node[inner sep=0pt] (UP1) [left = 1cm of D2]{};
    \node[inner sep=0pt] (UP2) [right = 1.5cm of D3] {};

    \node[inner sep=0pt] (C1) [below right = 0.2cm and 0.2cm of U11]{};
    \node[inner sep=0pt] (C2) [above = 3cm of C1] {};

    \node[inner sep=0pt] (CC1) [right = 2.cm of C1]{};
    \node[inner sep=0pt] (CC2) [right = 2.cm of C2] {};

        \begin{feynman}
        \draw[-, thick, black, line width=0.5mm, momentum=\(k\)] (U1) to (U2);
        
     \draw[-, thick, black, line width=0.5mm, momentum=\(k+q \ \ \ \ \ \ \ \)] (U2) to (UM);
     \draw[-, thick, black, line width=0.5mm,momentum=\(\ \ \ \ \ \ p+q-k\)] (UM) to (U3);


     \draw[-, thick, blue, line width=0.5mm] (CC1) to (CC2);

     \draw[-, thick, black, line width=0.5mm] (U4M) to (U44);
     \draw[-, thick, black, line width=0.5mm] (U4M) to (U444);

     \draw[-, thick, black, line width=0.5mm, reversed momentum=\(k\)] (U4) to (U11);

     \draw[-, thick, black, line width=0.5mm] (U4M) to (U44);

     \draw[-, thick, black, line width=0.5mm, momentum=\(p-k\)] (U3) to (U4M);
     
     \draw[-, thick, electron-purple, line width=0.5mm] (UP1) to (UP2);

     \draw[-, thick, black, line width=0.5mm, reversed momentum=\(p-2k \ \ \ \ \)] (UM) to (U111);

     \draw[-, thick, black!70!white, dotted, line width=0.4mm] (D2) to (U2);
     \draw[-, thick, black!70!white, dotted, line width=0.4mm] (U3) to (D3);
    	
    	\end{feynman}


	\path[draw=black, fill=black] (U2) circle[radius=0.05];
 \path[draw=black, fill=black] (UM) circle[radius=0.05];
 \path[draw=black, fill=black] (U3) circle[radius=0.05];
 \path[draw=black, fill=black] (U4M) circle[radius=0.05];

\end{tikzpicture}}}=\frac{1}{2^6}\int \frac{\mathrm{d}^d k}{(2\pi)^d} \frac{\tilde{\delta}^+(k)\tilde{\delta}^+(p/2-k)\tilde{\delta}^+(p+q-k)}{(k-p)^2 (k+q)^2}.
\end{align}
We see that the extra winding edge affects momentum conservation, but even if it is cut twice only one Dirac delta function appears in the above. This concludes the discussion of the examples.
\section{MP cross-sections and discontinuities of Feynman diagrams}
\label{sec:MPI_and_discs}

In the previous section, we have shown that a notion of embedding can be used to group infrared-finite contributions together. In this section, we will show that the relation between embeddings and interference diagrams is even sharper: \emph{interference diagrams naturally arise by taking discontinuities of vacuum graph embeddings}. This provides an extended version of Cutkosky's theorem~\cite{Cutkosky:1960sp}: not only interference diagrams can be seen as discontinuities of forward-scattering diagrams, but forward-scattering diagrams can be seen as discontinuities of vacuum graph embeddings. Thus, instead of analytically computing the interference diagrams contributing to the cross-section, we are free to analytically compute the discontinuities of the vacuum graph embeddings. Both problems are easily approachable by usual analytic integration methods. 

This identification is much more than a simple piece of mathematical trivia: it opens up the possibility of computing MP cross-sections as discontinuities of Feynman diagrams, an object that is widely studied in the literature. Discontinuities are both a core object of investigation for those that are interested in learning about the analytic structure of Feynman diagram~\cite{Abreu:2014cla,Abreu:2017enx,Hannesdottir:2022xki,Bourjaily:2020wvq} and the target of computational frameworks, such as Reverse Unitarity~\cite{Anastasiou:2002qz,Anastasiou:2002yz,Anastasiou:2003yy,Anastasiou:2013mca,Anastasiou:2013srw}, that have proven to be extremely effective in computing cross-sections at high orders and for complicated processes. In this sense, this identification, possible in virtue of the ``simplest clustering'' choice, potentially allows for the deployment of the MP formalism at the scale required to realistically meet the need of predictions of a collider.

We structure this section as follows: after having introduced in sect.~\ref{sec:dis_kinematics} the kinematic properties of deep inelastic scattering, we continue in sect.~\ref{sec:embeddings_anal} to studying the leading-virtuality expansion of embeddings and the analytic properties of the coefficients that arise through such expansion. We use these constraints to compute double discontinuities of the embedding. We then continue in sect.~\ref{sec:interference_diags} to a seemingly unrelated topic, that of computing interference diagrams within the Reverse Unitarity formalism. Finally, in sect.~\ref{sec:double_optical_thm}, we relate the two topics that were previously discussed by showing that interference diagrams can be written as iterated discontinuities of embeddings.

We will be focused on the computation of the embeddings in $\embdis$, but we don't expect particular changes in applying what discussed in this section to higher winding embeddings.

\subsection{DIS kinematics and virtuality expansion}
\label{sec:dis_kinematics}

As also discussed in sect.~\ref{sec:setup}, DIS kinematics are determined in terms of two momenta $p$, $q$. 
$q$ is space-like, $q^2<0$, while $p$ and $p+q$ are timelike $p^2,(p+q)^2>0$. 
We furthermore introduce the virtuality $Q^2=-q^2$ and the dimensionless Bjorken variable $x$ through
\begin{equation}
    \frac{(p+q)^2}{Q^2}=\frac{1-x}{x}\,.
\end{equation}
This definition differs from the parton model definition by higher-virtuality terms in $p^2/Q^2$. Generally speaking, the embeddings have branch-cuts at $(n_1 p+n_2 q)^2=0$, $n_1,n_2\in \mathbb{Z}$ inside the region $p^2,(p+q)^2>0$. For example, for embeddings in $\embdis$, there are non-analiticities in the invariants $(p+q)$, $(p-q)^2$, $(p+2q)^2$.
We can rewrite them in terms of $p^2,Q^2$ and $1-x$:
\begin{equation}
\frac{(p+n q)^2}{Q^2}=n^2\frac{\frac{1}{n}-x}{x}+\frac{p^2}{Q^2}.
\end{equation}
We should note that if one restricts their attention to class one and two embeddings only, i.e. the minimal set of diagrams required to make the parton model, then $p^2$ and $1-x$ are sufficient to capture all non-analiticities.

\subsection{Analytic properties of embeddings at leading virtuality}
\label{sec:embeddings_anal}

We start by investigating the analytic properties of embeddings. Since the computation of embeddings can be turned into that of Feynman diagrams, we can use the usual technology devoted to the integration of Feynman diagrams, such as integration-by-parts identities and differential equations. Once that is done, we can ask what are the analytic properties of embeddings, and in particular the distribution of poles and branch-cuts in the kinematic region of interest.

\subsubsection{Embeddings at leading virtuality}

 Let $I_{\embs}$ be the integral corresponding to the embedding $\embs=(G,\mathbf{w},a_1)$. We would like to look into its analytic properties. The precise details of them depend based on whether the embedding $\Gamma$ is in class one, two, three or if it is in class four. At next-to-leading order, all embeddings (through their definition in terms of $V_\Gamma$) can be written by partial-fractioning, numerator reduction and IBP identities as sums of scalar triangle and bubble topologies:
 \begin{equation}
I_{\embsI}\left(p,q\right)=\frac{\displaystyle\sum_{i=1}^{N_T} t_i(p,q)T\left(\left\{p_j^{(i)}(p,q)\right\}_{j=1}^3\right)+\sum_{i=1}^{N_B} b_i(p,q)B\left(p^{(i)}(p,q)\right)}{[p^2]^n[(p+q)^2]^m}.
 \end{equation}
 The coefficients $t_i(p,q),b_i(p,q)$ are rational functions of the invariants, $p^2,q^2,p\cdot q$. The momenta $p_j^{(i)}(p,q)$ and $p^{(i)}(p,q)$ are linear combinations of $p$ and $q$ with integer or rational coefficients, depending on normalisation. We are interested in the leading-virtuality expansion terms of embeddings. In particular, we focus on the behaviour in the region where 
\begin{equation}
\label{eq:expansion_region}
\frac{p^2}{(n_1p+n_2q)^2},\frac{p^2}{Q^2}\sim\lambda,
\end{equation}
with $\lambda$ infinitesimally small, $n_2\neq 0,$ i.e. in the region where $p^2$ is much smaller than all the other invariants. In turn, this expansion is not valid when $(n_1p+n_2q)^2\sim p^2$. While eq.~\eqref{eq:expansion_region} expresses the condition needed to treat the most general case, in the following we will only be interested with the cases $n_1=n_2=1$ and $n_1=1$, $n_2=2$. 

Typical expansion-by-regions~\cite{Chetyrkin:1988zz,Gorishnii:1989dd,etde_6475136,Beneke:1997zp,Smirnov:2002pj,Jantzen:2011nz} arguments apply here. One may take the expression of the embedding in terms of masters, and expand the masters at leading order in $p^2$, and find, at next-to-leading order:
\begin{equation}
I_{\embsI}(p,q)= (p^2)^{-1}H_{\embsI}(p^2,(p+q)^2,q^2)+(-p^2)^{-1-\epsilon}C_{\embsI}(p^2,(p+q)^2,q^2),
\end{equation}
where $H_{\embsI}$ and $C_{\embsI}$, the hard and collinear regions, are analytic in $p^2$\footnote{We use here in the following that IR singularities in physical theories are logarithmic. This implies that even diagrams with self-energy insertions, $n=2$, $m=1$ or $n=1$, $m=2$, must scale at worse like $(-p^2)^{-1}$ and $(x-1)^{-1}$ at the integrated level.}. Obtaining this expression may involve cancellations across different masters. In particular, we may take the expression above and evaluate $p^2=0$ when it appears as an argument of these two functions:
\begin{equation}
I_{\embsI}(p,q)= (p^2)^{-1}H_{\embsI}((p+q)^2,q^2)+(-p^2)^{-1-\epsilon}C_{\embsI}((p+q)^2,q^2)+\mathcal{O}((p^2)^{0}),
\end{equation}
where we abbreviated $H_{\embsI}((p+q)^2,q^2)=H_{\embsI}(0,(p+q)^2,q^2)$ and $C_{\embsI}((p+q)^2,q^2)=C_{\embsI}(0,(p+q)^2,q^2)$. Anything that is higher order will contribute at sub-leading virtuality. The hard and collinear functions, for fixed $q^2<0$, still present non-analiticities in $x$. We now plan to manifest them.

We can manifest the analytic properties of the hard and collinear functions in $x$ by carefully plugging the expression for the region integrals of the masters.  Whenever the masters are expanded, their hard regions will end up contributing to the hard region of the full graph, while their collinear regions will end up contributing to the collinear region of the full graph. Hence, we focus on the bubble and triangle topologies. The bubble is completely trivial, and the only problems could come from the triangle, whose leading-virtuality expansion can be determined purely in terms of hypergeometric functions. Writing
\begin{equation}
T(p_1,p_2,p_3)=T_h(p_2,p_3)+(-p_1^2)^{-\epsilon}T_c(p_2,p_3)+\mathcal{O}(p_1^2),
\end{equation}
the hard function is given by is the two-mass triangle, which reduces to bubbles after the application of integration-by-parts identities (see appendix~\ref{sec:hard_region_triangle}). For what concerns the collinear function $T_c$, the analytic integration is slightly more involved. However, it can be performed in light-cone coordinates. We give the result in appendix~\ref{sec:collinear_region_triangle}. Using these two results, we can constrain the analytic structure of the hard and collinear coefficients. We have
\begin{align}
H_{\embsI}((p+q)^2,q^2)=
h_0(x,q^2)+\frac{h_1(x,q^2)}{x-1}+h_2(x,q^2)(x-1)^{-1-\epsilon},
\end{align}
where $h_0(x,q^2),h_1(x,q^2)$ and $h_2(x,q^2)$ are analytic for any $x\in(0,1]$. When $n=m=0$, one can check that $h_0=h_1=h_2=0$. The analytic structure of the collinear coefficient is significantly more complicated. For the purposes of compactness, we write it in a form valid for all $n$ and $m$:
\begin{align}
C_{\embsI}&((p+q)^2,q^2)=c_0(x,q^2)+\frac{c_1(x,q^2)+c_2(x,q^2)(x-1)^{-\epsilon}}{1-x}+c_3(x,q^2)(2x-1)^{-1-\epsilon} \nonumber \\
&+c_4(x,q^2)(x-1)^{-\epsilon}(1-2x)^{-1-\epsilon}\Theta(1-2x)+c_5(x,q^2)\frac{(x-1)^{\epsilon}}{1-2x}+\frac{c_6(x,q^2)}{1-2x},\label{eq:collinear_function}
\end{align}
where the $c_i$ are analytic for $x\in(0,1]$. For diagrams that only contribute at sub-leading virtuality, of course, one has $c_i=0$ for all $i=1,..,5,6$. \emph{For diagrams in classes one to three, one has $c_3=c_4=c_5=c_6=0$}.

\subsubsection{Iterated discontinuities of embeddings}
\label{sec:iterated_disc}

We are interested in studying iterated discontinuities of leading-virtuality embeddings and their relation with interference diagrams. In particular, we are interested in taking iterated discontinuities in $p^2$ and $(1-x)$. We consider the following definition of discontinuity:
\begin{equation}
\label{eq:disc_definition}
\text{disc}_{s} f(s)= \lim_{\varepsilon\rightarrow 0} (f(s-\imath\varepsilon)-f(s+\imath\varepsilon))\,.
\end{equation}
Iterated discontinuities are somewhat tricky to define: one can think of them as singling out symbols of generalised polylogarithms or in terms of a contour in the space of complexified momenta. The situation, in our case, is made even more complicated by the non-analiticities in $(\tfrac{p}{2}+q)^2$, since:
\begin{equation}
\label{eq:deformed_by_both}
(\tfrac{p}{2}+q)^2=Q^2\left[\frac{1/2-x}{x}+\frac{p^2}{4Q^2}\right]=Q^2\frac{(1/2-x)}{x}+\mathcal{O}(p^2).
\end{equation}
is deformed both by $(1-x)\rightarrow (1-x)+i\varepsilon$ and by $p^2\rightarrow p^2+i\varepsilon$. However, at leading virtuality, as we have seen in the previous section, all the non-analiticities are captured by the invariants $p^2$, $1-x$, $1-2x$, so there are no invariants that are deformed both by $\text{disc}_{p^2}$ and $\text{disc}_{1-x}$. Even more, the fact that we have manifested all the scalings in $p^2$, $(1-x)$ and $(1-2x)$ and expressed everything else in terms of functions that are analytic at the branch-cut location, makes the calculation of the discontinuity a matter of using the following two relations:
\begin{align}
\text{disc}_{s}(-s)^{-\alpha-\beta\epsilon}&=2\imath(-1)^{\alpha-1}\sin(\beta\pi\epsilon)s^{-\alpha-\beta\epsilon}\Theta(s), \\
\text{disc}_{s}\frac{1}{s}&=2\pi \imath\delta(s)=\tilde{\delta}(s),
\end{align}
The only extra rule concerns the discontinuity acting on a product of non-analytic functions. There is only one such case:
\begin{align}
\label{eq:disc_product}
&\text{disc}_{1-x} \frac{(x-1)^{-\epsilon}}{1-2x}=-2\imath\left[\text{PV}\left[\frac{1}{1-2x}\right]\sin(\pi\epsilon) +\imath\pi \delta(1-2x)\cos(\pi\epsilon)\right](1-x)^{-\epsilon} 
\end{align}
Note that since we plan to eventually integrate this quantity between $(0,1)$, we have $\text{PV}[(1-2x)^{-1}]=[(1-2x)^{-1}]_+$. Using these facts, we can straight-forwardly evaluate discontinuities of $I_{\text{LT}}$. In what follows, we will be particularly concerned with the real part of the double discontinuity of $I$ in $p^2$ and $(1-x)$. The discontinuity in $p^2$ is especially trivial due to the simplicity of the virtuality expansion; the hard region has a pole in $p^2$ and the collinear region has a branch-cut in $p^2$ that is entirely captured by its $(-p^2)^{-1-\epsilon}$ scaling:
\begin{align}
\label{eq:embedding_double_disc}
\text{disc}_{p^2}\text{disc}_{1-x}I_{\embsI}(p^2,q^2,1-x)&=\tilde{\delta}(p^2)\text{disc}_{1-x}H_{\embsI}(1-x,q^2) \nonumber \\
&+2\imath\sin(\pi\epsilon)(p^2)^{-1-\epsilon}\text{disc}_{1-x}C_{\embsI}(1-x,q^2).
\end{align}
We are left with evaluating the discontinuities in $(1-x)$ of the hard and collinear coefficients. This is a bit more complicated due to the presence of non-analiticities starting at $x=1/2$. However, the fact that we were able to expose them in eq.~\eqref{eq:collinear_function} considerably simplifies the task. For the hard function, we have
\begin{align}
\label{eq:embedding_double_disc_p}
\text{disc}_{1-x}H_{\embsI}(1-x,q^2)=
h_1(x,q^2)\tilde{\delta}(1-x)+2\imath h_2(x,q^2)\sin(\pi\epsilon)(1-x)^{-1-\epsilon},
\end{align}
while for the collinear function
\begin{align}
\label{eq:embedding_double_disc_pp}
&\text{disc}_{1-x}C_{\embsI}(1-x,q^2)=
\tilde{\delta}(1-x)c_1(x,q^2) \nonumber \\
&+2\imath\sin(\pi\epsilon)\Bigg[c_2(x,q^2)(1-x)^{-1-\epsilon}+c_3(x,q^2)(2x-1)^{-1-\epsilon}\Theta(2x-1) \nonumber \\
&-c_4(x,q^2)(1-x)^{-\epsilon}(1-2x)^{-1-\epsilon}\Theta(1-2x)-c_5(x,q^2)(1-x)^{\epsilon}\text{PV}\left[\frac{1}{1-2x}\right]\Bigg].
\end{align}
Note that the imaginary part in eq.~\eqref{eq:disc_product} cancels, leaving only the principal value of the pole, and that $c_6$ disappears once the discontinuity is taken: both of these facts are a consequence of cancellations between the discontinuity of the term in $c_5$ and $c_6$. As we will soon see, the cross-section can be expressed in terms of the double discontinuity of eq.~\eqref{eq:embedding_double_disc}. Before we can lay that formula out, however, we need to have a closer look at the interference diagrams that contribute to the cross-section.

\subsection{Direct calculation of interference diagrams within Reverse Unitarity}
\label{sec:interference_diags}

The most direct approach to the analytic calculation of the contributions of individual interference diagrams is within the Reverse Unitarity approach. We start by characterising the relevant cuts. For this purpose, we define the cut operator $\text{Cut}_{q}$ that, when acting on a loop diagram, sums all the Cutkosky cuts such that the external momentum $q$ flows through them. Letting the one-loop virtual diagram of interest be
\begin{equation}
V_{\embsI}=\int \frac{\mathrm{d}^d k}{(2\pi)^d} \frac{1}{\prod_{j=1}^n(k-p_j)^2}\,.
\end{equation}
The cut operator is easily defined as
\begin{equation}
\text{Cut}_q V_{\embsI} =\sum_{\substack{i,l=1 \\ p_i-p_l\propto q}}^n \int \frac{\mathrm{d}^d k}{(2\pi)^{d}} \frac{\tilde{\delta}^+(k-p_i)\tilde{\delta}^-(k-p_l)}{\prod_{\substack{j=1 \\ j\neq i,l}}^n(k-p_j)^2}.
\end{equation}
Since the external momenta of the loop diagrams we are interested in are linearly dependent, in general the sum will have more than one non-vanishing term. In the case of linearly independent external momenta, instead, the sum always features only one term at one loop.

The iteration of the cut operator selects two Cutkosky cuts, one having $q_1$ flowing through it, the other $q_2$. However, not all couplets of cuts are valid: only those that leave three connected components upon deletion of the cut edges (see the discussion of iterated cutting rules in ref.~\cite{Capatti:2022mly}). In turn, this implies that the two cuts must share one edge:
\begin{align}
\label{eq:cut_two}
\text{Cut}_{q_1}\text{Cut}_{q_2} V_{\embsI}&=\sum_{\sigma\in\{\pm 1 \}}\sum_{\substack{i,l,m=1 \\ p_i-p_l\propto q_1 \\ p_i-p_m\propto q_2}}^n \int \frac{\mathrm{d}^d k}{(2\pi)^{d}} \frac{1}{\prod_{\substack{j=1 \\ j\neq i,l,m}}^n(k-p_j)^2} \nonumber \\
&\times\tilde{\delta}^{\sigma}(k-p_i)\tilde{\delta}^{-\sigma}(k-p_l)\tilde{\delta}^{-\sigma}(k-p_m).
\end{align}
Given the embedding $I_{\embsI}$, the cut diagrams that belong to the equivalence class identified by the embedding itself can be written as
\begin{align}
&W_{\Gamma}(p,q)=\Bigg[\tilde{\delta}^{(n-1)}(p^2)\tilde{\delta}^{(m-1)}((p+q)^2)V_{\embsI}(p,q)+\tilde{\delta}^{(m-1)} ((p+q)^2)\frac{\text{Cut}_{p}V_{\embsI}(p,q)}{[p^2]^n}\nonumber
 \\
 &+\tilde{\delta}^{(n-1)}(p^2)\frac{(\text{Cut}_{p+q}+\text{Cut}_{\tfrac{p}{2}+q})V_{\embsI}(p,q)}{[(p+q)^2]^m}+\frac{\text{Cut}_{p}(\text{Cut}_{p+q}+\text{Cut}_{\tfrac{p}{2}+q})V_{\embsI}(p,q)}{ [p^2]^n [(p+q)^2]^m}\Bigg]\label{eq:cut_two},
\end{align}
where
\begin{align}
\tilde{\delta}^{(n)}(x)=\begin{cases}
\displaystyle\frac{2\pi \imath}{n!}\frac{\mathrm{d}^{n}}{\mathrm{d}x^n}\delta(x)\quad &\text{if }n\ge0 \\
\displaystyle 0 \quad &\text{if } n<0
\end{cases}.
\end{align}
 $n,m$ are the embedding-dependent quantities defined in sect.~\ref{sec:four_types_of_embeddings}. $W_\Gamma(p,q)$ is the sum of all interference diagrams arising from the embedding $\Gamma$ defined in eq.~\eqref{eq:w_interference}. 
The cut integrals can be computed within a Reverse Unitarity approach. By performing numerator reduction and applying integration-by-parts identities, the sum of interference diagrams associated with an embedding reduces to a set of (cut) masters, and we find
\begin{align}
W_{\Gamma}(p,q)&=\Bigg[a\tilde{\delta}^{(n-1)}(p^2)\tilde{\delta}^{(m-1)}((p+q)^2)B(q)+\tilde{\delta}^{(m-1)}((p+q)^2)\frac{b\slashed{B}(p)}{[p^2]^n} \nonumber \\
 &+\tilde{\delta}^{(n-1)}(p^2)\frac{c\slashed{B}(p+q)}{[(p+q)^2]^m} 
 +\frac{d_1\slashed{T}(p,q,-p-q)+d_2\slashed{T}(p,p+2q,-2p-2q)}{[p^2]^n[(p+q)^2]^m} \nonumber\\
 &+\frac{d_3\slashed{T}(p,2q,-p-2q)+d_4\slashed{T}(2p,-p-q,-p+q)}{[p^2]^n[(p+q)^2]^m}\Bigg]\label{eq:cut_two},
\end{align}
where $a,b,c,d_i$ are rational functions of the invariants $p^2,p\cdot q, q^2$. The masters are easily computed in closed form at all orders in the dimensional regulator (see appendix~\ref{sec:masters1loop}). We may now easily take the virtuality expansion in $p^2$ for those terms that do not multiply $\delta(p^2)$. Doing so, one arrives at the expression:
\begin{align}
&W_{\Gamma}(p,q)=\tilde{\delta}(p^2)\slashed{H}_{\embsI}((p+q)^2,q^2)+(p^2)^{-1-\epsilon}\slashed{C}_{\embsI}((p+q)^2,q^2)+\mathcal{O}((p^2)^0).
\end{align}
Direct comparison of this expansion with eq.~\eqref{eq:embedding_double_disc_p} and eq.~\eqref{eq:embedding_double_disc_pp} reveals that $\slashed{H}_{\embsI}((p+q)^2,q^2)=\text{disc}_{1-x}H_{\embsI}(1-x,q^2)$ and $\slashed{C}_{\embsI}((p+q)^2,q^2)=2\imath\sin(\pi\epsilon)\text{disc}_{1-x}C_{\embsI}(1-x,q^2)$. This naturally leads us to the statement of the main result in the next section.

\subsection{A ``doubled'' optical theorem}

\label{sec:double_optical_thm}

We now present a result that ties together the discussion of discontinuities of embeddings and the computation of interference diagrams within Reverse Unitarity: we find that, in the virtuality expansion, the partonic tensor, thought of the sum of all interference diagrams contributing to the process, can be written as the iterated discontinuity of the embedding. 

\subsubsection{Partonic tensor as an iterated discontinuity}

Having detailed the analytic properties of embeddings in sec.~\ref{sec:embeddings_anal} and those of interference diagrams in sec.~\ref{sec:interference_diags}, we are ready to establish a relationship between the two. We have, by direct comparison of the NLO coefficients in eq.~\eqref{eq:embedding_double_disc},\eqref{eq:embedding_double_disc_p}\eqref{eq:embedding_double_disc_pp} and the Reverse Unitarity calculation:
\begin{equation}
\label{eq:discs_and_int}
\boxed{W_{\Gamma}(p,q)=\frac{\text{disc}_{p^2}\text{disc}_{1-x}I_{\embsI}(p^2,q^2,1-x)}{\text{av}(\q)}+\mathcal{O}((p^2)^{0})}.
\end{equation}
At leading-virtuality, interference diagrams can be obtained as double discontinuities of embeddings. For deep inelastic scattering, it generalises the optical theorem and the work of Cutkosky: not only interference diagrams are discontinuities of forward-scattering diagrams, but forward-scattering diagrams are themselves discontinuities of embeddings. 

By analogous arguments, the iterated discontinuity may also be re-written in terms of discontinuities of $V_{\embsI}$:
\begin{align}
\label{eq:disc_and_discs}
&\text{disc}_{p^2}\text{disc}_{1-x}I_{\embsI}=\Bigg[\tilde{\delta}^{(n-1)}(p^2)\tilde{\delta}^{(m-1)}((p+q)^2)V_{\embsI}(p,q)+\tilde{\delta}^{(m-1)}((p+q)^2)\frac{\text{disc}_{p^2}V_{\embsI}(p,q)}{[p^2]^n}\nonumber
 \\
 &+\tilde{\delta}^{(n-1)}(p^2)\frac{\text{disc}_{1-x}V_{\embsI}(p,q)}{[(p+q)^2]^m}+\frac{\text{disc}_{p^2}\text{disc}_{1-x}V_{\embsI}(p,q)}{[p^2]^n[(p+q)^2]^m}\Bigg]+\mathcal{O}((p^2)^0).
\end{align}
This expression clarifies the relationship between the cut operators and the discontinuities, when they act on the loop diagram $V_{\embsI}$. Eq.~\eqref{eq:disc_and_discs} can also simply derived if one assumes, or checks explicitly, that $V_{\embsI}$, the loop integral, does not have poles in $p^2$ and $(p+q)^2$. 

Let us stress, again, that the only reason eq.~\eqref{eq:discs_and_int} only holds at leading virtuality is due to the non-analiticities starting at $x=1/2$, and in particular to the appearance of non-analiticities in the new invariants, such as that of eq.~\eqref{eq:deformed_by_both}, that are deformed by both discontinuities. In fact, one could more precisely say that eq.~\eqref{eq:discs_and_int} holds at each fixed order in the virtuality expansion, but not for the full result: at the level of the embedding, the act of taking a discontinuity and that of expanding in virtuality \emph{do not commute}. An extremely simple example of this fact is the function $(p^2-m^2)^{-1}$. Its discontinuity in $p^2$ is non-zero, unless we expand in small $p^2$, truncate, which gives a polynomial, and only then take the discontinuity. This fact should not be particularly disconcerting: if we think of interference diagrams as the ``ground truth'', eq.~\eqref{eq:discs_and_int} is simply stating their relationship with the virtuality-expanded embedding. In other words, even assuming that the computation of the cross-section at all orders in virtuality carried any physical meaning, one could perform it by computing the interference diagrams one-by-one, simply letting go of their relationship with the double discontinuity of the embedding. 

It is striking that the existence of a generalised optical theorem should be tied to a notion of a virtuality expansion, and there is indeed the possibility that this cross-talk is only an artifact of our inability to see a type of cancellation that would remove the problems we identified. For current lack of more precise answers, and given that leading-virtuality is anyway what we could, at best, think of as a physically meaningful quantity, we will defer a more detailed investigation to future work.

\subsubsection{MP cross-section and the KLN theorem}

Recall that the partonic tensor $W_{\q}(p,q)$ is related to the partonic cross-section by eq.~\eqref{eq:x_sec_decomp}. In other words, in order to obtain an object that gives the full cross-section by simple contraction with the leptonic tensor, we first have to integrate $W_{\q}(p,q)$ over $p^2$ taking all values from $0$ to the infrared regulator $\Lambda^2$:
\begin{align}
W_{\q}(x,\Lambda^2,q^2)&=W_{\q}^{(0)}(x,\Lambda^2,q^2)+\ObigLambda \\
&=\frac{1}{4\pi}\int_0^{\Lambda^2} \frac{\mathrm{d}p^2}{2\pi \imath} W_{\q}^{(0)}(p,q)+\ObigLambda\,.
\end{align}
This integration is especially simple if we plug in the expression for $W_{\q}(p,q)$ in terms of hard and collinear functions. Indeed, since the hard and collinear coefficients do not depend on $p^2$, integration in $p^2$ reduced to the integral of $(p^2)^{-1-\epsilon}$ and $\delta(p^2)$, which are trivial. Using eq.~\eqref{eq:embedding_double_disc}, we obtain
\begin{align}
4\pi\text{av}(\q)W_{\q}^{(0)}(x,\Lambda^2,q^2)&=\tilde{\delta}(p^2)\text{disc}_{1-x}H(1-x,q^2)\nonumber \\
&-\frac{\sin(\pi\epsilon)(\Lambda^2)^{-\epsilon}}{\pi \epsilon}\text{disc}_{1-x}C(1-x,q^2)\label{eq:wq_H_C}+\ObigLambda.
\end{align}
where $H$ and $C$ are the sum of hard and collinear functions for each embedding $\embsI$ (the integration in $p^2$ can be performed per-embedding, of course). Note that, in order to achieve this step, we needed to know the dependence of $W_{\q}(p,q)$ on $p^2$ at all orders in the dimensional regulator.

We are now ready to expand the result in $\epsilon$: perhaps unsurprisingly, and similarly to the integration in $p^2$, we also need to be careful. In particular, we should think of the discontinuities of hard and collinear functions as distributions. When expanding eq.~\eqref{eq:embedding_double_disc_p} and eq.~\eqref{eq:embedding_double_disc_pp} in $\epsilon$, we have to be careful to substitute
\begin{equation}
\label{eq:plus_distr}
\frac{\Theta(\pm(1-nx))}{[\pm(1-nx)]^{1+\epsilon}}=-\frac{1}{\epsilon}\delta(1-nx)\pm\sum_{i=1}^\infty \frac{\epsilon^i}{i!}\left[\frac{\log(\pm(1-nx))^i}{(1-nx)}\right]_+\Theta(\pm(1-nx)).
\end{equation}
Having done that, the partonic tensor is now expressed in terms of an $\epsilon$ expansion. In virtue of the KLN theorem, we have that
\begin{equation}
\label{eq:KLN}
4\pi\text{av}(\q)W_{\q}^{(0)}(x,\Lambda^2,q^2)=\sum_{n=0}^\infty a_n \epsilon^n,
\end{equation}
implying that the result is finite in the $\epsilon\rightarrow 0$ limit\footnote{In general, the finiteness statement also requires UV renormalisation. In the case of NLO DIS no UV renormalisation procedure is needed, since the electric charge does not receive QCD corrections. Furthermore, contrary to the parton model, and due to the infrared-ultraviolet re-arrangement at play in self-energy diagrams, we also do not need to manually account for the introduction of field renormalisation factors at external legs (see refs.~\cite{Capatti:2022tit,Capatti:2023omc} for a general discussion of this mechanism).}. 

The anti-climactic finiteness result of eq.~\eqref{eq:KLN} deserves to be expanded upon. The first observation is that it is obvious, given eq.~\eqref{eq:discs_and_int}: the embedding is infrared finite, and taking discontinuities of it should not create poles. More precisely, the fact that we related the partonic tensor to the iterated discontinuity of the leading-virtuality embedding allows us to see the infrared finiteness of the cross-section and the cancellations across interference diagrams \emph{in terms of a deformed version of the pole-cancellation mechanism that incurs across the different region integrals in an expansion-by-regions approach}. To be more specific, we expect the hard and collinear functions $H$ and $C$ to have single collinear poles that are created by the virtuality expansion. Of course, the diagrams that participate to $H$ and $C$ may also have double poles associated to soft singularities, but these poles cancel in the sum of diagrams \emph{within} $H$ and $C$. However, $H/p^2+(p^2)^{-1-\epsilon}C$ starts at a finite order, which implies the poles of $H$ and $C$ are equal and opposite in signs. The same must hold for $\text{disc}_{1-x}H$ and $\text{disc}_{1-x}C$ and hence, using eq~\eqref{eq:wq_H_C}, the partonic tensor itself must be finite. Finally, \emph{observe that the parton model result is simply given by $\text{disc}_{1-x}H$}.

\subsection{Examples}

We now detail the calculation of the interference diagrams first introduced in sect~\ref{sec:embedding_examples} by two means: first, through direct computation of the cut diagrams within the Reverse Unitarity formalism (sect.~\ref{sec:interference_diags}), second, by computing the iterated discontinuity of the embedding. The two approaches give the same results (sect.~\ref{sec:embeddings_anal}). 

\subsubsection{Triangle diagram}

Let us start by discussing the case of the embedding in fig.~\ref{fig:emb1_ex}. The embedding reads
\begin{equation}
I_{\Gamma_1}=\frac{T(p,q,-p-q)}{p^2(p+q)^2}. 
\end{equation}
When $q^2<0$, the triangle is non-analytic for $(p+q)^2>0$ and $p^2>0$. We can expand the triangle in $p^2$ and obtain
\begin{equation}
I_{\Gamma_1}=\frac{T_h(p,q)+(p^2)^{-\epsilon}T_c(p,q)}{p^2(p+q)^2}+\mathcal{O}((p^2)^0),
\end{equation}
where hard and collinear functions are given in appendix~\ref{sec:collinear_region_triangle}. Using those expressions, we find:
\begin{align}
I_{\Gamma_1}&=\frac{h_1(x,q^2)}{p^2(1-x)}+\frac{h_2(x,q^2)}{p^2(x-1)^{1+\epsilon}} +\frac{c_1(x,q^2)}{(1-x)(-p^2)^{1+\epsilon}}+\frac{c_2(x,q^2)}{(-p^2)^{1+\epsilon}(x-1)^{1+\epsilon}}+\mathcal{O}((p^2)^0),
\end{align}
with
\begin{align}
(2\pi)^{4-2\epsilon} h_1(x,q^2)&=\imath\frac{2\epsilon-1}{\epsilon}\frac{\pi^{2-\epsilon}\Gamma(1-\epsilon)^2 \Gamma(\epsilon)}{\Gamma(2-2\epsilon)}x^2 (-q^2)^{-2-\epsilon}\,, \\
(2\pi)^{4-2\epsilon} h_2(x,q^2)&=\imath\frac{2\epsilon-1}{\epsilon}\frac{\pi^{2-\epsilon}\Gamma(1-\epsilon)^2 \Gamma(\epsilon)}{\Gamma(2-2\epsilon)}x^{2+\epsilon} (-q^2)^{-2-\epsilon}\,, \\
(2\pi)^{4-2\epsilon} c_1(x,q^2)&=-\frac{\imath x^2\pi^{2-\epsilon}\csc(-\epsilon\pi)^2}{(-q^2)^2\Gamma(1+\epsilon)\Gamma(1-2\epsilon)} \, {}_2F_1\left[1,2\epsilon,1+\epsilon,1-x\right]\,,\\
(2\pi)^{4-2\epsilon} c_2(x,q^2)&=-\frac{\imath\pi^{4-\epsilon}\csc(-\epsilon\pi)^2}{(-q^2)^2\Gamma(1 - \epsilon)}x^{2-\epsilon}.
\end{align}
We note that $h_1$ may be seen as the hard region in the expansion $p^2\sim(p+q)^2\sim\lambda$, while $h_2$ is the anti-collinear region, $c_1$ is the collinear region and $c_2$ the soft region, provided we keep all orders in $(p+q)^2$ but only the zeroth order in $p^2$. Taking the iterated discontinuity:
\begin{align}
&\text{disc}_{p^2}\text{disc}_{1-x}I_{\Gamma_1}=\tilde{\delta}(p^2)\tilde{\delta}(1-x)h_1(x,q^2)+2\imath \tilde{\delta}(p^2)\sin(\pi\epsilon)\frac{h_2(x,q^2)}{(1-x)^{1+\epsilon}} \nonumber \\
&+2\imath\tilde{\delta}(1-x)\sin(\pi\epsilon)\frac{c_1(x,q^2)}{(p^2)^{1+\epsilon}}+(2\imath)^2\sin(\pi\epsilon)^2\frac{c_2(x,q^2)}{(p^2)^{1+\epsilon}(1-x)^{1+\epsilon}}+\mathcal{O}((p^2)^0).
\end{align}
Integrating in $\mathrm{d}p^2/(2\pi \imath)$ from $0$ to $\Lambda^2$ and dividing by the flux factor $4\pi$ gives:
\begin{align}
&\frac{1}{4\pi}\int_0^{\Lambda^2} \frac{\mathrm{d}p^2}{2\pi \imath}\text{disc}_{p^2}\text{disc}_{1-x}I_{\Gamma_1}=\frac{\imath}{2}\delta(1-x)h_1(x,q^2)+\imath\sin(\pi\epsilon)\frac{h_2(x,q^2)}{2\pi(1-x)^{1+\epsilon}} \nonumber \\¨
&-\imath\delta(1-x)\sin(\pi\epsilon)(\Lambda^2)^{-\epsilon}\frac{c_1(x,q^2)}{2\pi\epsilon}-\imath\sin(\pi\epsilon)^2(\Lambda^2)^{-\epsilon}\frac{c_2(x,q^2)}{2\pi^2 \epsilon(1-x)^{1+\epsilon}}+\ObigLambda.
\end{align}
Expanding in $\epsilon$ according to the distributional rule of eq.~\eqref{eq:plus_distr} yields precisely the result of eq.~\eqref{eq:expanded_triangle} up to the coupling factor $\lambda^4$.

\subsubsection{Box diagram}

We will now show that the interference diagrams of fig.~\ref{fig:emb2_interference} can be obtained as double discontinuities of the integrand associated to the embedding of fig.~\ref{fig:emb2_ex}, which we identified with the box diagram of fig.~\ref{fig:emb2_feyn}. The embedding is partial-fractioned into triangles:
\begin{equation}
I_{\Gamma_2}=\frac{B(p,q,-p,-q)}{p^2}=\frac{T(p,q,-p-q)+T(-p,q,p-q)}{p^2(p\cdot q)}.
\end{equation}
Although the discussion can also be carried here at all orders in virtuality, let us expand for pedagogical purposes. At leading virtuality, be may substitute the triangles by their collinear approximation of appendix~\ref{sec:collinear_region_triangle}. The triangle $T(-p,q,p-q)$ is clearly analytic for $x\in (0,1)$, since $(p-q)^2=q^2(1+x)<0$ is euclidean. Hence all non-analiticies in $x$ arise from the triangle $T(p,q,-p-q)$. We can expose them and leave implicit all that is analytic for $x\in (0,1)$. We find:
\begin{align}
I_{\Gamma_2}=\frac{(h_1(x,q^2)+h_2(x,q^2)(x-1)^{-\epsilon})+(-p^2)^{-\epsilon}(c_1(x,q^2)+c_2(x,q^2)(x-1)^{-\epsilon})}{p^2}+\mathcal{O}((p^2)^0),
\end{align}
with
\begin{align}
(2\pi)^{4-2\epsilon}h_2(x,q^2)&=2\imath\frac{2\epsilon-1}{\epsilon}\frac{\pi^{2-\epsilon}\Gamma(1-\epsilon)^2 \Gamma(\epsilon)}{\Gamma(2-2\epsilon)}x^{2+\epsilon} (-q^2)^{-2-\epsilon}\,, \\
(2\pi)^{4-2\epsilon}c_2(x,q^2)&=-2\imath\frac{\pi^{4-\epsilon}\csc(-\epsilon\pi)^2}{(-q^2)^2\Gamma(1 - \epsilon)}x^{2-\epsilon}\,.
\end{align}
Taking discontinuities in $p^2$ and $1-x$ is trivial and isolates the coefficients $h_2$ and $c_2$, which is why we did not bother to specify the other two. In particular:
\begin{equation}
\label{eq:box_double_disc}
\text{disc}_{p^2}\text{disc}_{1-x}I_{\Gamma_2}=2\sin(\pi\epsilon)\frac{\tilde{\delta}(p^2)h_2(x,q^2)-2\imath(p^2)^{-1-\epsilon}\sin(\pi\epsilon)c_2(x,q^2)}{(1-x)^{\epsilon}}+\mathcal{O}((p^2)^0)
\end{equation}
We can compare this result from that computed by direct computation of cut diagrams within Reverse Unitarity. We can write the interference diagrams of fig.~\ref{fig:emb2_interference} as cuts of the box diagram of fig.~\ref{fig:emb2_feyn}:
\begin{equation}
W_{\Gamma_2}=\raisebox{-1.2cm}{\scalebox{0.8}{\begin{tikzpicture}

    \node[inner sep=0pt] (A1) {};
    \node[inner sep=0pt] (A2) [below=2cm of A1] {};
    \node[inner sep=0pt] (A3) [right=2cm of A2] {};
    \node[inner sep=0pt] (A4) [above =2cm of A3] {};

    \node[inner sep=0pt] (C1) [above=0.3cm of A1] {};
    \node[inner sep=0pt] (C2) [left=0.3cm of A1] {};

    \node[inner sep=0pt] (CCC1) [below=0.3cm of A3] {};
    \node[inner sep=0pt] (CCC2) [right=0.3cm of A3] {};

    \node[inner sep=0pt] (CC1) [above right=0.3cm and 1cm of A1] {};
    \node[inner sep=0pt] (CC2) [below right=0.3cm and 1cm of A2] {};

    \node[inner sep=0pt] (E1) [above left=0.5cm and 0.5cm of A1] {};
    \node[inner sep=0pt] (E2) [below left=0.5cm and 0.5cm of A2] {};
    \node[inner sep=0pt] (E3) [below right=0.5cm and 0.5cm of A3] {};
    \node[inner sep=0pt] (E4) [above right=0.5cm and 0.5cm of A4] {};

\begin{feynman}
 \draw[-, thick, black, line width=0.4mm] (A2) to (A1);
 \draw[-, thick, black, line width=0.4mm] (A3) to (A2);
 \draw[-, thick, black, line width=0.4mm] (A4) to (A3);
 \draw[-, thick, black, line width=0.4mm] (A1) to (A4);

 \draw[-, thick, black, line width=0.4mm, red] (C1) to (C2);
 \draw[-, thick, black, line width=0.4mm, red] (CCC1) to (CCC2);
 \draw[-, thick, black, line width=0.4mm, blue] (CC1) to (CC2);

  \draw[-, thick, black, line width=0.4mm,  momentum=\(p\)] (E1) to (A1);
 \draw[-, thick, black, line width=0.4mm, momentum=\(q\)] (E2) to (A2);
 \draw[-, thick, black, line width=0.4mm, reversed momentum=\(-p\)] (A3) to (E3);
 \draw[-, thick, black, line width=0.4mm,momentum=\(-q\)] (E4) to (A4);
\end{feynman}
    
	\path[draw=black, fill=black] (A1) circle[radius=0.05];
    \path[draw=black, fill=black] (E1) circle[radius=0.05];
    \path[draw=black, fill=black] (A2) circle[radius=0.05];
    \path[draw=black, fill=black] (A3) circle[radius=0.05];
    \path[draw=black, fill=black] (A4) circle[radius=0.05];

\end{tikzpicture}}}+\raisebox{-1.2cm}{\scalebox{0.8}{\begin{tikzpicture}

    \node[inner sep=0pt] (A1) {};
    \node[inner sep=0pt] (A2) [below=2cm of A1] {};
    \node[inner sep=0pt] (A3) [right=2cm of A2] {};
    \node[inner sep=0pt] (A4) [above =2cm of A3] {};

    \node[inner sep=0pt] (C1) [above right=0.3cm and 0.8cm of A1] {};
    \node[inner sep=0pt] (C2) [below left=0.8cm and 0.3cm of A1] {};

    \node[inner sep=0pt] (CC1) [above right=0.3cm and 1cm of A1] {};
    \node[inner sep=0pt] (CC2) [below right=0.3cm and 1cm of A2] {};

    \node[inner sep=0pt] (E1) [above left=0.5cm and 0.5cm of A1] {};
    \node[inner sep=0pt] (E2) [below left=0.5cm and 0.5cm of A2] {};
    \node[inner sep=0pt] (E3) [below right=0.5cm and 0.5cm of A3] {};
    \node[inner sep=0pt] (E4) [above right=0.5cm and 0.5cm of A4] {};

\begin{feynman}
 \draw[-, thick, black, line width=0.4mm] (A2) to (A1);
 \draw[-, thick, black, line width=0.4mm] (A3) to (A2);
 \draw[-, thick, black, line width=0.4mm] (A4) to (A3);
 \draw[-, thick, black, line width=0.4mm] (A1) to (A4);

 \draw[-, thick, black, line width=0.4mm, red] (C1) to (C2);
 \draw[-, thick, black, line width=0.4mm, blue] (CC1) to (CC2);

  \draw[-, thick, black, line width=0.4mm,  momentum=\(p\)] (E1) to (A1);
 \draw[-, thick, black, line width=0.4mm, momentum=\(q\)] (E2) to (A2);
 \draw[-, thick, black, line width=0.4mm, reversed momentum=\(-p\)] (A3) to (E3);
 \draw[-, thick, black, line width=0.4mm,momentum=\(-q\)] (E4) to (A4);
\end{feynman}
    
	\path[draw=black, fill=black] (A1) circle[radius=0.05];
    \path[draw=black, fill=black] (E1) circle[radius=0.05];
    \path[draw=black, fill=black] (A2) circle[radius=0.05];
    \path[draw=black, fill=black] (A3) circle[radius=0.05];
    \path[draw=black, fill=black] (A4) circle[radius=0.05];

\end{tikzpicture}}}+\raisebox{-1.2cm}{\scalebox{0.8}{\begin{tikzpicture}

    \node[inner sep=0pt] (A1) {};
    \node[inner sep=0pt] (A2) [below=2cm of A1] {};
    \node[inner sep=0pt] (A3) [right=2cm of A2] {};
    \node[inner sep=0pt] (A4) [above =2cm of A3] {};

    \node[inner sep=0pt] (C1) [below left=0.3cm and 0.8cm of A3] {};
    \node[inner sep=0pt] (C2) [above right=0.8cm and 0.3cm of A3] {};

    \node[inner sep=0pt] (CC1) [above right=0.3cm and 1cm of A1] {};
    \node[inner sep=0pt] (CC2) [below right=0.3cm and 1cm of A2] {};

    \node[inner sep=0pt] (E1) [above left=0.5cm and 0.5cm of A1] {};
    \node[inner sep=0pt] (E2) [below left=0.5cm and 0.5cm of A2] {};
    \node[inner sep=0pt] (E3) [below right=0.5cm and 0.5cm of A3] {};
    \node[inner sep=0pt] (E4) [above right=0.5cm and 0.5cm of A4] {};

\begin{feynman}
 \draw[-, thick, black, line width=0.4mm] (A2) to (A1);
 \draw[-, thick, black, line width=0.4mm] (A3) to (A2);
 \draw[-, thick, black, line width=0.4mm] (A4) to (A3);
 \draw[-, thick, black, line width=0.4mm] (A1) to (A4);

 \draw[-, thick, black, line width=0.4mm, red] (C1) to (C2);
 \draw[-, thick, black, line width=0.4mm, blue] (CC1) to (CC2);

  \draw[-, thick, black, line width=0.4mm,  momentum=\(p\)] (E1) to (A1);
 \draw[-, thick, black, line width=0.4mm, momentum=\(q\)] (E2) to (A2);
 \draw[-, thick, black, line width=0.4mm, reversed momentum=\(-p\)] (A3) to (E3);
 \draw[-, thick, black, line width=0.4mm,momentum=\(-q\)] (E4) to (A4);
\end{feynman}
    
	\path[draw=black, fill=black] (A1) circle[radius=0.05];
    \path[draw=black, fill=black] (E1) circle[radius=0.05];
    \path[draw=black, fill=black] (A2) circle[radius=0.05];
    \path[draw=black, fill=black] (A3) circle[radius=0.05];
    \path[draw=black, fill=black] (A4) circle[radius=0.05];

\end{tikzpicture}}}.
\end{equation}
Note again that the upper-right external propagator should be included. Applying partial fractioning relations, we find:
\begin{equation}
W_{\Gamma_2}=\frac{2\pi \imath\delta(p^2)}{p\cdot q}\raisebox{-1.2cm}{\scalebox{0.8}{\begin{tikzpicture}

    \node[inner sep=0pt] (A1) {};
    \node[inner sep=0pt] (A2) [below right=1.5cm and 0.9cm of A1] {};
    \node[inner sep=0pt] (A3) [below left=1.5cm and 0.9cm of A1] {};

    \node[inner sep=0pt] (E1) [above =0.71cm of A1] {};
    \node[inner sep=0pt] (E2) [below right=0.5cm and 0.5cm of A2] {};
    \node[inner sep=0pt] (E3) [below left=0.5cm and 0.5cm of A3] {};

    \node[inner sep=0pt] (CC1) [above=0.75cm of A2] {};
    \node[inner sep=0pt] (CC2) [below left=0.5cm and 0.75cm of A2] {};

\begin{feynman}
 \draw[-, thick, black, line width=0.4mm] (A1) to (A2);
 \draw[-, thick, black, line width=0.4mm] (A2) to (A3);
 \draw[-, thick, black, line width=0.4mm] (A1) to (A3);
 \draw[-, thick, blue, line width=0.4mm] (CC1) to (CC2);

  \draw[-, thick, black, line width=0.4mm,  momentum=\(p\)] (E1) to (A1);
 \draw[-, thick, black, line width=0.4mm, reversed momentum=\(-p-q\)] (A2) to (E2);
 \draw[-, thick, black, line width=0.4mm, reversed momentum=\(q\)] (A3) to (E3);
\end{feynman}
    
	\path[draw=black, fill=black] (A1) circle[radius=0.05];
    \path[draw=black, fill=black] (A2) circle[radius=0.05];
    \path[draw=black, fill=black] (A3) circle[radius=0.05];

\end{tikzpicture}}}+\frac{1}{p^2 (p\cdot q)}\raisebox{-1.2cm}{\scalebox{0.8}{\begin{tikzpicture}

    \node[inner sep=0pt] (A1) {};
    \node[inner sep=0pt] (A2) [below right=1.5cm and 0.9cm of A1] {};
    \node[inner sep=0pt] (A3) [below left=1.5cm and 0.9cm of A1] {};

    \node[inner sep=0pt] (E1) [above =0.71cm of A1] {};
    \node[inner sep=0pt] (E2) [below right=0.5cm and 0.5cm of A2] {};
    \node[inner sep=0pt] (E3) [below left=0.5cm and 0.5cm of A3] {};

    \node[inner sep=0pt] (CC1) [above=0.75cm of A2] {};
    \node[inner sep=0pt] (CC2) [below left=0.5cm and 0.75cm of A2] {};

    \node[inner sep=0pt] (C1) [below right=0.3cm and 0.7cm of A1] {};
    \node[inner sep=0pt] (C2) [below left=0.3cm and 0.7cm of A1] {};

\begin{feynman}
 \draw[-, thick, black, line width=0.4mm] (A1) to (A2);
 \draw[-, thick, black, line width=0.4mm] (A2) to (A3);
 \draw[-, thick, black, line width=0.4mm] (A1) to (A3);
 \draw[-, thick, blue, line width=0.4mm] (CC1) to (CC2);
 \draw[-, thick, red, line width=0.4mm] (C1) to (C2);

  \draw[-, thick, black, line width=0.4mm,  momentum=\(p\)] (E1) to (A1);
 \draw[-, thick, black, line width=0.4mm, reversed momentum=\(-p-q\)] (A2) to (E2);
 \draw[-, thick, black, line width=0.4mm, reversed momentum=\(q\)] (A3) to (E3);
\end{feynman}
    
	\path[draw=black, fill=black] (A1) circle[radius=0.05];
    \path[draw=black, fill=black] (A2) circle[radius=0.05];
    \path[draw=black, fill=black] (A3) circle[radius=0.05];

\end{tikzpicture}}}.
\end{equation}
Now, in order to apply integration-by-parts identities, we need to treat differently the triangle evaluated at $p^2=0$, which reduces to a bubble, from the triangle evaluated at general kinematics, which does not reduce. In particular:
\begin{align}
W_{\Gamma_2}&=2\pi \imath\frac{(1-2\epsilon)\delta(p^2)}{2\epsilon \, p\cdot (p+q)(p\cdot q)}\raisebox{-0.75cm}{\scalebox{0.8}{\begin{tikzpicture}

    \node[inner sep=0pt] (A1) {};
    \node[inner sep=0pt] (A2) [right=1.5cm of A1] {};

    \node[inner sep=0pt] (E1) [left =0.7cm of A1] {};
    \node[inner sep=0pt] (E2) [right=0.7cm of A2] {};

    \node[inner sep=0pt] (C1) [above right=1cm and 0.75cm of A1] {};
    \node[inner sep=0pt] (C2) [below right=1cm and 0.75cm of A1] {};

\begin{feynman}

 \draw[-, thick, black, line width=0.4mm, half left] (A1) to (A2);
 \draw[-, thick, black, line width=0.4mm, half right] (A1) to (A2);
 \draw[-, thick, blue, line width=0.4mm] (C1) to (C2);

  \draw[-, thick, black, line width=0.4mm,  momentum=\(p+q\)] (E1) to (A1);
  \draw[-, thick, black, line width=0.4mm] (E2) to (A2);

\end{feynman}
    
	\path[draw=black, fill=black] (A1) circle[radius=0.05];
    \path[draw=black, fill=black] (A2) circle[radius=0.05];

\end{tikzpicture}}}+\frac{1}{p^2(p\cdot q)}\raisebox{-1.2cm}{\scalebox{0.8}{\begin{tikzpicture}

    \node[inner sep=0pt] (A1) {};
    \node[inner sep=0pt] (A2) [below right=1.5cm and 0.9cm of A1] {};
    \node[inner sep=0pt] (A3) [below left=1.5cm and 0.9cm of A1] {};

    \node[inner sep=0pt] (E1) [above =0.71cm of A1] {};
    \node[inner sep=0pt] (E2) [below right=0.5cm and 0.5cm of A2] {};
    \node[inner sep=0pt] (E3) [below left=0.5cm and 0.5cm of A3] {};

    \node[inner sep=0pt] (CC1) [above=0.75cm of A2] {};
    \node[inner sep=0pt] (CC2) [below left=0.5cm and 0.75cm of A2] {};

    \node[inner sep=0pt] (C1) [below right=0.3cm and 0.7cm of A1] {};
    \node[inner sep=0pt] (C2) [below left=0.3cm and 0.7cm of A1] {};

\begin{feynman}
 \draw[-, thick, black, line width=0.4mm] (A1) to (A2);
 \draw[-, thick, black, line width=0.4mm] (A2) to (A3);
 \draw[-, thick, black, line width=0.4mm] (A1) to (A3);
 \draw[-, thick, blue, line width=0.4mm] (CC1) to (CC2);
 \draw[-, thick, red, line width=0.4mm] (C1) to (C2);

  \draw[-, thick, black, line width=0.4mm,  momentum=\(p\)] (E1) to (A1);
 \draw[-, thick, black, line width=0.4mm, reversed momentum=\(-p-q\)] (A2) to (E2);
 \draw[-, thick, black, line width=0.4mm, reversed momentum=\(q\)] (A3) to (E3);
\end{feynman}
    
	\path[draw=black, fill=black] (A1) circle[radius=0.05];
    \path[draw=black, fill=black] (A2) circle[radius=0.05];
    \path[draw=black, fill=black] (A3) circle[radius=0.05];

\end{tikzpicture}}} \nonumber \\
&=2\pi \imath\frac{(1-2\epsilon)\delta(p^2)\slashed{B}((p+q)^2)}{2\epsilon \, p\cdot (p+q)(p\cdot q)}+\frac{\slashed{T}(p^2,q^2,(p+q)^2)}{p^2(p\cdot q)}
\end{align}
The cut masters of sect.~\ref{sec:masters1loop} can now we plugged in this expression and virtuality-expanded, and the result expressed in terms of $p^2,x,q^2$. We find the result of eq.~\eqref{eq:box_double_disc}. The equivalence, much like for the triangle diagram, actually holds at all orders in virtuality. We need the next example to see why the virtuality expansion is really needed to find this general relation.
The integrated contribution reads:
\begin{equation}
\frac{1}{4\pi}\int_0^{\Lambda^2} \frac{\mathrm{d}p^2}{2\pi \imath}\text{disc}_{p^2}\text{disc}_{1-x}I_{\Gamma_2}=\sin(\pi\epsilon)\frac{\displaystyle h_2(x,q^2)+(\Lambda^2)^{-\epsilon}\sin(\pi\epsilon)\frac{c_2(x,q^2)}{\pi \epsilon}}{2\pi \imath (1-x)^{\epsilon}}+\ObigLambda.
\end{equation}
Once expanded in $\epsilon$, we obtain:
\begin{equation}
\frac{1}{4\pi}\int_0^{\Lambda^2} \frac{\mathrm{d}p^2}{2\pi \imath}\text{disc}_{p^2}\text{disc}_{1-x}I_{\Gamma_2}=\frac{x^2}{(4\pi q^2)^2}\log\left(\frac{\Lambda^2 x^2}{q^2}\right)+\mathcal{O}(\Lambda^2,\epsilon),
\end{equation}
which concludes the computation.

\subsubsection{Pentagon diagram}

The most involved example we study relates to the embedding of fig.~\ref{fig:emb3_ex}. As we have seen, its integrand is the same as the pentagon of fig.~\ref{fig:emb3_feyn}. Upon partial fractioning, the pentagon diagram may be expressed in terms of triangles:
\begin{align}
I_{\Gamma_3} =&-\frac{T(p,q,-p-q)}{p^2 q\cdot (p+q)}+\frac{T(q,p+q,-p-2q)}{4q\cdot (p+q)(\tfrac{p}{2}+q)^2} \nonumber \\
&+\frac{T(\tfrac{p}{2},q,-\tfrac{p}{2}-q)+T(\tfrac{p}{2},\tfrac{p}{2}+q,-p-q)}{8p^2(\tfrac{p}{2}+q)^2}
\end{align}
We see immediately that, at leading virtuality, the triangle $T(q,p+q,-p-2q)$ does not contribute. For what concerns the remaining three triangles, we should expand them in $p^2$ and determine the hard and collinear region of the full embedding in the process. Writing
\begin{align}
I_{\Gamma_3} =\frac{1}{p^2}H_{\Gamma_3}(q^2,(p+q)^2)+(-p^2)^{-1-\epsilon}C_{\Gamma_3}(q^2,(p+q)^2)+\mathcal{O}((p^2)^0), 
\end{align}
we have
\begin{align}
H_{\Gamma_3}(q^2,(p+q)^2)={}&\frac{T_h(q,-\tfrac{p}{2}-q)+T_h(-p-q,\tfrac{p}{2}+q)-8T_h(q,-p-q)}{8 q\cdot (p+q)}=0\,,\nonumber\\
C_{\Gamma_3}(q^2,(p+q)^2)={}&\frac{2^{2\epsilon}T_c(q,-\tfrac{p}{2}-q)+2^{2\epsilon}T_c(-p-q,\tfrac{p}{2}+q)-8T_c(q,-p-q)}{8 q\cdot (p+q)}\,.
\end{align}
The expression for the collinear functions of appendix~\ref{sec:collinear_region_triangle} may be plugged in the above. We are interested, in line with the general discussion of sect.~\ref{sec:embeddings_anal}, to identify the non-analiticities of the collinear function. We find
\begin{align}
(2\pi)^d C_{\Gamma_3}&(q^2,(p+q)^2)=-2\imath\frac{ \pi^{3-\epsilon}x^2\csc(\pi\epsilon)\Gamma(1-\epsilon)}{(q^2)^2(1-2x)^2\Gamma(2-2\epsilon)}\Bigg[4^{\epsilon} {}_2F_1[1,1-\epsilon,2-2\epsilon,(1-2x)^{-1}] \nonumber \\
&+4^{\epsilon} {}_2F_1[1,1-\epsilon,2-2\epsilon,(2x-1)^{-1}]+\frac{(1-2x)}{x}{}_2F_1[1,1-\epsilon,2-2\epsilon,x^{-1}]\Bigg]
\end{align}
Observe that all three hypergeometric functions are evaluated at arguments that introduce non-analiticities in the entire range $x\in(0,1)$. Taking discontinuities of the above is now easy, especially using the standard transformation property detailed in eq.~\eqref{eq:hypergeometric_anal} and the rule of eq.~\eqref{eq:disc_product}. An interesting (though expected) aspect to notice is the cancellation, in $\text{disc}_{1-x}C_{\Gamma_3}$, of the Dirac delta term $\delta(1-2x)$, analogous to that in eq.~\eqref{eq:disc_product}. A cut computation could never generate true poles at $p^2$, $1-2x$ and $1-x$, and the cancellation of their residues in $I_{\pent}$ is crucial for realising the equivalence between the two approaches.  

In summary, using eq.~\eqref{eq:hypergeometric_anal} (including its equivalent when the argument of the hypergeometric function is $\pm(1-2x)^{-1}$) and eq.~\eqref{eq:disc_product}, we may write:
\begin{align}
&(2\pi)^d\text{disc}_{1-x}C_{\Gamma_3}(q^2,(p+q)^2)=4\frac{\pi ^{4-\epsilon} x^2  \csc (\pi  \epsilon) 
   }{(q^2)^2\Gamma
   (1-\epsilon)}\Big( 2^{\epsilon}x^{-\epsilon} (1-2x)^{-1-\epsilon} \Theta(1-2x) \nonumber \\
   &+ \text{PV}\left[\frac{1}{1-2x}\right] (1-x)^{-\epsilon} x^{-\epsilon}
   - 2^{\epsilon}(2 x-1)^{-1-\epsilon} (
   x-1)^{-\epsilon} \Theta(2x-1)\Big).\label{eq:pentagon_disc_comp}
\end{align}
Finally we may evaluate the discontinuity in $p^2$:
The discontinuity in $p^2$ is easily evaluated:
\begin{align}
\label{eq:pentagon_disc_comp_psq}
\text{disc}_{p^2}\text{disc}_{1-x}I_{\Gamma_3} =-2\imath\sin(\pi\epsilon)(p^2)^{-1-\epsilon}C_{\Gamma_3}(q^2,(p+q)^2)+\mathcal{O}((p^2)^0). 
\end{align}
On the cut computation side, we start from eq.~\eqref{eq:cut_two}, which in this case ($n=m=0$) reads:
\begin{align}
\label{eq:cut_two}
&W_{\Gamma_3}=\text{Cut}_{p}(\text{Cut}_{p+q}+\text{Cut}_{\tfrac{p}{2}+q})I_{\Gamma_3}(p,q).
\end{align}
The right-hand side is easily written out in terms of cuts of the pentagon. In particular:
\begin{align}
\text{Cut}_{p}\text{Cut}_{p+q}\raisebox{-1cm}{\scalebox{0.5}{\input{Submission/diagrams/discs_and_cuts/pentagon_ex/pentagon_labelled}}}&=\raisebox{-1cm}{\scalebox{0.5}{\input{Submission/diagrams/discs_and_cuts/pentagon_ex/pentagon_cut1}}}+\raisebox{-1cm}{\scalebox{0.5}{\input{Submission/diagrams/discs_and_cuts/pentagon_ex/pentagon_cut2}}}\nonumber \\
&+\raisebox{-1cm}{\scalebox{0.5}{\input{Submission/diagrams/discs_and_cuts/pentagon_ex/pentagon_cut3}}}+\raisebox{-1cm}{\scalebox{0.5}{\input{Submission/diagrams/discs_and_cuts/pentagon_ex/pentagon_cut4}}}
\end{align}
This gives the first four out of the six interference diagrams in fig.~\ref{fig:emb3_interference}. The remaining two are singled out by the other combination of cut operators, namely:
\begin{align}
\text{Cut}_{p}\text{Cut}_{\tfrac{p}{2}+q}\raisebox{-1cm}{\scalebox{0.5}{\input{Submission/diagrams/discs_and_cuts/pentagon_ex/pentagon_labelled}}}&=\raisebox{-1cm}{\scalebox{0.5}{\input{Submission/diagrams/discs_and_cuts/pentagon_ex/pentagon_cut5}}}+\raisebox{-1cm}{\scalebox{0.5}{\input{Submission/diagrams/discs_and_cuts/pentagon_ex/pentagon_cut6}}}.
\end{align}
These two cut pentagon diagrams are identified with the last two interference diagrams of fig.~\ref{fig:emb3_interference}. Their computation goes along the usual lines: we partial fraction the six cut pentagon diagrams, which gives six triangle diagrams, but only four distinct ones. In particular, after partial fractioning, the cut operators act on triangles only, giving us an expression that only depends on the master topologies:
\begin{align}
W_{\Gamma_3}&=\frac{\slashed{T}(p,q,-p-q)}{p^2q\cdot (p+q)}+\frac{\slashed{T}(\tfrac{p}{2},\tfrac{p}{2}+q,-p-q)}{8p^2(\tfrac{p}{2}+q)^2}\nonumber \\
&+\frac{\slashed{T}(\tfrac{p}{2},q,-\tfrac{p}{2}-q)}{8p^2(\tfrac{p}{2}+q)^2}\label{eq:cut_pentagon_comp}. 
\end{align}
The cut masters can then be substituted, virtuality-expanded, and written in terms of $p^2,x,q^2$: we find the combined result of eq.~\eqref{eq:pentagon_disc_comp} and eq.~\eqref{eq:pentagon_disc_comp_psq}. Each of the three terms of eq.~\eqref{eq:pentagon_disc_comp} is mapped to a double cut topology in eq.~\eqref{eq:cut_pentagon_comp}. One interesting aspect is that the comparison between the two equations fixes the prescription of the pole at $(1-2x)$ of eq.~\eqref{eq:cut_pentagon_comp} to be a principal value prescription.

The final steps of integration and expansion in $\epsilon$ follow as usual:
\begin{equation}
\frac{1}{4\pi}\int_0^{\Lambda^2} \frac{\mathrm{d}p^2}{2\pi \imath}\text{disc}_{p^2}\text{disc}_{1-x}I_{\Gamma_3}=\sin(\pi\epsilon)(\Lambda^2)^{-\epsilon}\frac{C_{\Gamma_3}(q^2,(p+q)^2)}{4\pi^2\epsilon}+\ObigLambda
\end{equation}
and, expanding in $\epsilon$:
\begin{align}
\frac{1}{4\pi}\int_0^{\Lambda^2} &\frac{\mathrm{d}p^2}{2\pi \imath}\text{disc}_{p^2}\text{disc}_{1-x}I_{\Gamma_3}=-\frac{x^2}{4\pi(-q^2)^2}\Bigg[\left[\frac{1}{1-2x}\right]_+\Big(\log\left(1-x\right)\Theta(1-2x) \nonumber\\
&+\log\left(x\right)\Theta(2x-1)\Big)+\left[\frac{\log(|1/2-x|)}{|1-2x|}\right]_+\Bigg]+\ObigLambda.
\end{align}
We see that, contrary to the two other cases, there are no logarithms in $\Lambda^2$. In fact, embedding belonging to class four as defined in sect.~\ref{sec:four_types_of_embeddings} will never contribute to the result with logarithms of the infrared regulator.

\section{Results}
\label{sec:results}

We now proceed to presenting and discussing the results of the computation of leading and next-to-leading order structure functions in the multi-parton model.

\subsection{Leading order}

We start by checking that the MP model gives the parton model result at leading order. This should happen regardless of any scheme change. A more subtle discussion concerns the inclusion of ``higher-winding'' diagrams. 

\partitle{Parton model calculation} The leading order partonic tensor in the MP model equals that computed within the parton model. Indeed, at zeroth order, we find
\begin{align}
W_q^{\mu\nu}(p,q)&=\tilde{\delta}(p^2)\text{Tr}[\imath\slashed{p}(-\imath e_q\gamma^\mu)\imath(\slashed{p}+\slashed{q})(-\imath e_q\gamma^\nu)]\tilde{\delta}((p+q)^2)\nonumber\\
&=\tilde{\delta}(p^2)4e_q^2(2p^\mu p^\nu +p^\mu q^\nu+p^\nu q^\mu -g^{\mu\nu}p\cdot q)\tilde{\delta}((p+q)^2).
\end{align}
It satisfies the Ward identity:
\begin{equation}
q_\nu W_q^{\mu\nu}(p,q)=4e_q^2(p+q)^2\tilde{\delta}(p^2)\tilde{\delta}((p+q)^2)=0.
\end{equation}
Using projectors, we find:
\begin{align}
W_{1q}=-4 e_q^2 p\cdot q \tilde{\delta}(p^2)\tilde{\delta}((p+q)^2) =-\frac{p\cdot q}{2}W_{2q}.
\end{align}
Integrating the partonic tensor against the measure $\mathrm{d}\Pi_q$, we find the leading-order partonic tensor
\begin{equation}
W_{1q}(x,q^2)=\frac{1}{4\pi}\int \mathrm{d}\overline{\Pi} \, W_{1q}(p,q)= e_q^2\delta(x-1)=\frac{1}{4xq^2}W_{2q}(x,q^2).
\end{equation}
These values for the components of the projected partonic tensor match with those traditionally obtained within the parton model.

\partitle{Higher windings} One interesting extension of the leading-order computation involves the inclusion of higher-winding terms. It is well-known that a rigorous application of the KLN theorem introduces the possibility for particles to wind back and forth in time. To be more specific, allowing for multi-partonic initial-states and for disconnected amplitudes (more precisely, for spectator particles) imply that, at a fixed order, a countably-infinite amount of diagrams has to be included, one differing from the other by the amount of spectator particles. This infinite sum may be divergent and not alternating~\cite{Lavelle_2006}, but convergence under certain conditions has been shown~\cite{Frye:2018xjj,khalil2017completediagrammaticimplementationkinoshitaleenauenberg,Akhoury1997}. 

In the leading-order MP cross-section, any diagram with additional windings contributed to a partonic tensor with different quark content. Furthermore, the partonic tensor with quark content $\mathfrak{q}=\pm n$, $|n|\ge 1$ is related to the parton model partonic tensor by the following relation
\begin{equation}
W_{\mathfrak{q}=\pm n}^{\mu\nu}=\frac{1}{n^4}W_{q}^{\mu\nu}(\tfrac{p}{n},q),
\end{equation}
where the factor $n^{-4}$ arises when solving the momentum-conservation imposing the sum of the momenta of the $n$ initial state quarks to be $p$. In other words:
\begin{align}
W_{\mathfrak{q}=\pm n}^{\mu\nu}(p,q)={}&\frac{1}{4\pi n^4}\left[2\left(g^{\mu\nu}-\frac{q^\mu q^\nu}{q^2}\right)q^2+\frac{8}{n^2}P^\mu(p,q)P^\nu(p,q)\right] \nonumber\\
&\times \tilde{\delta}\left(\frac{2p\cdot q}{n}+q^2\right)\tilde{\delta}\left(\frac{p^2}{n^2}\right),
\end{align}
with $P^\mu(p,q)=p^\mu-\frac{p\cdot q}{q^2}$. For each $n$, the leading-order partonic tensor $W_{\mathfrak{q}=\pm n}^{\mu\nu}(p,q)$ contains a single diagram, so we can easily visualise it. For example:
\begin{equation}
W_{\mathfrak{q}=3}^{\mu\nu}(p,q)=\raisebox{-1.2cm}{\scalebox{0.8}{\begin{tikzpicture}

    \node[inner sep=0pt] (U1) {};
    \node[inner sep=0pt] (U2) [above right = 1cm and 1cm of U1] {};
    \node[inner sep=0pt] (U3) [right = 2cm of U2] {};
    \node[inner sep=0pt] (U4) [below right = 1cm and 1cm of U3] {};

    \node[inner sep=0pt] (U1D1) [below=0.5cm of U1] {};
    \node[inner sep=0pt] (U1D2) [below=1.5cm of U1] {};

    \node[inner sep=0pt] (U4D1) [below=1cm of U4] {};
    \node[inner sep=0pt] (U4D2) [below=1.5cm of U4] {};

    \node[inner sep=0pt] (U12) [above right = 0.45cm and 0.45cm of U1] {};
    \node[inner sep=0pt] (U23) [right = 0.75cm of U2] {};

    \node[inner sep=0pt] (D2) [above = 1cm of U2]{};
    \node[inner sep=0pt] (D3) [above = 1cm of U3] {};


    \node[inner sep=0pt] (C1) [above right = 1cm and 1.2cm of U2]{};
    \node[inner sep=0pt] (C2) [below right = 2.5cm and 1.2cm of U2] {};

        \begin{feynman}

    	\draw[fermion, thick, black, line width=0.5mm, momentum=\(p_1\)] (U1) to (U2);
     \draw[fermion, thick, black, line width=0.5mm] (U2) to (U3);
     \draw[fermion, thick, black, line width=0.5mm, momentum=\(p_1\)] (U3) to (U4D2);
     \draw[fermion, thick, black, line width=0.5mm, momentum=\(p_1\)] (U1D1) to (U4);
     \draw[fermion, thick, black, line width=0.5mm, momentum=\(p_1\)] (U1D2) to (U4D1);
     

     \draw[-, thick, black!70!white, photon, line width=0.4mm] (U2) to (D2);
     \draw[-, thick, black!70!white, photon, line width=0.4mm] (U3) to (D3);

     \draw[-, thick, blue, line width=0.4mm] (C1) to (C2);
     \end{feynman}

	\path[draw=black, fill=black] (U2) circle[radius=0.05];
 \path[draw=black, fill=black] (U3) circle[radius=0.05];

\end{tikzpicture}
}}
\end{equation}
The permutations associated with all possible attachments of the spectators cancel against the factorial normalisation of multiparticle states. The spectator with momentum $p_1$ ``winds around'' two times before scattering with the photon. The simplest clustering criterion imposes that the sum of the momenta of the three initial states equals $p$, $3p_1=p$. Since there are three initial-state quarks, this diagram contributes to the partonic tensor with quark number $\q=3$. By reversing the fermion flow, we would obtain a diagram contributing to the partonic tensor with quark number $\q=-3$. 

Projecting the partonic tensor to structure functions and integrating against the measure $\mathrm{d}\Pi_{\mathrm{p}}$:
\begin{align}
F_{2\mathfrak{q}=\pm n }(x,q^2)=\frac{4}{n^3}e_q^2\delta(x-\tfrac{1}{n})=2xF_{1q}(x,Q^2), \quad F_{L\mathfrak{q}=\pm n}(x,q^2)=0
\end{align}
We see that each $n$-winding contribution is localised at $x=1/n$. At higher orders, the winding contributions will introduce new branch-cuts instead, which end up overlapping with the original one. Summing over $n$ now yields a perfectly well-defined quantity. The full, convolved, leading-order cross-section for deep inelastic scattering now reads:
\begin{align}
\frac{\mathrm{d}\sigma}{\mathrm{d}y\mathrm{d}q^2}&=\frac{4\pi\alpha^2}{(-q^2)^2}[1+(1-y)^2]\sum_{\substack{\mathfrak{q}=-\infty \\ \mathfrak{q}\neq 0}}^\infty\int \mathrm{d}x f_{\mathfrak{q}}(x) F_{1\mathfrak{q}}(x,q^2)\nonumber \\ 
&=\frac{8\pi\alpha^2}{(-q^2)^2}[1+(1-y)^2]\sum_{\substack{\mathfrak{q}=-\infty \\ \mathfrak{q}\neq 0}}^\infty\frac{f_{\mathfrak{q}}(1/\mathfrak{q})}{\mathfrak{q}^2}.
\end{align}
In order to retrieve the parton model result, we need to impose some constraints on $f_{\mathfrak{q}}$. One of the many ways to get the equivalence is to ask that $f_{\mathfrak{q}}(1/\mathfrak{q})=\mathfrak{q}^{\alpha}f_{q}(1)$. $f_q(1)$ is then the standard parton distribution function for a quark, up to a normalisation factor $\zeta(2-\alpha)$.

\subsection{Next-to-leading order calculation}

In this section, we detail the results of the NLO calculation of the deep inelastic scattering cross-section and its comparison with traditional results obtained within the parton model. In the following, we will refer to the structure functions as computed in the MP model as $F_{i\q }$, $i=1,2,L$, while we will denote structure functions computed in the parton model as $F_{i\q}^{\text{PM}}$. All structure functions given in this section include the constribution of all embeddings in $\embdis$.

\partitle{General tests of the computation} In order to validate the result, we tested that \textbf{a)} the Ward identity holds at all orders in virtuality, by verifying that $q^\mu q^\nu W_{\mu\nu}(p,q)=0$ at the IBP-level and after the substitution of the master topologies. Furthermore, since our pipeline provides a way to compute contributions from the parton model on their own, we checked that \textbf{b)} our pipeline, when excluding the extra MP diagrammatic contributions, reproduces the well-known results of~\cite{Bardeen:1978yd} (also reported in~\cite{Vermaseren_2005,Ellis:1996mzs}). These parton model contributions cancel their infrared poles with new diagrammatic contributions in the MP model, and we indeed verified that \textbf{c)} such IR cancellations happen, and in particular that they take place embedding by embedding. This set of checks (and in particular \textbf{b)} and \textbf{c)}) provides a strong confirmation of the computation of some of the diagrammatic contributions, but leaves out all those embeddings that do not contain parton model diagrams, i.e. those belonging to class three and four (since interference diagrams arising from these embeddings do not undergo IR cancellations with parton model diagrams). One last test of contributions for all embeddings is that \textbf{d)} they should be integrable across the full range $x\in(0,1)$.

\partitle{Longitudinal structure functions in the MP model} We start by giving the expressions for the longitudinal structure functions: we have:
\begin{align}
F_{Lq}(x)&=F_{\bar{q}L}(x)=F_{Lq}^{\text{PM}}(x)=F_{L\bar{q}}^{\text{PM}}(x)=e_q^2\frac{\alpha_s}{2\pi}x\left[2C_F x\right], \\
F_{Lg}(x)&=F_{Lg}^{\text{PM}}(x)=e_q^2\frac{\alpha_s}{2\pi}x\left[4 T_F x(1-x)\right], \\ 
F_{Lqq}(x)&=F_{L\bar{q}\bar{q}}(x)=0.
\end{align}
In these expressions $C_F$ and $T_F$ are color factors.
They match those computed in the parton model, \emph{ensuring that a scheme-change between the two models exists for $q$ and $g$ quark contents}. We now proceed to writing down $F_{2\q }$.

\partitle{Gluonic $F_{2g}$ in the MP model} For the gluonic contributions, it reads
\begin{equation}
\label{eq:gluonic_MP}
F_{2g}=e_q^2\frac{\alpha_s}{2\pi}x\left[ P_{gq}(x)\log\left(\frac{-q^2}{\Lambda^2}\right)+C_{2g}(x)\right], \quad P_{gq}(x)=x^2+(1-x)^2,
\end{equation}
with
\begin{equation}
C_{2g}(x)=2T_F\left[-(1+3x(x-1))-P_{gq}(x)\log\left(x\right)\right].
\end{equation}
We can compare this with the parton model expression for the coefficient function, which reads
\begin{equation}
C_{2g}^{\text{PM}}(x)=T_F\left[-(1+8x(x-1))+P_{gq}(x)\log\left(\frac{1-x}{x}\right)\right].
\end{equation}
The main difference that jumps to the eye is the absence of the $\log(1-x)$ logarithm. 

\partitle{Quark $F_{2q}$ in the MP model} We now proceed to $F_{2q}$. We write its contribution as
\begin{align}
&F_{2q}=e_q^2\frac{\alpha_s}{2\pi}x\left[ P_{qq}(x)\log\left(\frac{-q^2}{\Lambda^2}\right)+C_{2q}(x)|_{\text{class-}1,2}+C_{2q}(x)|_{\text{class-}3,4}\right], \label{eq:fermionic_MP}\\ 
&P_{qq}(x)=C_F\left[\frac{3}{2}\delta(1-x)+\frac{1+x^2}{(1-x)_+}\right].
\end{align}
The coefficient functions $C_{2q}(x)|_{\text{class-}1,2}$ and $C_{2q}(x)|_{\text{class-}3,4}$ contain the contributions of the minimal set of diagrams needed to make the parton model finite and all the remaining ones, respectively. These latter set of diagrams \emph{do not} contribute to the scale dependence. Regardless, we find for classes 1 and 2
\begin{align}
C_{2q}(x)|_{\text{class-}1,2}&=C_F \Bigg[\left(1+\frac{2\pi^2}{3}\right)\delta(1-x)+2\frac{1+x^2}{1-x}\log(x)-(1+3x)+\frac{3}{2}\frac{1}{(1-x)_+}\Bigg]\,.
\end{align}
For classes 3 and 4 we find
\begin{align}
&C_{2q}(x)|_{\text{class-}3,4}= C_F\Bigg[2 \log ^2(2) \delta(1-x)+\Theta (2 x-1)+\frac{x
   \left(5 x^2+3\right) }{2
   \left(x^2-1\right)}\log\left(\frac{1+x}{x}\right)\nonumber \\
   &+\frac{\left(5 x^2-4
   x+1\right)}{x-1}\left(\Theta (1-2 x) \log
   \left(\frac{\tfrac{1}{2}-x}{1-x}\right)+\Theta (2
   x-1) \log \left(\frac{x-\tfrac{1}{2}}{
   x}\right)\right)\Bigg].
\end{align}
We see new logarithms appear instead, $\log(1+x)$, which is infinitely differentiable in $x\in(0,1)$ and, more importantly, $\log(|2x-1|)$, which has an integrable singularity at $x=1/2$. The contribution from embeddings in class three and four is significantly more involved than that from embedding one and two; however, it is still integrable over the entire range $x\in(0,1)$. 

\partitle{Double quark $F_{2qq}$ in the MP model} We finally give the result for the $qq$ structure function:
\begin{align}
\label{eq:qq_MP}
&F_{2qq}^{\text{MP}}(x)=\frac{ e_q^2 \alpha_s C_F x}{2 \pi  (2N_c)}\Bigg[ 
  \frac{1}{2}(1 - 2 x) \left(\log\left(\frac{\tfrac{1}{2} - x}{1 - x}\right) \Theta(1 - 2 x) + 
    \log\left(\frac{x-\frac{1}{2}}{ x}\right) \Theta(2 x-1)\right) \nonumber\\
    &-\frac{1}{2}\frac{\log(2 - 2 x)}{1 - 2 x}\Theta(
      1 - 2 x)-\frac{1}{2}\frac{\log(2 x)}{1 - 2 x}\Theta(2 x-1)+\frac{1}{2}\left[\frac{\log(|1 - 2 x|)}{1 - 2 x}\right]_+-\Theta(2 x-1)\Bigg] .
\end{align}
This contribution is also integrable over the full range $x\in(0,1)$, and is suppressed by $2N_c$ due to the modified averaging factor for the $qq$ contributions. Only class four embeddings contribute to $F_{2qq}$.

\partitle{Scale dependence} An important feature of the result is its scale dependence. As highlighted in eq.~\eqref{eq:gluonic_MP} and eq.~\eqref{eq:fermionic_MP}, the scale dependence of the MP result is the same as that of the parton model. In particular, $F_{2q}$ and $F_{2g}$ both depend logarithmically on $\Lambda^2/Q^2$ through the usual Altarelli-Parisi splitting kernels. Conversely, $F_{2qq}$, as given in eq.~\eqref{eq:qq_MP}, does not depend on $\Lambda^2$. \emph{We find that at NLO contributions from all class three and four embeddings do not have infrared logarithms, although they do contribute to the finite part}. This may be seen as a direct consequence of the specific KLN cancellation pattern involved in these contributions, which makes it so that each of the four contributions in eq.~\eqref{eq:disc_and_discs} is free of initial-state singularities (although that is not the case for individual interference diagrams). Let us see this for class four embeddings: as we have seen, for a class four embedding $\Gamma$, it holds that
\begin{equation}
W_\Gamma(p,q)=(p^2)^{-1-\epsilon}C_\Gamma(1-x,q^2),
\end{equation}
i.e. the hard function vanishes. In order for $W^{\mu\nu}_\Gamma(p,q)$ to be a finite distribution, we must thus have that $C_\Gamma(1-x,q^2)=\mathcal{O}(\epsilon^0)$, meaning that $C_\Gamma$ vanishes for $\epsilon\rightarrow 0$. In turn, this implies that 
\begin{equation}
\label{eq:gamma4}
W_\Gamma(p,q)=-\frac{\delta(p^2)}{\epsilon}C_\Gamma(1-x,q^2)+\mathcal{O}(\epsilon^0),
\end{equation}
which, integrated in $p^2$ from $0$ to $\Lambda^2$, gives no logarithms at the finite order, but it will give a finite part that is localised at $p^2=0$.

\partitle{Class three and four embeddings} In the presentation of the results, we explicitly divided the contribution to coefficients functions from embeddings in classes one and two from that of class three and four embeddings. Class three and four embeddings introduce new analytic structures, such as branch-cuts starting at $x=1/2$, that are unusual and not commonly seen in the context of the parton model. This, together with the fact that they are actually not strictly needed to make the parton model finite, could be an argument for their exclusion. Nonetheless, this intuition should be backed by more rigorous arguments. One possibility is that contact term such as those of eq.~\eqref{eq:gamma4}, which do not arise from diagrams having a propagator with momentum $p$ being cut, should be excluded, in alignment with the idea that diagrams with multiple initial-state partons can have infinitesimal, but non-vanishing, virtuality.

\partitle{Higher windings} At NLO, the discussion of higher windings is a bit subtler. The set of embeddings one should work with is, in principle, infinite: at any fixed perturbative order, one needs amplitudes with a fixed amount of vertices but with an arbitrary number of spectators. In the results above, we have only computed a gauge invariant and IR-finite subset, corresponding to the embeddings given as ancillary material. 

We start by observing that higher winding embedding are all in class four. Hence, a physical argument that excludes the contribution of class four embeddings would at the same time eliminate the problem of higher winding. If instead one plans to account for the contribution of all embeddings with arbitrary windings, some type of formal truncation or resummation procedure is needed. Moreover, the specifics of this resummation or truncation procedure should be somewhat consistent with physical principles. 

We believe that a general solution to the problem of higher windings can only be found by exploring at the same time its relation with expected notions of universality, e.g. by comparison of the NLO DIS computation performed in this paper with an NLO Drell-Yan computation. 

\section{First glance beyond next-to-leading order MP deep inelastic scattering}
\label{sec:beyond_NLO_DIS}

In this section we explore the applicability of the model at higher orders and for hadronic collisions. For what concerns the first objective, we detail KLN cancellations for a selected two-loop embedding relevant to the NNLO calculation of the deep inelastic scattering cross-section. For what concerns the second, we extend show how to extend the simplest clustering criterion in a Drell-Yan toy model.

\subsection{Discontinuities and KLN cancellations for a two-loop example}
\begin{figure}
    \centering
    \input{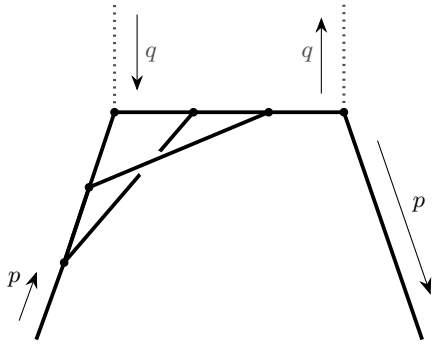}
    \caption{The three-point integral, including the additional propagator $\frac{1}{(p+q)^2}$.}
    \label{fig:ThreeptIntegral}
\end{figure}

Let us outline the computation at NNLO by illustrating the IR cancellations for the diagram depicted in  Figure~\ref{fig:ThreeptIntegral} that will arise in an NNLO computation.
As at NLO we need to compute 4 classes
\begin{align}
    \mathcal{C}_1={}&\frac{1}{(p+q)^2}\int_0^{\Lambda^2}\frac{\mathrm{d}p^2}{p^2}\operatorname{disc}_{p^2}\operatorname{disc}_{(p+q)^2}F_4(p^2,(p+q)^2,q^2)\,,\\
    \mathcal{C}_2={}&\delta((p+q)^2)\int_0^{\Lambda^2}\frac{\mathrm{d}p^2}{p^2}\operatorname{disc}_{p^2} F_4(p^2,0,q^2)\,,\\
    \mathcal{C}_3={}&\frac{1}{(p+q)^2}\operatorname{disc}_{(p+q)^2}F_4(0,(p+q)^2,q^2)\,,\\
    \mathcal{C}_4={}&\delta((p+q)^2)F_4(0,0,q^2)\,.
\end{align}
All integrals are special cases of the integral 
\begin{equation}
F_4(p_1^2,p_2^2,p_3^2)=\int\frac{\mathrm{d}^D\ell_1\mathrm{d}^D\ell_2}{(\imath \pi^{D/2})^2}\frac{1}{\ell_1^2\ell_2^2(\ell_1+p_1)^2(\ell_2+p_2)^2(\ell_1-\ell_2+p_1)^2(\ell_2-\ell_1+p_2)^2}\,.
\end{equation}
Note that this expression is fully symmetric in its arguments.
This integral has been studied frequently in the literature, for example in refs.~\cite{Usyukina:1992jd,Chavez:2012kn}. We will only compute in an expansion about $\Lambda^2= 0$, which means we can expand in $p^2$.

As for the NLO case we will need this integral in various
configurations of the external lines, where some of them are on-shell and in addition we need to compute discontinuities. All necessary information can be obtained from the off-shell expression. 
It is convenient to derive differential equations for the integral with respect to the kinematic parameters. 
\subsubsection{Double discontinuity}
For the double discontinuity we directly evaluate the required cuts in the limit of small $p^2$ using the method of regions.
In this way we can obtain expressions that are exact in the dimensional regulator thus allowing to perform the integration in $p^2$ and the extraction of the distributional part in $(p+q)^2$. 
When expanding in $\epsilon$ we find a finite expression
\begin{align}
&\operatorname{disc}_{p^2}\operatorname{disc}_{(p+q)^2}F_4(p^2,(p+q)^2,q^2)\nonumber\\
={}& -\frac{8 \pi^2}{3 \left((p+q)^2 - q^2\right)^2} \Bigg( 
  \pi^2 
  + 6 \log(p^2) \left[ \log((p+q)^2) - \log(-q^2) \right] \nonumber\\
& \quad 
  - 12 \log((p+q)^2) \log\left(1 - \frac{(p+q)^2}{q^2} \right)
  - 6 \log((p+q)^2) \log(-q^2)\nonumber \\
& \quad 
  + 12 \log\left(1 - \frac{(p+q)^2}{q^2} \right) \log(-q^2)
  + 6 \log^2(-q^2)
  - 6 \operatorname{Li}_2\left( \frac{(p+q)^2}{q^2} \right)
\Bigg)\,.
\end{align}
This agrees with the result in Ref.~\cite{Usyukina:1992jd}, expanded in small $p^2$.
Note, however, that even though the discontinuity is finite in $d=4$ the contribution $\mathcal{C}_1$ is not. This is due to the additional integration and due to the distributional nature of the result. In fact, we find that the result is of order $\mathcal{O}(\epsilon^{-4})$.
This is due to cancellation patterns between different regions of the form 
\begin{equation}
    (1-x)^{-1+\epsilon}-(1-x)^{-1+2\epsilon}=\frac{1}{2\epsilon}\delta(1-x)+\mathcal{O}(\epsilon^{0})\,.
\end{equation}
A naive expansion would instead result in a (non-integrable) expression of order $\mathcal{O}(\epsilon)$. This pattern is repeated for both the $p^2$ and $x$ integration, and hence we observe a $\mathcal{O}(\epsilon^{-4})$ enhancement with respect to the finite integrand.

\subsubsection{Single discontinuity}
We need the integral with a single discontinuity and a on-shell leg.
For this we can specify $p_1^2=0$ and $p_2^2>0$, $p_3^2<0$ and then obtain the two cases needed for the problem at hand by setting $p_1=p,p_2=p+q$ and $p_1=p+q,p_2=p$ respectively. 
We proceed by computing univariate differential equations in $z=p_2^2/p_3^2$. The boundary conditions can be fixed by the method of regions e.g. in the limit $p_2\to 0$. From this we can obtain the discontinuity in a form, e.g. for case 2
\begin{equation}
    F_4(0,p_2^2,p_3^2)=(-p_3^2/\mu^2)^{-2\epsilon}f_0(p_2^2/p_3^2)+(-p_3^2/\mu^2)^{-\epsilon}(p_2^2/\mu^2)^{-\epsilon}f_1(p_2^2/p_3^2)+(p_2^2/\mu^2)^{-2\epsilon}f_2(p_2^2/p_3^2)\,.\label{eq:TwoOffShell}
\end{equation}
Where $f_i$ are analytic at $p^2=0$.
From this expression, it is straightforward to obtain the discontinuity.
In order to perform the $p^2$ integration as required for case 2, we expand the functions $f_i$ in small arguments. Keeping the leading order all integrations are trivial\footnote{More generally we can work out the full integrals by building appropirate subtraction terms. In this way we obtain an integrable expression in terms of polylogs that can be readily integrated in terms of Generalized polylogs. We refrain from giving more details as the leading order suffices for the discussion at hand.}.
For case 3 we set $p_1^2=p^2=0$,$p_2^2=(p+q)^2>0$ and expand the exponential factors using the distributional identity. 
\begin{align}
    \frac{1}{(p+q)^2}\operatorname{disc}_{(p+q)^2}F_4(0,(p+q)^2,q^2)={}&\sin(\pi\epsilon)(-q^2)^{-\epsilon}((p+q)^2)^{-1-\epsilon}f_1(x)\nonumber\\
    {}&+\sin(2\pi\epsilon)((p+q)^2)^{-1-2\epsilon}f_2(x)\,.
\end{align}
Note that this expression is homogeneous in the $q$-expansion.

\subsubsection{On-shell integral}
Finally, we also need the integral with two on-shell legs. Given that this is the hard region of the integral we can obtain it from Eq.~\eqref{eq:TwoOffShell}
\begin{equation}
    \mathcal{C}_4=\delta((p+q)^2)F_4(0,0,q^2)=\delta((p+q)^2)(-q^2/\mu^2)^{-2\epsilon}f_0(0)\,.
\end{equation}
Interestingly, the contribution from the double discontinuity is finite, but when integrated over $p^2$ and carefully treating dependence on $(p+q)^2$ in a distributional sense, we find a pole of order $\mathcal{O}(\epsilon^{-4})$.
Adding up all the contributions, we observer cancellations of IR poles, leaving a finite expression
\begin{align}
    I={}&\mathcal{C}_1+\mathcal{C}_2+\mathcal{C}_3+\mathcal{C}_4\nonumber\\
    ={}&\frac{1}{(q^2)^3}\left[I_0(x)+I_1(x)\log(-q^2/\Lambda^2)+I_2(x)\log^2(-q^2/\Lambda^2)+\mathcal{O}(\Lambda^2/q^2)\right]
\end{align}
where
\begin{align}
    I_0={}&\frac{29}{60} \pi^4 \, \delta(1 - x) + \frac{2\pi^2}{3} \left[ \frac{\log(1 - x)}{1 - x} \right]_+
- \frac{2}{3  (1 - x)} \Bigg[
\pi^2 (x^3 - 1) \log(1 - x) 
\nonumber\\
&\qquad\qquad + x^3 \log x \left( -5 \pi^2 + 6 \log^2 x \right) 
+ 12 x^3 \log x \, \operatorname{Li}_2(x)
\Bigg]\,,
\\
I_1={}&\frac{2}{3 (x-1)} \left(6 x^3 \operatorname{Li}_2(x)-2 \pi ^2 x^3+9 x^3 \log ^2(x)-6 x^3 \log (1-x) \log (x)+\pi ^2\right)
\nonumber\\
&+ \frac{2}{3} \pi^2 \left[ \frac{1}{1 - x} \right]_+\,,
\\
I_2={}&\frac{\pi^2}{3} \, \delta(1 - x) 
+ \frac{2}{(1 - x)} \left[
(x^3-1) \log(1 - x) - x^3 \log x
\right] 
- 2\left[ \frac{\log(1 - x)}{1 - x} \right]_+\,.
\end{align}
These expressions are integrable and can be convoluted against an appropriate test function, such as a parton distribution function. Note that the renormalization scale $\mu$ has canceled along with the poles and the only regulator dependence is through the cutoff $\Lambda$.

\subsection{Simplest clustering in the Drell-Yan process}
\label{sec:DY}
A natural question concerning the content of this paper is how one would plan to extend the formalism presented here to processes involving the scattering of two protons. However we plan to approach the broad question of how multi-partonic initial-states can be used in the context of proton collisions, we are forced to address the issue of the long-range interactions that involve the protons subject to collision. We will try to start answering this question by describing one way to do this that is entirely compatible with the simplest clustering criterion which, it bears repeating, states that the only constraints from the clustering of the initial state particles into one or two protons should be achieved by momentum-conservation constraints alone. 

In the following, for simplicity, we will focus on a scalar version of the Drell-Yan process, $j(P_1) j(P_2) \rightarrow\phi^\star+X$. Here, in accordance with the KLN theorem, we consider two initial-state jets made of scalars. Hence interference diagrams involve initial-state cuts with any number of massless scalars. We start by focusing on one embedding, namely the equivalence class containing the following interference diagrams:
\begin{equation}
\raisebox{-1.1cm}{\scalebox{0.9}{\input{Submission/diagrams/beyond_NLO/DY_triangle_1}}}, \ \raisebox{-1.1cm}{\scalebox{0.9}{\input{Submission/diagrams/beyond_NLO/DY_triangle_2}}}, \ \raisebox{-1.1cm}{\scalebox{0.9}{\input{Submission/diagrams/beyond_NLO/DY_triangle_3}}}, \ \raisebox{-1.1cm}{\scalebox{0.9}{\input{Submission/diagrams/beyond_NLO/DY_triangle_4}}}.
\end{equation}
Since in this class only the initial-state cut varies, we have reversed the usual way we display forward-scattering diagrams, by keeping the final-state-instead of the initial-state, on the utmost left and right. Indeed, the dashed particle, cut by the blue line, is the final-state of-shell scalar $\phi^\star$ with mass $m_\star$. Of course, this same embedding also appears in the process $\phi^\star\rightarrow j(P_1) j(P_2)$, with the meaning of initial and final-state cuts being swapped.

We would now like to construct the jets from the particles crossed by the red cut. A naive extension of the simplest clustering criterion would be the following: given a certain number of initial-state particles, labelled by momenta $p_1,...,p_n$, where $n=2$ or $n=3$ in the interference diagrams above, we construct all bipartitions of these particles into two jets with momenta $P_1$ and $P_2$, and impose that the momenta of the particles in the first/second jet sum up to $P_1$/$P_2$. These partitions amount to a sectoring of the kinematic space of the initial-state particles. When $n=2$, we can only have $p_1=P_1, \ p_2=P_2$ or $p_1=P_2, \ p_2=P_1$. Hence, we can assign the following momentum labels to the interference diagrams
\begin{equation}
\raisebox{-1.3cm}{\scalebox{0.9}{\input{Submission/diagrams/beyond_NLO/DY_triangle_1_label}}} , \quad \raisebox{-1.3cm}{\scalebox{0.9}{\input{Submission/diagrams/beyond_NLO/DY_triangle_4_label}}}, \quad (P_1\leftrightarrow P_2). \nonumber
\end{equation}
When $n=3$, instead, we have the following sectors:
\begin{align}
&\Omega_1^{\text{c}}=\{\{p_i\}_{i=1}^3 \ | \ P_1=p_1+p_2, \ P_2=p_3\}, \ \ \Omega_2^{\text{c}}=\{\{p_i\}_{i=1}^3 \ | \ P_1=p_1, \ P_2=p_2+p_3\}, \\
&\Omega_3^{\text{c}}=\{\{p_i\}_{i=1}^3 \ | \ P_1=p_1+p_3, \ P_2=p_2\}, \ \ \Omega_4^{\text{c}}=\{\{p_i\}_{i=1}^3 \ | \ P_2=p_1+p_2, \ P_1=p_3\}, \\
&\Omega_5^{\text{c}}=\{\{p_i\}_{i=1}^3 \ | \ P_2=p_1, \ P_1=p_2+p_3\}, \ \
\Omega_6^{\text{c}}=\{\{p_i\}_{i=1}^3 \ | \ P_2=p_1+p_3, \ P_1=p_2\}. 
\end{align}
Up to conjugation and $P_1\leftrightarrow P_2$ transformations, and considering the following conventional assignment of momenta:
\begin{equation}
\raisebox{-1.3cm}{\scalebox{0.9}{\input{Submission/diagrams/beyond_NLO/DY_triangle_2_conv_label}}},\nonumber
\end{equation}
these are all the corresponding valid momentum labellings up to re-routing ok $k$
\begin{equation}
\Omega_1^{\text{c}}: \raisebox{-1.3cm}{\scalebox{0.9}{\input{Submission/diagrams/beyond_NLO/DY_triangle_2_label1}}}, \quad \Omega_2^{\text{c}}: \raisebox{-1.3cm}{\scalebox{0.9}{\input{Submission/diagrams/beyond_NLO/DY_triangle_2_label2}}}, \quad \Omega_3^{\text{c}}: \raisebox{-1.3cm}{\scalebox{0.9}{\input{Submission/diagrams/beyond_NLO/DY_triangle_2_label3}}}.  \nonumber 
\end{equation}
It may not surprise the reader familiar with clustering algorithms that, constructing the cross-section for these interference diagrams and integrating in $P_1^2$ and $P_2^2$ from $0$ to $\Lambda^2$ in an analogous (but doubled!) way to deep inelastic scattering, does not give an infrared-finite result. The reason is that the region in which the ``vertical particle'' is soft is double-counted. In the language of simplest clustering, we must include the case in which particles belong to both jets at the same time. We will soon see that in the virtuality expansion this region will indeed force certain particles to be soft. In the mean-time, let us define these regions as
\begin{align*}
&\Omega_1^{\text{s}}=\{\{p_i\}_{i=1}^3 \ | \ P_1=p_1+p_2, \ P_2=p_2+p_3\}, \\ 
&\Omega_2^{\text{s}}=\{\{p_i\}_{i=1}^3 \ | \ P_1=p_2+p_1, \ P_2=p_1+p_3\}, \\
&\Omega_3^{\text{s}}=\{\{p_i\}_{i=1}^3 \ | \ P_1=p_2+p_3, \ P_2=p_3+p_1\}. \\
&\Omega_4^{\text{s}}=\{\{p_i\}_{i=1}^3 \ | \ P_2=p_1+p_2, \ P_1=p_2+p_3\}, \\ 
&\Omega_5^{\text{s}}=\{\{p_i\}_{i=1}^3 \ | \ P_2=p_2+p_1, \ P_1=p_1+p_3\}, \\
&\Omega_6^{\text{s}}=\{\{p_i\}_{i=1}^3 \ | \ P_2=p_2+p_3, \ P_1=p_3+p_1\}.
\end{align*}
Up to conjugation and $P_1\leftrightarrow P_2$ transformations, they correspond to the following labellings
\def\stack#1{\begin{subarray}{c}#1\end{subarray}}

\begin{equation}
\Omega_1^{\text{s}}: \raisebox{-1.3cm}{\scalebox{0.9}{\input{Submission/diagrams/beyond_NLO/DY_triangle_2_label1_s}}}, \ \ \Omega_2^{\text{s}}: \raisebox{-1.3cm}{\scalebox{0.9}{\input{Submission/diagrams/beyond_NLO/DY_triangle_2_label2_s}}}, \ \ \Omega_3^{\text{s}}: \raisebox{-1.3cm}{\scalebox{0.9}{\input{Submission/diagrams/beyond_NLO/DY_triangle_2_label3_s}}}.  \nonumber 
\end{equation}
We are now ready to construct the cross-section. The contribution of an interference diagram to the cross-section is written by inserting the Feynman rules of the diagram, including cutting rules. Then, the unconstrained momentum $k$ is integrated over, and the momenta $P_1$ and $P_2$ are chosen to be aligned with the $z$ axis such that $P_1=(\sqrt{x_1^2 s/4 + m_1^2},0,x_1 \sqrt{s}/2)$, $P_2=(\sqrt{x_2^2 s/4 + m_2^2},0,-x_2 \sqrt{s}/2)$. Finally, $m_1^2$ and $m_2^2$ are integrated from $0$ to $\Lambda_1^2$ and $0$ to $\Lambda_2^2$ respectively. 
Letting
\begin{equation}
\mathrm{d}\Pi_{HH}= \prod_{i=1}^2\mathrm{d}m_i^2\Theta(\Lambda_i^2-m_i^2)
\end{equation}
we have
\begin{align}
\raisebox{-1.cm}{\scalebox{0.75}{\input{Submission/diagrams/beyond_NLO/DY_triangle_1_label}}}=\delta(x_1x_2s-m_\star^2)\int \mathrm{d}\Pi_{HH}\tilde{\delta}(m_1^2)\tilde{\delta}(m_2^2)T(P_1,P_2,-P_1-P_2),
\end{align}
where $T$ is the triangle diagram. Another example is:
\begin{align}
\raisebox{-1.cm}{\scalebox{0.75}{\input{Submission/diagrams/beyond_NLO/DY_triangle_2_label1}}}=\delta(x_1x_2s-m_\star^2)\int \mathrm{d}\Pi_{HH}\tilde{\delta}(m_2^2)\frac{\text{disc}_{m_1^2}T(P_1,P_2,-P_1-P_2)}{m_1^2},
\end{align}
we see that the simplest clustering criterion, again, gives us discontinuities. Concerning an example for the soft region, a careful look at the routing corresponding to $\Omega_1^{\text{s}}$ reveals that:
\begin{align}
\raisebox{-1.cm}{\scalebox{0.75}{\input{Submission/diagrams/beyond_NLO/DY_triangle_2_label1_s}}}=\delta(x_1x_2s-m_\star^2)\int \mathrm{d}\Pi_{HH}\frac{\text{disc}_{m_1^2,m_2^2}T(P_1,-P_2,-P_1+P_2)}{m_1^2 m_2^2},
\end{align}
Note the sign reversal in front of $P_2$ in the arguments of the triangle. These integrals can be carries out and the virtuality expansion in $\Lambda_i$ taken. At this point, many of the sectors we defined previously will automatically drop out as they are sub-leading virtuality, such as $\Omega_2^\mathrm{s}$, $\Omega_3^\mathrm{s}$, $\Omega_3^{\text{c}}$ and those obtained from these ones by swapping $P_1\leftrightarrow P_2$. For example, $\Omega_3^{\text{c}}$ is written as
\begin{equation}
\raisebox{-1.0cm}{\scalebox{0.75}{\input{Submission/diagrams/beyond_NLO/DY_triangle_2_label3}}}=\delta(x_1x_2s-m_\star^2)\int \mathrm{d}\Pi_{HH}\tilde{\delta}(m_2^2)\text{disc}_{m_1^2}B(P_1+P_2,P_2,-P_2,-P_1-P_2),
\end{equation}
where $B$ is the box diagram, which for the given kinematics gives a contribution starting at $\Lambda_1^2$ once integrated in $m_1^2$. In summary, including only the leading-virtuality diagram labellings, and dividing by $2$ to average over $P_1\leftrightarrow P_2$ symmetry, the cross-section can be written as
\begin{align}
\sigma=&2\delta(x_1x_2s-m_\star^2)\int \mathrm{d}\Pi_{HH}\Bigg( \tilde{\delta}(m_1^2)\tilde{\delta}(m_2^2)\text{Re}[T(P_1,P_2,-P_1-P_2)]\nonumber \\
+&\tilde{\delta}(m_2^2)\frac{\text{disc}_{m_1^2}T(P_1,P_2,-P_1-P_2)}{m_1^2}
+\tilde{\delta}(m_1^2)\frac{\text{disc}_{m_2^2}T(P_1,P_2,-P_1-P_2)}{m_2^2} \nonumber\\
+&\frac{\text{disc}_{m_1^2,m_2^2}T(P_1,-P_2,-P_1+P_2)}{m_1^2 m_2^2}\Bigg)+\mathcal{O}\left(\frac{\Lambda_1^2}{m_\star^2},\frac{\Lambda_2^2}{m_\star^2}\right).
\end{align}
Note that the cross-section expresses as a sum of iterated discontinuities, one for each of the two incoming protons. If one also interprets the final-state cut as a discontinuity (although it may be more instructive to do so for a diagram that includes final-state QCD radiation), then \emph{the inclusive partonic cross-section expresses as a triple discontinuity}.

Explicit evaluation of all the integrals gives a finite quantity:
\begin{equation}
\sigma=\delta(z -m_\star^2/s)\frac{\pi^2-6\log(\Lambda_1^2/m_\star^2)\log(\Lambda_2^2/m_\star^2)}{48 s m_\star^2},
\end{equation}
with $z=x_1x_2$. This concludes our preliminary investigation of the implementation of the simplest clustering criterion in Drell-Yan.

\section{Outlook}

We first briefly summarise the main aspects discussed in this paper:
\begin{itemize}
\item We have defined a perturbative model that includes the effect of multi-partonic interactions at leading-virtuality within the deep inelastic scattering process. In order to do so, we accounted for all diagrams with an arbitrary number of initial and final state partons. We have clustered initial state partons into a jet using momentum-conservation to set the sum of the momenta of the partons to that of the incoming jet and setting an upper-bound for the invariant mass of the jet. We have called this clustering the ``simplest clustering''. We have argued that, if we are interested in leading terms in the expansion of the invariant mass of the jet, the ``simplest clustering'' criterion arises naturally. We have also discussed how to extend the simplest clustering criterion to the Drell-Yan process.

\item We have discussed how diagrammatic contributions to the cross-section defined in this way can be classified in non-overlapping subsets of infrared finite contributions. Diagrams in the same infrared-finite class can be obtained from different cuts of the same object, the embedding. We have explained how to generate non-isomorphic embeddings. Then, we have provided a classification of embedding into four types. Finally, we have shown that there is a natural way to associate an loop integrand and a loop integral to each embedding.

\item We have shown that instead of computing all the interference diagrams sharing the same embedding, one may instead compute a double discontinuity of the embedding's loop integral. We have reformulated the KLN theorem in terms of iterated discontinuities of embeddings. We looked at the analytic properties of these iterated discontinuities passing through expansion-by-regions arguments.

\item We have presented the results at NLO for the inclusive deep inelastic scattering process. We found that the result differs from that of the parton model by a scheme change, since the longitudinal structure functions evaluate to the same in the two models. The scale dependence of $F_2$ is also the same in the two models. However, the finite pieces differ. New analytic structures appear for the fermionic structure function. We also computed the contribution to the cross-section arising from the simultaneous scattering of two quarks/anti-quarks with the electron. This contribution cannot be accounted for in the parton model. 
\end{itemize}

What about future work? A few possibilities come to mind:
\begin{itemize}

\item One of the first conceptual developments that should follow this paper is the systematic inclusion within the formalism of the infinite tower of higher-winding embeddings. Their treatment is unsatisfactory at the current stage of our work, although they can each be computed by the same methods discussed in this paper. Solving this problem would likely involve one of three routes: a) resumming them, b) truncating the expansion in winding (under the assumption that they contribute less and less with the increase in winding numbers) or c) arguing by physical principles that they can be entirely removed (it may indeed be that the most physically-sensible formalism entails the inclusion of only the minimal set of embeddings containing the parton model). Regardless, we see an NLO Drell-Yan calculation as the perfect setup to address this question completely.

\item Relating to the previous point, an obvious extension of this calculation is the NLO correction to the Drell-Yan cross-section. The question would be: is the scheme change term necessary to turn MP DIS into the parton model DIS the same that is needed to turn MP DY into the parton model DY? If the answer to this question is negative, it constitutes a first proof that the two models produce different predictions. 

\item In general, the applicability of the model for general processes at NLO should be studied. If automation at NLO could be achieved, it could provide a unique opportunity to test predictions from the parton model, and thus give a handle on the estimation of theoretical uncertainties. Extension to NNLO should also be investigated.

\item A generic simplest clustering algorithm could be applied to cluster final states too. If successful, such an endeavour would allow to write differential cross-sections in terms of iterated discontinuities: one discontinuity for each colliding proton, and one for each observed final-state jet. This way of casting the problem of phase-space integration would make it especially suited to analytic integration.

\item One interesting endeavour is the study of power-corrections within the MP model. In principle, power-corrections in the MP model can be computed, including their scaling in $\Lambda^2/Q^2$, although one should still be careful to assess the extent of their physical significance.  
These calculations could be used to test and extend previous work stating that the inclusion of MP interactions within a KLN-like formalism pushes power-corrections to order $(\Lambda^2/Q^2)^2$ in the Drell-Yan process~\cite{Akhoury_1996,Akhoury:1995sp}. 

\end{itemize}

This list does not exhaust completely the possibilities for future research, although it states the ones we find most compelling.

\acknowledgments The work presented in this paper is also the result of countless discussion the authors had with high-energy physics researchers. An exhaustive list would be prohibitive. We thank Valentin Hirschi for his continuous support and help throughout the project. We also thank him and Ben Ruijl for what concerns diagram generation. We thank Einan Gardi for pointing out the relation between the integrals appearing in the MP model and iterated discontinuities of Feynman diagrams, and Pier Monni and Thomas Gehrmann for their questions on the extension of the MP model to hadronic collisions, which directly motivated our writing of sect.~\ref{sec:DY}. At different stages of the development of the project, the authors had illuminating discussions on the topics of this paper with Samuel Abreu, Babis Anastasiou, Thomas Becher, Fabrizio Caola, Stefano Forte, Alexander Huss, Lorenzo Magnea, Bernhard Mistleberger, Hua-Sheng Shao, George Sterman, Gherardo Vita. We warmly thank Thomas Becher, Thomas Gehrmann and Lorenzo Magnea for their comments on a draft of this paper. This work was initiated at the Aspen Center for Physics, which is supported by National Science Foundation grant PHY-2210452. The work of Z.C. and L.H. is supported by the Swiss National Science Foundation (SNSF) under grant number PCEFP2\_203335. The work of M.H. is supported by the Department of Energy, Contract DE-AC02-76SF00515.

\appendix

\section{One loop masters}

In this appendix we give the expressions for the masters required for the Reverse Unitarity calculation detailed in sect.~\ref{sec:interference_diags} as well as the region integrals required for the discontinuity calculation of sect.~\ref{sec:embeddings_anal}.

\subsection{Cut masters}
\label{sec:masters1loop}
As mentioned in sect.~\ref{sec:interference_diags}, the Reverse Unitarity calculation takes the virtual bubble $B(p^2)$, the cut bubble $\slashed{B}(p^2)$ and the doubly-cut triangle $\slashed{T}(p_1^2,p_2^2,p_3^2)$ as inputs. The virtual bubble is obtained straight-forwardly by Feynman parametrisation
\begin{align}
(2\pi)^d B(p^2)=\int \mathrm{d}^d k \frac{1}{k^2(p-k)^2}
&=\frac{\imath\pi^{\frac{d}{2}}\Gamma\left(\frac{d-2}{2}\right)^2 \Gamma\left(\frac{4-d}{2}\right)}{\Gamma\left(d-2\right)}(-p^2)^{\frac{d-4}{2}} \nonumber\\
&=\frac{\imath\pi^{2-\epsilon}\Gamma\left(1-\epsilon\right)^2 \Gamma\left(\epsilon\right)}{\Gamma\left(2-2\epsilon\right)}(-p^2)^{-\epsilon}\label{eq:virtual_bubble}
\end{align}
where we set $d=4-2\epsilon$. The cut bubble and the doubly cut triangle are obtained by direct integration in momentum space:
\begin{align}
(2\pi)^d\slashed{B}(p^2)=\int \mathrm{d}^d k \tilde{\delta}^+(k)\tilde{\delta}^+(p-k)&=-\frac{2^{4-d}(p^2)^{\frac{d-4}{2}}\pi^{\frac{d+3}{2}}}{\Gamma\left(\frac{d-1}{2}\right)}\Theta(p^0)\Theta(p^2) \nonumber\\
&=-\frac{2^{2\epsilon}(p^2)^{-\epsilon}\pi^{\frac{7}{2}-\epsilon}}{\Gamma\left(\frac{3}{2}-\epsilon\right)}\Theta(p^0)\Theta(p^2),
\end{align}
For the cut triangle we find
\begin{align}
(2\pi)^d\slashed{T}(p_1^2,p_2^2,p_3^2)&=\int \mathrm{d}^d k \, \tilde{\delta}^-(k)\tilde{\delta}^+(k+p_1)\tilde{\delta}^+(k-p_3)\nonumber \\
&=-4\imath\frac{(-p_1^2p_2^2p_3^2)^{\frac{d-4}{2}}}{\lambda(p_1^2,p_2^2,p_3^2)^{\frac{d-3}{2}}}\frac{\pi^{\frac{d}{2}+2}}{\Gamma\left(\frac{d}{2}-1\right)}\Theta(-p_1^0)\Theta(p_1^2)\Theta(p_3^0)\Theta(p_3^2)\nonumber\\
&=-4\imath\frac{(-p_1^2p_2^2p_3^2)^{-\epsilon}}{\lambda(p_1^2,p_2^2,p_3^2)^{\frac{1}{2}-\epsilon}}\frac{\pi^{4-\epsilon}}{\Gamma\left(1-\epsilon\right)}\Theta(-p_1^0)\Theta(p_1^2)\Theta(p_3^0)\Theta(p_3^2),\label{eq:double_disc}
\end{align}
where $\lambda(x,y,z)=x^2+y^2+z^2-2xy-2xz-2yz$ is the K\"allen function.

\subsection{Calculation of $T_h$}
\label{sec:hard_region_triangle}
The hard function $T_h$ required in the region expansion of sect.~\ref{sec:embeddings_anal} is simply the two-mass triangle, which IBP-reduces to bubbles:
\begin{equation}
T_h(p,q)=\frac{(d-3)[B(q)-B(p+q)]}{(d-4)p\cdot (p+q)}.
\end{equation}
Using the result for the bubble of eq.~\eqref{eq:virtual_bubble}:
\begin{equation}
(2\pi)^dT_h(p,q)=\imath \pi^{d/2} \frac{\Gamma(1 - \epsilon)^2 \Gamma(\epsilon)}{\Gamma(2 - 2 \epsilon)}\frac{(d-3)[(-q^2)^{\frac{d-4}{2}}-(-(p+q)^2)^{\frac{d-4}{2}}]}{(d-4)p\cdot (p+q)}.
\end{equation}

\subsection{Calculation of $T_c$}
\label{sec:collinear_region_triangle}

The collinear function $T_c$ required in the region expansion of sect.~\ref{sec:embeddings_anal} is computed in light-cone coordinates. It reads:
\begin{align}
(2\pi)^d(-p_1^2)^{-\epsilon}T_c(p_1,p_2)=&\int \mathrm{d}^d k \frac{1}{k^2(k-p_1)^2(2k^+ (n_+\cdot p_2) +p_2^2)}\,.
\end{align}
Where we have used light-cone variables, $k=k^+\hat{n}_++k^-\hat{n}_-+k_\perp$, $\hat{n}_{\pm}\cdot k_\perp=0$, $\hat{n}_{\pm}^2=0$ and $\hat{n}_+\cdot \hat{n}_-=2$. The integral $T_c$ corresponds to the leading collinear region in the $p_1^2$ expansion of the three-mass triangle. Changing to light-cone coordinates and performing contour integration in $k^-$, we have, assuming $p^+_1>0$
\begin{align}
(2\pi)^d(-p_1^2)^{-\epsilon}T_c=&\imath\pi\int_{0}^{1} \mathrm{d}t \int \mathrm{d}^{d-2}\vec{k}_\perp \frac{1}{(p_1^2 t (t-1)+|\vec{k}_\perp|^2)(2t(p_1\cdot p_2)+p_2^2)}.
\end{align}
where we have introduced $t=k^+/p_1^+$. The integral in $|\vec{k}_\perp|$ is trivial to perform in terms of a beta function. After integration in $|\vec{k}_\perp|$ and in the solid angle associated with the $\mathrm{d}^{d-2}\vec{k}_\perp$ integration, we obtain:

\begin{align}
(2\pi)^d(-p_1^2)^{-\epsilon}&T_c(p_1,p_2)=-\frac{\imath\pi^{d/2+1}(p_1^2)^{d/2-2}\csc((d-4)\pi/2)}{\Gamma(d/2-1)}\int_0^1 \mathrm{d}t \frac{\left[t (t-1)\right]^{d/2-2}}{2t(p_1\cdot p_2)+p_2^2} \nonumber \\
=&\frac{\imath\pi^{d/2+1}\csc(-\epsilon\pi)\Gamma(1-\epsilon)}{(-p_2^2)\Gamma(2-2\epsilon)}(-p_1^2)^{-\epsilon}{}_2F_1\left[1,1-\epsilon,2-2\epsilon,\frac{2p_1\cdot p_2}{-p_2^2}\right].
\end{align}
In the last step we have set $d=4-2\epsilon$. One may expose the non-analiticities in the argument of the hypergeometric function by usual means. For example, when $p_1=p$, $p_2=q$, the hypergeometric function is evaluated at $1/x$, which for $x\in(0,1)$ introduces non-analiticities:
\begin{align}
(2\pi)^d(-p^2)^{-\epsilon}T_c(p,q)=&\frac{\imath\pi^{d/2+1}\csc(-\epsilon\pi)\Gamma(1-\epsilon)}{(-q^2)\Gamma(2-2\epsilon)}(-p^2)^{-\epsilon}{}_2F_1\left[1,1-\epsilon,2-2\epsilon,\frac{1}{x}\right] \nonumber\\
=&-\frac{\imath\pi^{d/2+2}\csc(-\epsilon\pi)^2}{(-q^2)\Gamma(1 - \epsilon)}(-p^2)^{-\epsilon}x\Bigg[(x-1)^{-\epsilon} x^{-\epsilon} \nonumber \\
&+\frac{\Gamma(1-\epsilon)}{\Gamma(1-2\epsilon)\Gamma(1+\epsilon)}{}_2F_1\left[1,2\epsilon,1+\epsilon,1-x\right] \Bigg]\label{eq:hypergeometric_anal}.
\end{align}
Now the result is expressed in terms of a hypergeometric function that is analytic for $x\in(0,1)$.

\section{Double discontinuity of two-loop three-point master integrals at leading virtuality}
\label{sec:masters2loop}

One direct way of computing the iterated discontinuities of the two-loop three-point master integrals is by expressing them in terms of the different unitarity cuts of their underlying diagrams. To this end, let us define the following four cut integrals
\begin{align}
F_1(\alpha_1,\alpha_2)&=\int \frac{\mathrm{d}^d k \mathrm{d}^d l}{\pi^d} \delta^-(k) \delta^-(l) \frac{\delta^+(l-p_3)\delta^+(k+p_1)}{[(k+l+p_1)^2]^{\alpha_1}[(k+l-p_3)^2]^{\alpha_2}}, \\
F_2(\alpha_1,\alpha_2)&=\int \frac{\mathrm{d}^d k \mathrm{d}^d l}{\pi^d} \delta^-(k)\delta^-(l)  \frac{\delta^+(l-p_3)\delta^+(k+l+p_1)}{[(k+p_1)^2]^{\alpha_1}[(k+l-p_3)^2]^{\alpha_2}}, \\
F_3(\alpha_1,\alpha_2)&=\int \frac{\mathrm{d}^d k \mathrm{d}^d l}{\pi^d} \delta^-(k) \delta^-(l) \frac{\delta^+(k+l-p_3)\delta^+(k+l+p_1)}{[(k+p_1)^2]^{\alpha_1}[(l-p_3)^2]^{\alpha_2}}, \\
F_4(\alpha_1,\alpha_2)&=\int \frac{\mathrm{d}^d k \mathrm{d}^d l}{\pi^d} \delta^-(k)\delta^-(l)\frac{\delta^+(k+l-p_3)\delta^+(k+p_1)}{[(k+l+p_1)^2]^{\alpha_1}[(l-p_3)^2]^{\alpha_2}}.
\end{align}
where $\delta^\pm(q)=\delta(q^2)\Theta(\pm q^0)$. The double discontinuity of any of the three-point master integrals can be expressed as a linear combination of these four functions. In particular:
\begin{align}
\text{disc}_{p_1^2}&\text{disc}_{p_3^2}\raisebox{-1.1cm}{\scalebox{0.75}{\input{Submission/diagrams/appendix_masters/non_planar}}}=\raisebox{-1.1cm}{\scalebox{0.75}{\input{Submission/diagrams/appendix_masters/non_planar_cut2}}}+2\raisebox{-1.1cm}{\scalebox{0.75}{\input{Submission/diagrams/appendix_masters/non_planar_cut1}}}+2\raisebox{-1.1cm}{\scalebox{0.75}{\input{Submission/diagrams/appendix_masters/non_planar_cut3}}}\\
&+2\raisebox{-1.1cm}{\scalebox{0.75}{\input{Submission/diagrams/appendix_masters/non_planar_cut4}}} =F_1(1,1)+2(F_2(1,1)+F_3(1,1)+F_4(1,1)) \label{eq:disc1}.
\end{align}
\begin{align}
\text{disc}_{p_1^2}\text{disc}_{p_3^2}\raisebox{-1.1cm}{\scalebox{0.75}{\input{Submission/diagrams/appendix_masters/kite1}}}&=\raisebox{-1.1cm}{\scalebox{0.75}{\input{Submission/diagrams/appendix_masters/kite1_cut1}}}+\raisebox{-1.1cm}{\scalebox{0.75}{\input{Submission/diagrams/appendix_masters/kite1_cut2}}}
=F_2(0,1)+F_3(0,1) \label{eq:disc2}.
\end{align}
\begin{align}
\text{disc}_{p_1^2}\text{disc}_{p_3^2}\raisebox{-1.1cm}{\scalebox{0.75}{\input{Submission/diagrams/appendix_masters/kite2}}}&=\raisebox{-1.1cm}{\scalebox{0.75}{\input{Submission/diagrams/appendix_masters/kite2_cut2}}}+\raisebox{-1.1cm}{\scalebox{0.75}{\input{Submission/diagrams/appendix_masters/kite2_cut1}}}+\raisebox{-1.1cm}{\scalebox{0.75}{\input{Submission/diagrams/appendix_masters/kite2_cut3}}} \\
&=F_1(1,0)+F_2(1,0)+F_4(0,1). \label{eq:disc3}
\end{align}
\begin{align}
\text{disc}_{p_1^2}\text{disc}_{p_3^2}\raisebox{-1.1cm}{\scalebox{0.75}{\input{Submission/diagrams/appendix_masters/dunce_hat1}}}=\raisebox{-1.1cm}{\scalebox{0.75}{\input{Submission/diagrams/appendix_masters/dunce_hat1_cut1}}}=F_2(0,0).
\end{align}
\begin{align}
\text{disc}_{p_1^2}\text{disc}_{p_3^2}\raisebox{-1.1cm}{\scalebox{0.75}{\input{Submission/diagrams/appendix_masters/dunce_hat2}}}=\raisebox{-1.1cm}{\scalebox{0.75}{\input{Submission/diagrams/appendix_masters/dunce_hat2_cut1}}}=F_4(0,0).
\end{align}
One important check that this construction implies is pole cancellation. Discontinuities are finite as long as all the invariants are not zero, and thus
\begin{align}
F_1(1,1)+2(F_2(1,1)+F_3(1,1)+F_4(1,1))=\mathcal{O}(\epsilon^0), \\
F_2(0,1)+F_3(0,1)=\mathcal{O}(\epsilon^0), \\
F_1(1,0)+F_2(1,0)+F_4(0,1)=\mathcal{O}(\epsilon^0), \\
F_2(0,0),F_4(0,0)=\mathcal{O}(\epsilon^0).
\end{align}
We performed the integration at leading $p_1^2$ virtuality of the $F_i(\alpha_1,\alpha_2)$ functions:
\begin{align}
F_1(\alpha_1,\alpha_2)
&=\frac{(p_3^2)^{\frac{d-4}{2}}(p_1^2)^{\frac{d-4}{2}}}{\pi^2 (-2p_1\cdot p_3)^{\alpha_1+\alpha_2}}\left[\frac{\beta\left(\frac{d-2}{2}-\alpha_1,\frac{d-2}{2}-\alpha_2\right)}{\Gamma(d/2-1)}\right]^2,
\end{align}
\begin{align}
&F_2(\alpha_1,\alpha_2)=\frac{(p_3^2)^{d-3-\alpha_1}(p_1^2)^{d-3-\alpha_1}}{\pi(-2 p_1\cdot p_3)^{d-2+\alpha_2-\alpha_1}}\left[\frac{\beta\left(\frac{d-2}{2},\frac{d-2}{2}-\alpha_1\right)}{\Gamma\left(\frac{d}{2}-1\right)}\right]^2\csc\left(\alpha_1 \pi - \frac{d \pi}{
   2}\right)\times\nonumber\\
   &\times 
   \Bigg(-\frac{
    \Gamma\left(-1 + \frac{d}{2}\right) \phantom{}_2F_1\left[1 - \alpha_1, 
      2 + \alpha_2 - \frac{d}{2}, -\alpha_1 + \frac{d}{2}, \frac{p_3^2}{-2p_1\cdot p_3}\right]}{
    \Gamma(\alpha_1)\Gamma\left(- \alpha_1 + \frac{d}{2}\right)}+\left[\frac{-2p_1\cdot p_3}{p_3^2}\right]^{d/2-\alpha_1-1}\times \nonumber\\
    &\times\frac{
     \Gamma\left(-1 - \alpha_2 + \frac{d}{2}\right) \phantom{}_2F_1\left[
     3 + \alpha_1 + \alpha_2 - d, 2 - \frac{d}{2}, 2 + \alpha_1 - \frac{d}{2}, \frac{p_3^2}{-2p_1\cdot p_3}\right]}{
   \Gamma(-2 - \alpha_1 - \alpha_2 + d)\Gamma\left(2 + \alpha_1 - \frac{d}{2}\right)}\Bigg),
\end{align}
\begin{align}
&F_3(\alpha_1,\alpha_2)=-\frac{(-p_2^2)^{\frac{d-4}{2}}(p_3^2)^{d-3-\alpha_2}(p_1^2)^{d-3-\alpha_1}}{\pi(-2p_1\cdot p_3)^{\frac{3}{2}d-4}}\left[\frac{\beta\left(\frac{d-2}{2}, \frac{d-2}{2}-\alpha_1\right)}{\Gamma\left(\frac{d}{2}-1\right)}\right]^2e^{i \left(1 + \alpha_2 - \frac{d}{2}\right) \pi}\times \nonumber\\
&\times   \left(-i + \cot\left(\alpha_2 \pi - \frac{d \pi}{2}\right)\right)\frac{\Gamma\left[-1 + \frac{d}{2}\right]\phantom{}_2F_1\left[\alpha_1, 
   \frac{d}{2}-1-\alpha_2, -2 - \alpha_2 + d,\frac{p_3^2}{(-2p_1\cdot p_3)}\right]}{\Gamma\left[2 + \alpha_2 - \frac{d}{2}\right] \Gamma[-2 - \alpha_2 + d]},
\end{align}
\begin{align}
&F_4(\alpha_1,\alpha_2)=\frac{(p_3^2)^{d-3-\alpha_2}(p_1^2)^{\frac{d-4}{2}}}{\pi(-2p_1\cdot p_3)^{\frac{d}{2}-1+\alpha_1}}\left[\frac{\beta\left(\frac{d-2}{2}-\alpha_1,\frac{d-2}{2}-\alpha_2\right)}{\Gamma\left(\frac{d}{2}-1\right)}\right]^2 \left(-i + \cot\left(\alpha_2 \pi - \frac{d \pi}{2}\right)\right)\times \nonumber\\
&\times e^{i \left(1 + \alpha_2 - \frac{d}{2}\right) \pi}   \frac{\Gamma\left[-1 + \frac{d}{2}\right]\phantom{}_2F_1\left[2+\alpha_1-\frac{d}{2}, 
   \frac{d}{2}-1-\alpha_2, -2 - \alpha_2 + d,\frac{p_3^2}{(-2p_1\cdot p_3)}\right]}{\Gamma\left[2 + \alpha_2 - \frac{d}{2}\right] \Gamma[-2 - \alpha_2 + d]}.
   \end{align}
All four results truncate higher orders in $p_1^2$.
\section{Embedding tables and description of supplementary material}

\label{sec:emb_tables}

In this appendix we tabulate all the embeddings that have been computed in order to obtain the next-to-leading order structure functions for deep inelastic scattering. Each row corresponds to an embedding of a vacuum diagram. The first cell of each row provides the winding numbers with respect to a reference cycle basis. In other words, it allows to reconstruct the tuple $(G,\mathbf{w})$ belonging to $\embdis$. The chosen reference cycle bases for the two vacuum graphs are:

\begin{align*}
\mathcal{D}_{\text{DT}}&=\{d_1,d_2,d_3,d_4\}=\left\{\raisebox{-9mm}{\includegraphics[]{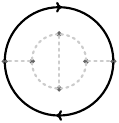}},\raisebox{-9mm}{\includegraphics[]{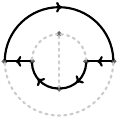}},\raisebox{-9mm}{\includegraphics[]{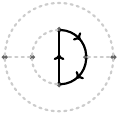}},\raisebox{-9mm}{\includegraphics[]{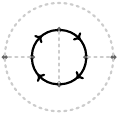}}\right\},\\
\mathcal{D}_{\text{SE}}&=\{c_1,c_2,c_3,c_4\}=\left\{\raisebox{-9mm}{\includegraphics[]{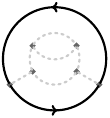}},\raisebox{-9mm}{\includegraphics[]{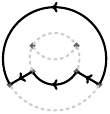}},\raisebox{-9mm}{\includegraphics[]{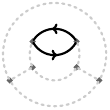}},\raisebox{-9mm}{\includegraphics[]{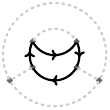}}\right\}
.
\end{align*}

The second and third cell of each row provide one selected forward-scattering diagram compatible with the embedding (namely one choice of rooted cut), with $a_1$ or $a_2$ being the initial-state electron, respectively. In other words, the second and third cell may be mapped to the two possible thruples $\Gamma=(G,\mathbf{w},a_1)$ and $\overline{\Gamma}=(G,\mathbf{w},a_2)$ obtained by supplementing the winding information with a choice of initial-state electron. The forward-scattering diagrams are routed according to the simplest clustering criterion, imposing that the sum of partons momenta in the initial state is $p$. By applying Feynman rules to each forward-scattering diagram (including the cutting rules for the initial-state particles), and promoting all cut propagators to ordinary Feynman propagators, we obtain the integrand of the embedding, $\tilde{I}_{\Gamma}$ and $\tilde{I}_{\overline{\Gamma}}$. The loop integration variable which in the main text is labelled as $k$ is, in the drawings, labelled by either $k_2$ or $k_3$, depending on the embedding.

We also provide the embedding integrand information in two separate \texttt{.m} files as ancillary material. Each file contains a list of embedding relating to one of the two vacuum graphs. Say they have been assigned the names 
\begin{center}
\texttt{dataSE=Import["path\_to\_file/dataSE.m"];} \\
\texttt{dataDT=Import["path\_to\_file/dataDT.m"];}
\end{center}
Each element of the list is an association list, corresponding to a defined embedding $(G,\mathbf{w})$. For example, calling 
\begin{center}
\texttt{dataSE[[1]][["embedding"]]} 
\end{center}
gives
\begin{center}
\texttt{ \{1,0,-1,1\} }
\end{center}
as output. Thus \texttt{dataSE[[1]]} corresponds to the embedding of the self-energy vacuum diagram with winding number $\mathbf{w}=(1,0,-1,1)^T$. The choices of initial-state electron $\Gamma$ and $\bar{\Gamma}$ can be obtained by instead calling 
\begin{center}
\texttt{dataSE[[1]][["a1"]]} 
\end{center}
and 
\begin{center}
\texttt{dataSE[[1]][["a2"]]}
\end{center} 
which yield two separate association tables (the entry is not present when there is no rooted cut for that embedding cutting the selected electron and satisfying $r_e\le 1$ for all $e$). In particular, 
\begin{center}
\texttt{dataSE[[1]][["a1"]][["Denominator"]]/.{k[x\_]:>k}}
\end{center}
 gives the propagator structure of $I_\Gamma$ as a product of \texttt{prop[m,p]}$=\frac{1}{p^2-m^2}$, 
 \begin{center}
\texttt{dataSE[[1]][["a1"]][["Numerator"]][["F2"]]/.{k[x\_]:>k}}\\
\texttt{dataSE[[1]][["a1"]][["Numerator"]][["FL"]]/.{k[x\_]:>k}}
\end{center}
  give the numerator of $I_\Gamma$ projected on the structure functions $F_2$ and $F_L$ (\texttt{G} and \texttt{eq} being the strong and electric coupling constants, \texttt{Nc} the number of colours, \texttt{TF} the normalisation of the $T^a$ matrices and \texttt{dim}$=d=4-2\epsilon$ the dimensional regulator) and 
  \begin{center}
  \texttt{dataSE[[1]][["a1"]][["Numerator"]][["zero"]]/.{k[x\_]:>k}}
  \end{center}
  gives the numerator of $I_\Gamma$ contracted with $q^\mu q^\nu$. The substitution rule \texttt{/.{k[x\_]:>k}} makes sure that in both propagators and numerators, the integration variable is labelled by \texttt{k}, instead of \texttt{k[2]} or \texttt{k[3]} depending on the embedding. Finally
  \begin{center}
  \texttt{dataSE[[1]][["a1"]][["Cut content"]]}
  \end{center}
  gives the cut content $\q_\Gamma$ of the embedding.

\newpage

\foreach \x in {1,...,6}
{%
\includepdf[pages={\x}]{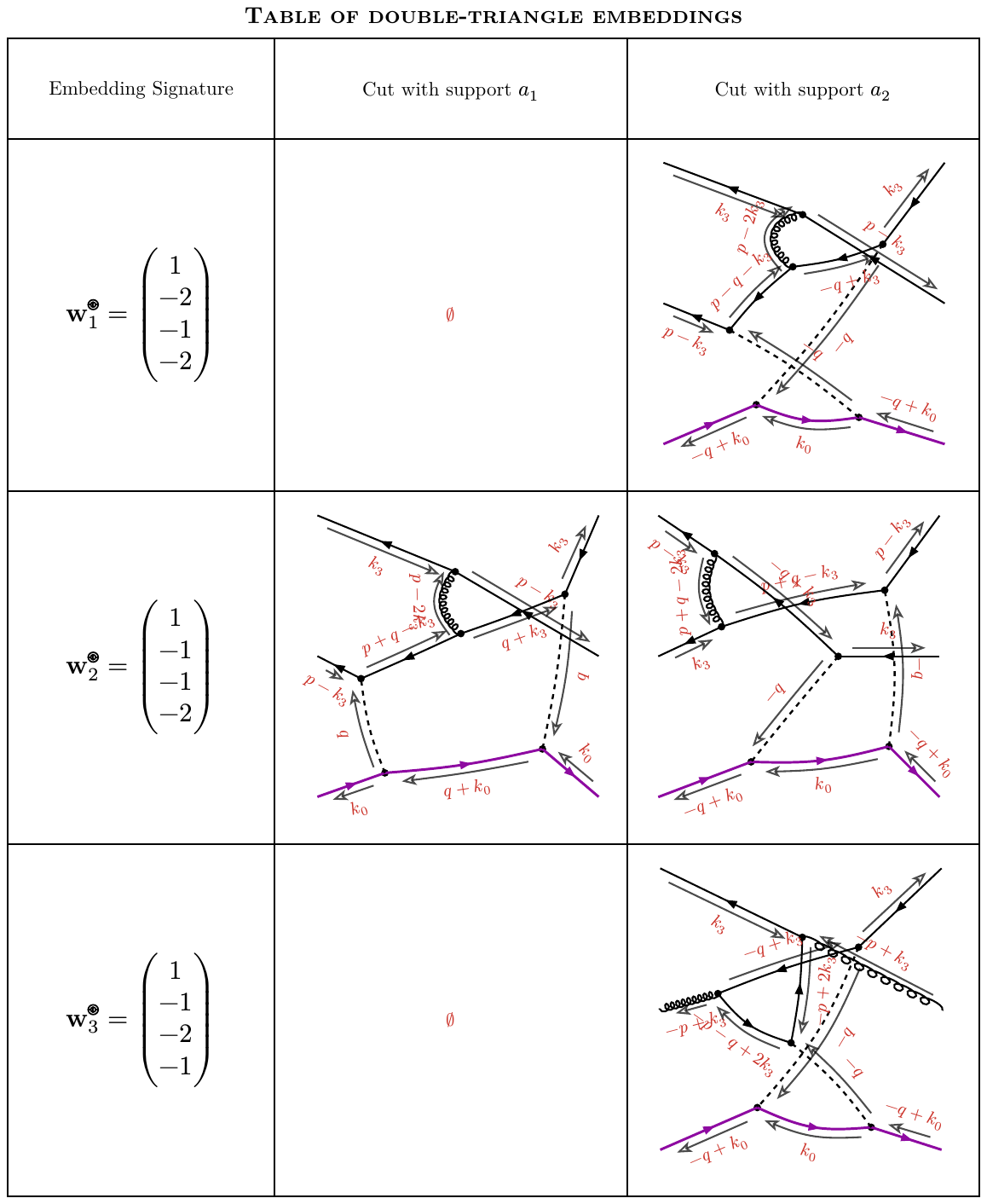}
}
\foreach \x in {1,...,9}
{%
\includepdf[pages={\x}]{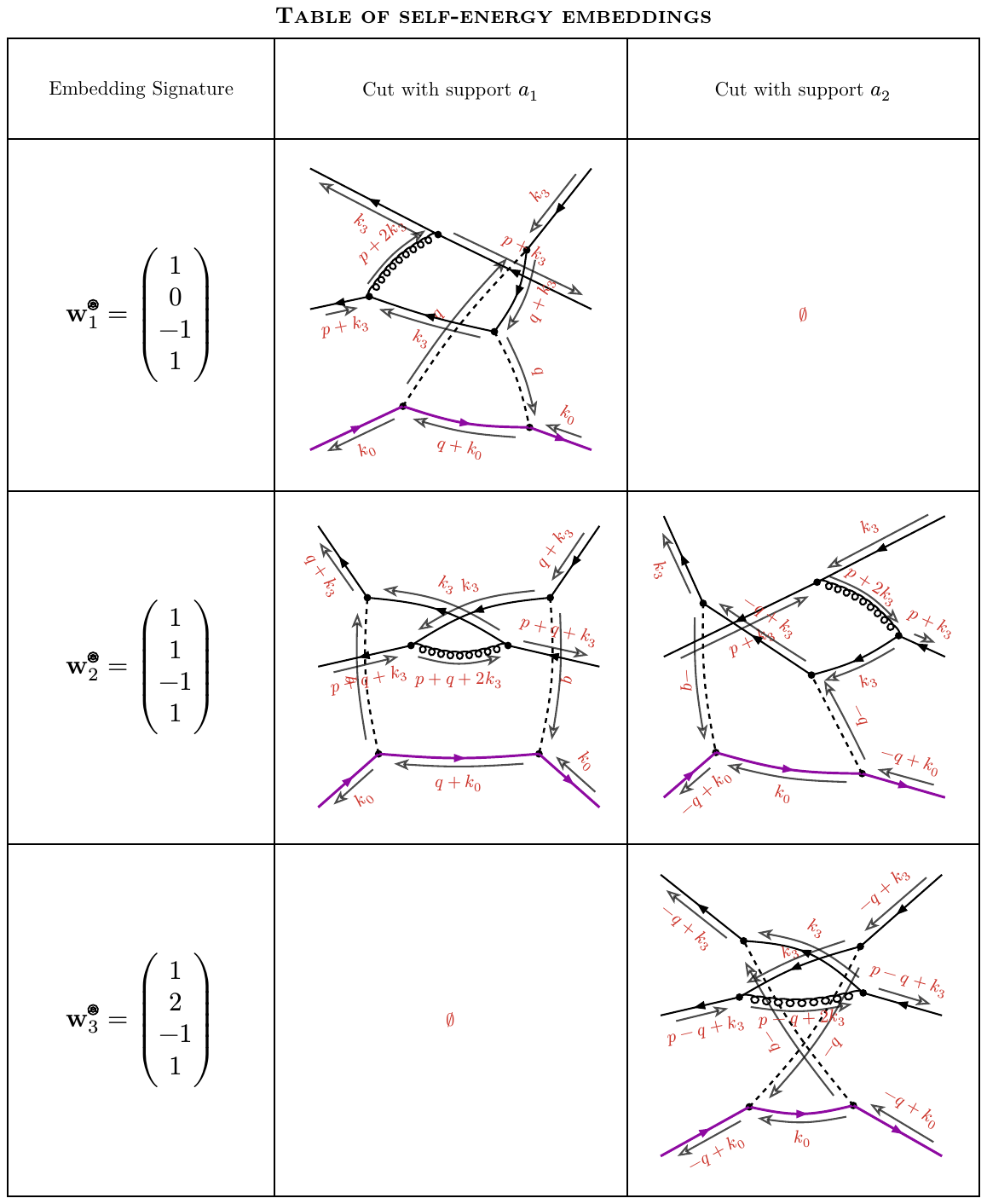}
}

\bibliography{Submission/biblio}

\end{document}